\documentclass[amsmath,amssymb,reprint,superscriptaddress,floatfix,nofootinbib]{revtex4-2}
\usepackage{graphicx}
\usepackage{enumitem}
\usepackage{changepage} 
\usepackage{multirow} 
\usepackage{hyperref}
\usepackage[normalem]{ulem}
\usepackage{float}
\usepackage{soul} 
\usepackage{amsmath} 
\usepackage{orcidlink}
\usepackage[normalem]{ulem}

\usepackage{url}
\usepackage{booktabs}
\usepackage{siunitx}

\usepackage[capitalise,noabbrev,nameinlink]{cleveref}
\hypersetup{
citecolor=blue,
 colorlinks=true,
 linkcolor=blue,
 filecolor=magenta,
 urlcolor=blue,
}

\Crefname{equation}{Eq.}{Eqs.}

\begin{document}

\title{
Building Neutron Stars with the MUSES Calculation Engine
}

\author{Mateus Reinke Pelicer\,\orcidlink{0000-0002-2189-706X}}
\affiliation{Center for Nuclear Research, Department of Physics, Kent State University, Kent, OH 44243, USA}
\author{Nikolas Cruz-Camacho\,\orcidlink{0009-0004-7870-0039}}
\affiliation{The Grainger College of Engineering, Illinois Center for Advanced Studies of the Universe, Department of Physics, University of Illinois Urbana-Champaign, Urbana, IL 61801, USA}
\author{Carlos Conde\,\orcidlink{0000-0002-5393-0565}}
\affiliation{The Grainger College of Engineering, Illinois Center for Advanced Studies of the Universe, Department of Physics, University of Illinois Urbana-Champaign, Urbana, IL 61801, USA}
\author{David Friedenberg\,\orcidlink{0009-0008-6766-5169}}
\affiliation{Cyclotron Institute, Texas A\&M University, College Station, TX 77843, USA}
\affiliation{Department of Physics and Astronomy, Texas A\&M University, College Station, TX 77843, USA}
\author{Satyajit Roy\,\orcidlink{0009-0007-5650-2308}}
\affiliation{Department of Physics and Astronomy, University of Tennessee,
Knoxville, TN 37996, USA}
\author{Ziyuan Zhang\,\orcidlink{0000-0003-4795-0882}}
\affiliation{Physics Department, Washington University in Saint Louis, 63130 Saint Louis, MO, USA}
\affiliation{McDonnell Center for the Space Sciences, Washington University in Saint Louis, Saint Louis, MO 63130, USA}
\author{T. Andrew Manning\,\orcidlink{0000-0003-2545-9195}}
\affiliation{National Center for Supercomputing Applications, University of Illinois Urbana-Champaign, Urbana, IL 61801, USA}
%

\author{Mark G.~Alford\,\orcidlink{0000-0001-9675-7005}}
\affiliation{Physics Department, Washington University in Saint Louis, 63130 Saint Louis, MO, USA}
\author{Alexander Clevinger\,\orcidlink{0000-0001-6478-7066}}
\affiliation{Center for Nuclear Research, Department of Physics, Kent State University, Kent, OH 44243, USA}
\author{Joaquin Grefa\,\orcidlink{0000-0001-7590-9364}}
\affiliation{Center for Nuclear Research, Department of Physics, Kent State University, Kent, OH 44243, USA}
\affiliation{Department of Physics, University of Houston, Houston, TX 77204, USA}
\author{Roland Haas\,\orcidlink{0000-0003-1424-6178}}
\affiliation{National Center for Supercomputing Applications, University of Illinois Urbana-Champaign, Urbana, IL 61801, USA}
\author{Alexander Haber\,\orcidlink{0000-0002-5511-9565}}
\affiliation{Physics Department, Washington University in Saint Louis, 63130 Saint Louis, MO, USA}
\author{Mauricio Hippert\,\orcidlink{0000-0001-5802-3908}}
\affiliation{Rio de Janeiro State University, Rio de Janeiro, RJ, 20550-013, BR}
\author{Jeremy W. Holt\,\orcidlink{0000-0003-4373-3856}}
\affiliation{Cyclotron Institute, Texas A\&M University, College Station, TX 77843, USA}
\affiliation{Department of Physics and Astronomy, Texas A\&M University, College Station, TX 77843, USA}
\author{Johannes Jahan\,\orcidlink{0009-0002-4557-4652}} 
\affiliation{Department of Physics, University of Houston, Houston, TX 77204, USA}
\author{Micheal Kahangirwe\,\orcidlink{0000-0001-9144-6240}}
\affiliation{Department of Physics, University of Houston, Houston, TX 77204, USA}
\author{Rajesh Kumar\,\orcidlink{0000-0003-2746-3956}}
\affiliation{Center for Nuclear Research, Department of Physics, Kent State University, Kent, OH 44243, USA}
\author{Jeffrey Peterson\,\orcidlink{0000-0002-6703-418X}}
\affiliation{Western Wyoming Community College, Rock Springs, WY 82901, USA}
\author{Hitansh Shah\,\orcidlink{0009-0008-1870-3157}}
\affiliation{Department of Physics, University of Houston, Houston, TX 77204, USA}
\author{Andrew W. Steiner\,\orcidlink{0000-0003-2478-4017}}
\affiliation{Department of Physics and Astronomy, University of Tennessee,
Knoxville, TN 37996, USA}
\affiliation{Physics Division, Oak Ridge National Laboratory, Oak Ridge, TN 37831, USA}
\author{Hung Tan\,\orcidlink{0000-0001-9101-048X}}
\affiliation{Syracuse University, Syracuse, NY, 13244, USA}
\author{Yumu Yang\,\orcidlink{0009-0001-8979-9343}}
\affiliation{The Grainger College of Engineering, Illinois Center for Advanced Studies of the Universe, Department of Physics, University of Illinois Urbana-Champaign, Urbana, IL 61801, USA}
\author{Volodymyr Vovchenko\,\orcidlink{0000-0002-2189-4766}}
\affiliation{Department of Physics, University of Houston, Houston, TX 77204, USA}
%
%
\author{Veronica Dexheimer\,\orcidlink{0000-0001-5578-2626}}
\affiliation{Center for Nuclear Research, Department of Physics, Kent State University, Kent, OH 44243, USA}
\author{Jorge Noronha\, \orcidlink{0000-0002-9817-0272}}
\affiliation{The Grainger College of Engineering, Illinois Center for Advanced Studies of the Universe, Department of Physics, University of Illinois Urbana-Champaign, Urbana, IL 61801, USA}
\author{Jaquelyn Noronha-Hostler\,\orcidlink{0000-0003-3229-4958}}
\affiliation{The Grainger College of Engineering, Illinois Center for Advanced Studies of the Universe, Department of Physics, University of Illinois Urbana-Champaign, Urbana, IL 61801, USA}
\author{Claudia Ratti\,\orcidlink{0000-0002-8335-567X}}
\affiliation{Department of Physics, University of Houston, Houston, TX 77204, USA}
\author{Nicol\'as Yunes\,\orcidlink{0000-0001-6147-1736}}
\affiliation{The Grainger College of Engineering, Illinois Center for Advanced Studies of the Universe, Department of Physics, University of Illinois Urbana-Champaign, Urbana, IL 61801, USA}

\begin{abstract}
Exploring the equation of state of dense matter is an essential part of interpreting the observable properties of neutron stars. We present here the first results for dense matter in the zero-temperature limit generated by the MUSES Calculation Engine, a composable workflow management system that orchestrates calculation and data processing stages comprising a collection of software modules designed within the MUSES framework.
The modules presented in this work calculate equations of state using algorithms spanning three different theories/models: (1) Crust Density Functional Theory, valid starting at low densities, (2) Chiral Effective Field Theory, valid around saturation density, and (3) the Chiral Mean Field model, valid beyond saturation density. 
Lepton contributions are added through the Lepton module to each equation of state, ensuring charge neutrality and the possibility of $\beta$-equilibrium.
Using the Synthesis module, we match the three equations of state using different thermodynamic variables and different methods. We then couple the complete equation of state to a novel full-general-relativity solver (QLIMR) module that calculates neutron star properties. 
We find that the matching performed using different thermodynamic variables affects differently the range obtained for neutron star masses and radii (although never beyond a few percent difference). We also investigate the universality of equation of state-independent relations for our matched stars. Finally, for the first time, we use the Flavor Equilibration module to estimate bulk viscosity and flavor relaxation charge fraction and rates (at low temperature) for Chiral Effective Field Theory and the Chiral Mean Field model.
\end{abstract}

\maketitle

\section{Motivation for MUSES}
\label{sec:MUSESmotivation}

Neutron stars are composed of multiple layers wherein different degrees of freedom and interactions dominate within a given layer \cite{Baym:2017whm}. 
The outer layers have low baryon number densities $n_B$, whereas the very core of neutron stars has extremely large $n_B$ that may be many times larger than the density of a nucleus (saturation density is approximately $n_{sat}\sim 0.16$ fm$^{-3}$).
The outer layers (referred to as the crust) are described using low-energy nuclear physics to correctly capture the properties of nuclei (below and above the neutron drip line) and free neutrons \cite{harrisonwheeler,Baym:1971pw}. The intermediate layers (or outer core) consist of interacting nucleons (protons and neutrons) \cite{NeutronSuperfluid} around saturation density $n_{sat}$ and include the nuclear liquid-gas phase transition.
At large enough $n_B\gtrsim 2\,n_{sat}$ that are found within the inner core of  neutron stars, it may be possible for strange baryons (hyperons) and resonances to appear, or even for quarks to become deconfined \cite{Ivanenko:1965dg,Ivanenko:1969gs,Collins:1974ky}. 
The nuclei/baryon/quark positive charge within the various layers of the neutron star (predominantly from protons) is  counterbalanced by leptons, fulfilling charge neutrality. Isolated neutron stars quickly reach chemical equilibrium, as they cool down and neutrinos escape within minutes of the supernova explosions that create them \cite{Burrows:1986me,Janka:2012wk}.

An important quantity for understanding the properties of neutron stars is the fraction of the (net-)electric charge density from nuclei, baryons, and quarks  $n_Q$ per baryon density $n_B$, i.e., $Y_Q=n_Q/n_B$, otherwise known as the charge fraction.
$Y_Q$ is also important for nuclear experiments, but in that case it is defined as the fraction of protons $Z$ over total nucleons $A$, i.e., $Y_Q=Z/A$.
The value of $Y_Q(n_B)$ (determined by the relevant nuclei/baryons/quarks\footnote{One can also define charge fraction as a ratio of quantum numbers, $Y_Q=Q/B$ where Q is the total electric charge of the system that does not include leptons and $B$ is the total number of baryons (where each quark contributes as $1/3$). We can imagine a system with nuclei and free nucleons such that $Q$ includes both the protons within a nucleus and the free protons, whereas $B$ includes all the nucleons both within and outside nuclei.} at that specific $n_B$) varies across the layers of a neutron star and is determined by $\beta$-equilibrium (where it is assumed that neutrinos are not trapped).

The strong force  governs the interactions between nuclei, nucleons, other baryons, and quarks, and is described by quantum chromodynamics (QCD). 
QCD cannot be calculated from first principles in the regime relevant for neutron stars \cite{Philipsen:2012nu} and, thus, either effective field theories or microscopic effective models are required to understand dense matter under these conditions. 
Because of the widely different degrees of freedom and interactions across the different layers of neutron stars, a single, uniform equation of state (EoS) approach does not yet exist. Instead, different regions inside neutron stars (sometimes a single layer or possibly multiple layers) are described with different and separate EoSs, and then matched together where their regimes of validity overlap. 

A major roadblock in understanding the neutron star EoS is that there are many different models, and each one of them has underlying free parameters that can significantly affect their predictions for neutron star properties. 
A single rendering of an EoS from tuned physical parameters cannot cover the entire allowable phase space for a specific model. 
Furthermore, many software packages that calculate EoSs are proprietary, such that they are not available to the wider public. Even when they are open source, they may not be user-friendly and/or they may only describe a small region of the star.
One approach to begin to tackle these challenges is to store EoS tables in large repositories. The largest one for astrophysics is called CompOSE \cite{Oertel:2016bki,Typel:2013rza,CompOSECoreTeam:2022ddl,compose}. CompOSE contains over 300 EoSs that cover different regions of the QCD phase diagram relevant to neutron stars. 
CompOSE has been very important for pushing forward the dense matter astrophysics field, giving a wider range of physicists across fields access to a broad range of neutron star EoSs using one standardized output \cite{Dexheimer:2022qhn}. 
However, because CompOSE is a database and does not include the EoS models themselves, one cannot vary free parameters within EoSs on CompOSE, nor can one edit or modify the EoS models, as new data furthers our knowledge. 

Furthermore, once EoSs have been obtained (regardless of the source), they need to be matched together across their regime of validity in order to describe entire neutron stars. 
Most importantly, EoS matching must be done in such a way as to preserve thermodynamic consistency, thermodynamic stability, and causality, i.e., the speed of sound squared is bounded between zero and the speed of light; see e.g.~App.~E of \cite{Cruz-Camacho:2024odu} for more details. 
Two main approaches are commonly used for bridging the different EoSs: a first-order phase transition or a smooth matching. 
Finally, the complete EoS, spanning from the crust to the core of neutron stars, is used to close Einstein's field equations and extract their macroscopic properties. For spherically-symmetric
and static stars, the Tolman–Oppenheimer–Volkoff (TOV) \cite{Tolman:1939jz,Oppenheimer:1939ne} equations are solved to generate mass-radius diagrams. Furthermore, the tidal deformability, an important property for gravitational wave studies, is obtained by analyzing the star's response to an external static tidal field through perturbation theory \cite{Hinderer:2007mb, Flanagan:2007ix}. Modeling rotating stars is more complex due to the interplay of rotation and gravity. While full numerical relativity simulations can accurately describe the behavior of rapidly rotating stars, these methods are computationally intensive. In this project, we use the Hartle–Thorne perturbative approach, which employs a slow-rotation expansion to estimate rotational properties with sufficient accuracy \cite{Hartle:1967he, Hartle:1968si} to describe neutron stars.

In this paper, we introduce the MUSES (Modular Unified Solver of the Equation of State) Calculation Engine, built to model the EoS across the QCD phase diagram (including inside neutron stars and during heavily ion collisions) with a variety of relevant models. The Calculation Engine (CE) is modular because it allows for a multitude of EoS descriptions (EoS modules) that can be easily exchanged with one another or with external input tables, and for numerous calculations of observables (Observable modules), given a unified EoS. 
The CE is unified because it combines the various modules to cover the entire QCD phase diagram from heavy-ion collisions to neutron stars. 
The EoS modules that currently make up the MUSES CE are the following: 
Crust Density Functional Theory (Crust-DFT, valid starting at low densities and at neutron star temperatures), 
Chiral Effective Field Theory ($\chi$EFT, valid around saturation density and at zero temperatures in the current version), 
Chiral Mean Field (CMF, valid beyond saturation density and at zero temperatures in the current version), 
Lepton (valid at any densities, but zero temperatures in the current version), 
4D Taylor-expanded Lattice (BQS, valid at low densities and high temperatures), 
Ising 2D $T'$-Expansion Scheme (TExS, valid at low densities and high temperatures), 
and Holographic (NumRelHolo, valid across a wide range of densities, but high temperatures). 
Each of these modules provides a prescription for the EoS, either for cold neutron stars (first 4 modules) or for heavy-ion collisions (last 3 modules), in a particular region of the QCD phase diagram. The EoSs generated by these modules (or external tables) are then combined to provide a unified description for a particular application in a user-defined region of the QCD phase diagram within the Synthesis module. The Observable modules that make up the MUSES CE are the following: 
Quadrupole moment, Tidal Love number, Moment of Inertia, Mass, and Radius  (QLIMR, valid for the calculation of stationary, axisymmetric and slowly-rotating neutron star observables in equilibrium), and Flavor Equilibration (Flavor Eq, valid for the calculation of out-of-equilibrium neutron star observables). Each of these modules acts on the unified EoS created by the synthesized EoS modules to calculate certain observables, like mass-radius curves, tidal deformabilities, bulk viscosities, and relaxation times. 

For the neutron-star EoS modules, we focus here on cold neutron stars ($T\sim0$ in the MeV scale) in $\beta$-equilibrium, which are described by a simple 1-dimensional (1D) EoS that can be used to calculate relevant thermodynamic properties as a function of $n_B$ or baryon chemical potential, $\mu_B$. In the neutron star crust, different nuclei dominate, reproducing nearly isospin symmetric matter (or, in terms of the charge fraction, $Y_Q\sim0.5$). In the dense regime inside neutron star core, matter possesses a large isospin asymmetry (related to large, negative charge chemical potential, $\mu_Q$, reproducing low $Y_Q$) and there are no constraints on the strangeness fraction. While in the crust and outer-core the contributions from protons are expected to dominate the nuclei/baryonic/quark electric charge, in the inner core, hyperons, $\Delta$ baryons, and deconfined quarks can also carry electric charge (of both positive and negative values). Overall, this leads to a positive charge density, $n_Q>0$ that is exactly balanced by the existence of leptons in the system, such that $n_Q^{\rm tot} = n_Q + n_l =0$. Finally, the number of strange particles is not conserved within neutron stars because they live long enough that they can undergo a significant number of weak decays. Because evolved neutron stars have (effectively) vanishing temperatures, no anti-hyperons appear, and therefore the strangeness density is negative, $n_S \leq 0$, because hyperons carry negative strangeness (and strangeness chemical potential $\mu_S=0$ in $\beta$ equilibrium). The MUSES CE models all of this physics through the different models mentioned above (Crust-DFT, $\chi$EFT, and CMF), which are matched smoothly through transition functions. Future work will extend modules that currently only have $T=0$ to
finite temperature, so that they can be used to study neutron star mergers. 

For the heavy-ion EoS modules, we focus on EoSs that reproduce lattice QCD at vanishing densities, the behavior of the quark gluon plasma at high temperatures, and the cross-over phase transition at low $\mu_B$. The modules differ, however, in how they treat phase transitions, critical points, and low temperatures.  The 
BQS module is a 4-dimensional model ($T,\mu_B,\mu_S,\mu_Q$ with baryon, strangeness, and electric charge chemical potentials) that is reconstructed from lattice QCD results with a Taylor series, incorporates hadrons at low temperatures (the hadron resonance gas model), and only includes a cross-over phase transition. The TExS module is a 2-dimensional model ($T,\mu_B$ with $\mu_S=\mu_Q=0$ or along strangeness neutrality and fixed $Y_Q$) that is reconstructed from lattice QCD results coupled to a 3D-Ising model, incorporates hadrons at low temperatures (the hadron resonance gas model), and includes a cross-over phase transition, followed by a QCD critical point and a first-order phase transition as one goes to higher $\mu_B$. 
The NumRelHolo is a 2-dimensional ($T,\mu_B$ with $\mu_S=\mu_Q=0$) holographic model that is tuned to reproduce lattice QCD thermodynamics at vanishing densities, without incorporating hadrons, but including a cross-over phase transition followed (depending on the parameters chosen) by a QCD critical point and first-order phase transition at larger $\mu_B$. 
Unlike in neutron stars, the hadron resonance gas phase includes all known light and strange hadrons and their resonances, which are important to reproduce lattice QCD results across the cross-over phase transition. Future work will include Thermal-FIST \cite{Vovchenko:2019pjl}, a hadron resonance gas model that can be calculated in 4D. At the moment, the heavy-ion collision EoSs are not connected to each other, but after the inclusion of Thermal-FIST they will be connected in 2D, 3D, or 4D, similarly to what is done in this work. 
In this paper, however, we place more emphasis on the neutron-star side of the MUSES CE, and leave a detailed study of the heavy-ion sector to future work. 

We not only develop and describe the MUSES CE in this paper, but we also use it to study physical properties of the resulting neutron stars. We investigate astrophysical properties, like the mass, moment of inertia, quadrupole moment and tidal Love number, and study how the matching procedure affects these observables. That is, we study the influence of different smooth matching procedures used to connect different 1D EoSs on stellar properties. We find that neutron star maximum masses and radii only vary by up to a few percent within a given matching procedure, while (as expected) EoS-independent relations remain insensitive to the matching scheme. We also study, for the first time, the bulk viscosity and the flavor equilibration relaxation time for two different EoSs across $n_B$.

This paper is organized as follows. We begin with an overview of the CE in Sec.~\ref{sec:MUSES_overview}. Then, in Sec.~\ref{sec:ns_eos_modules}, we provide an overview of the neutron star EoS modules of Crust-DFT in Sec.~\ref{sec:crustDFT}, $\chi$EFT in Sec.~\ref{sec:chiEFT}, CMF++ in Sec.~\ref{sec:CMF}, Lepton in Sec.~\ref{sec:lepton}, and Synthesis in Sec.~\ref{sec:synthesis}. In Sec.~\ref{sec:observable_modules} we outline the MUSES observation modules, with QLIMR presented in Sec.~\ref{sec:QLIMRmethods} and Flavor Equilibration in Sec.~\ref{sec:flavor}, where new results for the influence of the MUSES EoS on observables are shown. Finally, we conclude and discuss the implications of this work in Sec.~\ref{sec:conclusions}. Appendices \ref{app:cheft}, \ref{app1}, \ref{appQLIMR} are added to provide further details about this work. We use ``natural'' units throughout, where $\hbar=c=k_B=1$.

\section{Overview of MUSES Software Products}\label{sec:MUSES_overview}

This section provides a brief overview of MUSES software products both to motivate the rest of the paper and to introduce our complete framework for the description of neutron stars.

The MUSES project \cite{MUSES_WEBSITE} has three major software development goals: (1) create independent software modules that calculate EoSs, calculate physical observables, and/or process data; (2) define a software framework for integration and interoperability of these modules (and external tables); and (3) design and construct a management system to orchestrate the execution of composable, multi-module workflows. There is also (4) an effort to support interoperability with existing data formats.

These goals are met by creating a MUSES cyberinfrastructure with the following elements: 
\begin{enumerate}
\item \textbf{MUSES Modules.} 
Independent software packages that encode physics models to perform calculations that generate scientific data, including EoSs and derived physical observables. The description of the equations and the algorithmic design underlying these packages is the focus of this paper.
\item \textbf{MUSES Framework.} A specification and set of requirements for how to package a calculation script and its dependencies, define its programming interface (API), and publish its documentation. The framework is what allows these otherwise independent MUSES modules to interoperate in conformance with the MUSES development goals.  
\item \textbf{MUSES Calculation Engine.} A web application and workflow management system \cite{muses_calculation_engine_2025_v1} that orchestrates the execution of a set of MUSES modules according to an arbitrary directed acyclic graph (DAG) specified by the user. Each execution of a workflow is a ``job'' managed via the RESTful API. Jobs are added to a processing queue and run asynchronously across a pool of worker processes, efficiently leveraging parallelism within the constraints of the workflow's DAG to optimally utilize computing resources. Data can be piped between modules or input from uploaded files, and all generated output files can be downloaded for local analysis. 
\end{enumerate}

At the time of writing this paper, the EoS modules described below can be used in workflows executed by the CE:
\begin{itemize}[leftmargin=*]
\item \textbf{Neutron-star EoS Modules:} 
    \begin{itemize}
    \item \textbf{Crust Density Functional Theory (Crust-DFT)} calculates the EoS of neutrons and protons in equilibrium with an ensemble of nuclei (see \cite{Du:2018vyp,Du:2021rhq} for the theoretical groundwork and \cite{steiner_2025_14714273} for the open source C++ code and \cite{crust_dft_zenodo} for the EoS tables). At low $T$, the EoS can describe the neutron star crust, reproducing the nuclear liquid-gas phase transition. 
    \item \textbf{Chiral Effective Field Theory ($\chi$EFT)} provides an \textit{ab-initio} description of bulk hadronic matter consisting of nucleons applicable to the density region corresponding to the outer core of neutron stars (see \cite{Machleidt:2011zz,Drischler:2021kxf} for the theoretical groundwork, and \cite{zenodo_cheft} for the open source code). The $\chi$EFT EoS is calculated using many-body perturbation theory, being most applicable for $n_B\sim [0.5,1.5]~n_{sat}$, and extendable to $T \sim [0, 30]\ \mathrm{MeV}$ \cite{Wellenhofer:2014hya, Wellenhofer:2015qba}.
    \item \textbf{Chiral Mean Field (CMF++)} provides a description for dense matter, appropriate for the outer and inner core of neutron stars, with the entire baryon octet and decuplet, and light and strange quarks, leading to chiral symmetry restoration and deconfinement across $T$ and $\mu_B$ (see \cite{Dexheimer:2008ax,Dexheimer:2009hi} for the theoretical ground work, \cite{Cruz-Camacho:2024odu} for a detailed study of its properties at $T=0$ including a description of the C++ code, and \cite{zenodo_cmf} for the open source C++ code).
    \item \textbf{Lepton} calculates a free Fermi gas of leptons (electrons, muons, taus). It can be used alone or to fulfill charge neutrality and $\beta$-equilibrium in other EoSs. The module also allows one to include neutrinos and fix the lepton fraction. See \cite{pelicer_2025_14654137} for the open source code.
    \end{itemize}
\item \textbf{Heavy-Ion EoS Modules:}
    \begin{itemize}
    \item \textbf{4D Taylor-expanded lattice (BQS)} calculates the lattice QCD EoS, expanded as a Taylor series in powers of $\mu_i/T$, where $i=B,~Q,~S$ for baryon number, electric charge and strangeness. The three chemical potentials $\mu_i$ and the temperature can be changed independently, but the coverage in $\mu_i/T$ is limited by the range of validity of the lattice Taylor expansion (see \cite{Noronha-Hostler:2019ayj} for details, and \cite{jahan_2025_14639786} for the open source code). 
    \item \textbf{Ising 2D $\mathbf{T'}$-Expansion Scheme (Ising-2DTExS)} generates EoSs based on the novel $T'$-expansion scheme, introduced to obtain the lattice QCD EoS at finite $\mu_B$ with the broadest available coverage in the literature \cite{Borsanyi:2021sxv}, and with the inclusion of a critical point in the 3D Ising model universality class (see \cite{Kahangirwe:2024cny} for the theoretical groundwork). Users can change the location and strength of the critical point, or run it in lattice mode with no critical point (see \cite{kahangirwe_2025_14637802} for the open source code, and \cite{Vovchenko:2021yen} for the description of the Faá di Bruno method used to speed up calculations in the code).
    \item \textbf{Holographic (NumRelHolo)} is based on the gauge/gravity correspondence \cite{Maldacena:1997re}, and uses a bottom-up, non-conformal model in asymptotically anti-de-Sitter spacetime, constrained to reproduce lattice QCD thermodynamics at vanishing $\mu_B$ to predict the EoS across a wide range of $T$ and $\mu_B$, including the location of the QCD critical point and the first-order phase transition line. See \cite{DeWolfe:2010he,Critelli:2017oub, Grefa:2021qvt,Rougemont:2023gfz} for the associated theoretical groundwork, \cite{Hippert:2023bel} for the current capabilities of the open source code, \cite{yang_2025_14695243} for the open source code, and \cite{hippert_2024_13830379} for the open data set. 
    \end{itemize}
\end{itemize}

The above modules must be combined in their overlapping regime of validity, which is implemented through a special module, called 
\textbf{Synthesis}. In particular, this module combines different charge-neutral, $\beta$-equilibrium EoSs from MUSES EoS modules or external sources (converted to our MUSES standard format) smoothly, using different thermodynamical quantities, such as the speed of sound squared, the energy density, and the pressure. The module can also combine EoSs through a first-order phase transition with (Gibbs construction) or without (Maxwell construction) a mixed phase. See \cite{pelicer_2025_14654584} for the open source code.

\begin{figure}
    \centering
    \includegraphics[width=\linewidth,trim={4.5cm 0 4cm 0},clip]{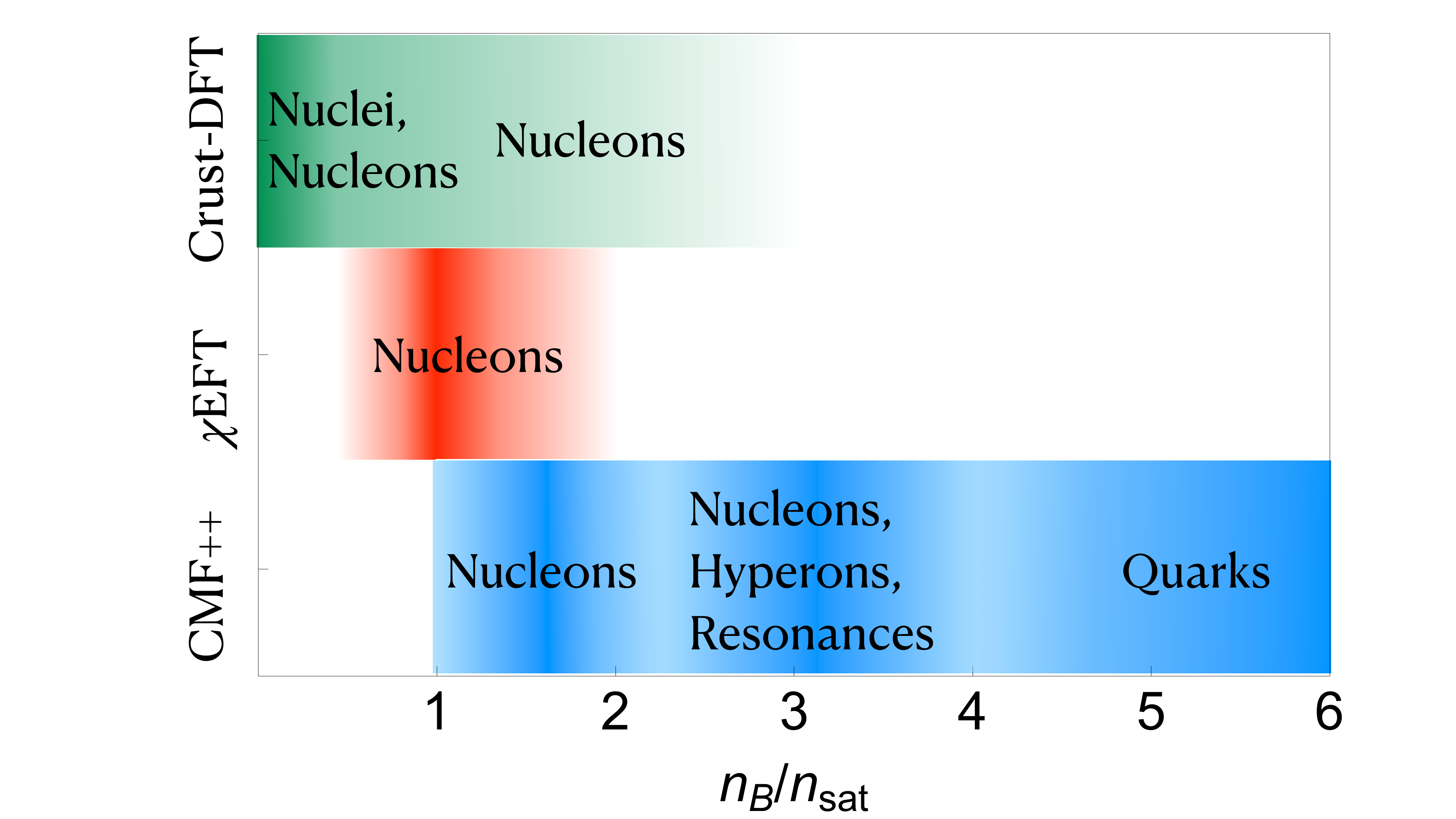} \\ 
\includegraphics[width=1.0\linewidth]{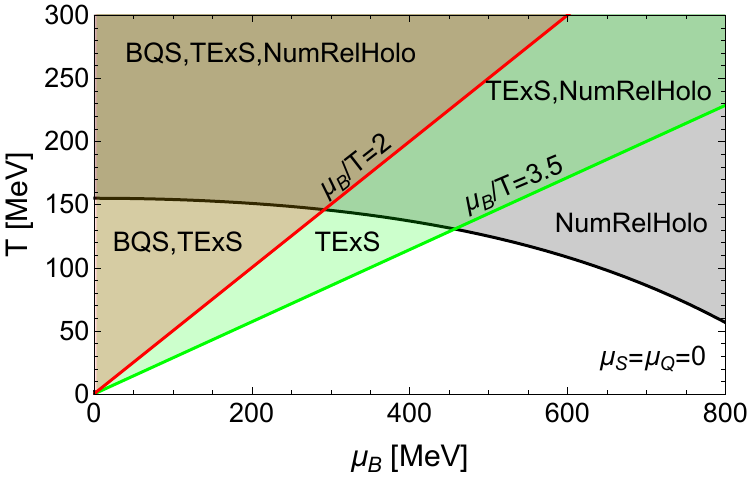}
    \caption{Range of validity for MUSES EoS modules in this paper. The neutron star EoSs are shown in the top panel and the heavy-ion collisions EoSs are shown in the bottom panel. }
    \label{fig:validity}
\end{figure}

Of the EoS modules, Crust-DFT, $\chi$EFT, CMF++, and Lepton can be used to model different layers of neutron stars, while Ising-2DTExS, BQS, and NumRelHolo can be used to describe the properties of the matter formed in ultra-relativistic heavy-ion collisions. 
In \Cref{fig:validity}, we show the estimated regime of validity for our neutron star EoS modules in the top panel and for our heavy-ion collision EoS modules in the bottom panel. 
The neutron star EoSs cover a wide range of $n_B$ relevant from the crust to the core of neutron stars. Due to the uncertainties in the relevant degrees of freedom at a given $n_B$, we cannot draw exact boundaries delimiting where one neutron-star EoS is valid or not, which is why the boundaries in the top figure fade out. 
However, we do know that below $n_{sat}$ and above $n_{drip} \sim 10^{-3}\, n_{sat}$ (i.e.,~$n_{drip} < n_B < n_{sat}$), the layers of neutron stars contain both nuclei and nucleons. 
Above $n_{sat}$ and below the deconfinement phase transition $n_{dec} > n_{sat}$ (i.e.,~$n_{sat} < n_ B< n_{dec}$), the neutron star layers must include nucleons (protons and neutrons), but may also have hyperons and resonances (like $\Delta$ baryons).\footnote{For the sake of roughly delimiting the boundaries of the layers of a neutron star in Fig.~\ref{fig:validity}, we have chosen $n_{dec} \sim 4\, n_{sat}$, but this should be taken with a huge grain of salt. The exact deconfinement number density is unknown, except that $n_{dec}>n_{sat}$, and less than $\sim 4\, n_{sat}$, where a simple calculation shows that nucleons start to overlap. 
Beyond that, quarks are expected to dominate, but possibly within a mixture of phases, a percolation, or a crossover. The CMF parametrization we discuss in this work predicts a steep first-order phase transition to quark matter at $T=0$ (on top of which we construct a mixed phase).}
We should note that, while $\chi$EFT calculations can be used to calculate properties of nuclei (see, e.g.,~\cite{Lee:2025req}), although the $\chi$EFT approach used in the MUSES CE is based on infinite-matter, many-body perturbation theory, which does not produce bound states (i.e., nuclei). 

The heavy-ion collisions EoSs have well-defined boundaries because the BQS and TExS are both based on series expansions, where uncertainty quantification is well-understood. 
Both are based on a $\mu_B/T$ expansion such that their regimes of validity are clearly shown in \Cref{fig:validity}.  
Since the hadron resonance gas is also expanded using a Taylor series, it breaks down quickly at large $\mu_B$ for low $T$. 
Both can already be used in hydrodynamics simulations (as was done in \cite{Plumberg:2024leb,Gardim:2024nyz}).    
NumRelHolo has an entirely different approach such that it only captures the quark-gluon plasma phase as well as around the phase transition, but does not include hadrons. 
Thus, it will need to be merged with a hadron resonance gas model first before being used in simulations. 
For heavy-ion collisions, we show only the specific slice in the QCD phase diagram of $\mu_S=\mu_Q=0$ where all three modules are valid. However, the BQS module is also valid at finite $\mu_S\neq 0$ and $\mu_Q\neq 0$.
 
MUSES EoSs satisfy standard constraints on the EoS coming from nuclear physics and astrophysics, as reviewed recently by the MUSES collaboration~\cite{MUSES:2023hyz}, including constraints from lattice QCD, perturbative QCD, $\chi$EFT, heavy-ion collisions, low-energy nuclear physics, and observations from neutron stars and their mergers.

\begin{figure}[t!]
    \centering
    \includegraphics[width=\linewidth]{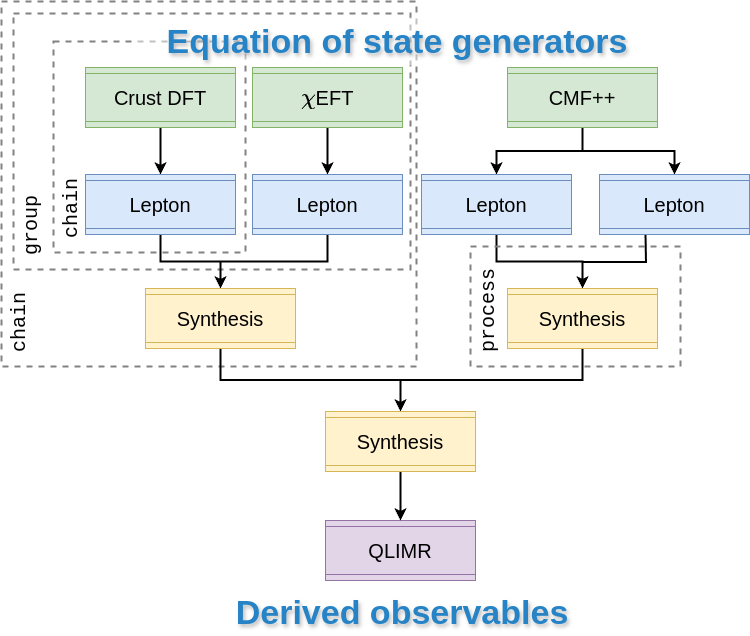}
    \caption{Workflows within the CE that create (Crust-DFT) crust, ($\chi$EFT) outer core, and (CMF++) inner core EoSs, turn them into charge neutral EoSs in $\beta$-equilibrium (Lepton), which are then combined (Synthesis) and used to calculate neutron star observables (QLIMR). 
    }
    \label{fig:workflows}
\end{figure}

With the wide range of applications for the EoSs produced by MUSES, different formats of output are necessary for the wider physics community. One such format that is common within the astrophysical community is the CompOSE format. CompOSE is an existing EoS repository containing thermodynamic, microscopic, compositional and astrophysical data for hundreds of EoSs, all with a common format \cite{Dexheimer:2022qhn}. Within MUSES, the CompOSE package in \texttt{Python} can produce output to fit these format specifications. The package contains functions to linearly interpolate data in regular grids of $(T,n_B,Y_Q)$, with options to follow recommended grid sizing from CompOSE or to specify a custom regular grid. It also includes functions to properly format output files that allow EoSs to be submitted to the CompOSE repository.  This package can be easily extended to support additional output formats. CompOSE files can also be converted to the standard MUSES format shown in \Cref{tab:eos_format} for use as inputs to MUSES workflows involving other modules to calculate EoSs and/or physical observables.

The CE also includes modules that connect theory to data and modules that
can be used in other theoretical frameworks (e.g., numerical relativity or relativistic viscous fluid dynamics). The relevant ones for this paper are summarized below.
\begin{itemize}
\item \textbf{Flavor Equilibration (Flavor Eq)} calculates the equilibrium charge fraction, the flavor (isospin) relaxation rate, and the frequency-dependent bulk viscosity of nucleonic matter \cite{Alford:2023gxq,Alford:2024xfb} (see \cite{zenodo_flavor_equil} for the open source code).
\item \textbf{Quadrupole moment, Tidal Love number, Moment of Inertia, Mass, and Radius  (QLIMR)} calculates macroscopic stellar properties using the QLIMR module to compute masses, radii and the I-Love-Q relations of slowly-rotating compact stars~\cite{Ravenhall1994,Bejger:2002ty,Yagi:2014qua} (see \cite{conde_ocazionez_2024_14525356} for the open source code).

\end{itemize}
For this paper, all observable modules are focused on neutron star physics at $T\sim0$, but later releases will include connections to heavy-ion collision observables across $(T,\mu_B,\mu_S,\mu_Q)$ and neutron star merger EoSs across $(T,n_B,Y_Q)$, eventually connecting heavy-ion collision EoSs to those from neutron star mergers.

\begin{figure}[t!]
    \centering
    \includegraphics[width=.65\linewidth]{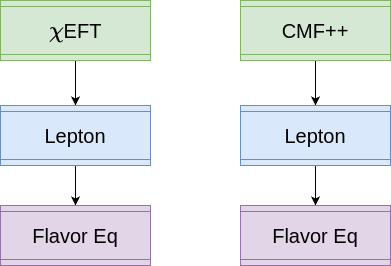}
    \caption{Workflows within the CE that create an ($\chi$EFT OR CMF++) EoS, turn it into a charge neutral EoS (Lepton), and use it to compute out-of-equilibrium effects (Flavor Equilibration).
    }
    \label{fig:workflows_flavor}
\end{figure}

Within the CE, one can define different workflows that connect the EoS modules together.  \Cref{fig:workflows,fig:workflows_flavor} show two examples at $T\sim 0$ that we explore in this paper, but we emphasize that many other possible workflows are possible and can be specified by the user.  
In the first set of workflows, as shown in  \Cref{fig:workflows}, the output of each EoS module is read by the lepton module, which adds lepton contributions. 
The CMF++ model can produce two separate phases (hadronic and quark), which have the lepton contributions added separately, before calculating stability in the Synthesis module (see subsection III-D of Ref.~\cite{Cruz-Camacho:2024odu}). The Synthesis module also combines Crust-DFT and $\chi$EFT results, which are then connected to CMF++ results.
To describe chemically-equilibrated neutron stars, the Lepton module reduces the initial 2-dimensional results of each EoS ($\mu_B$ and $\mu_Q$ or $n_B$ and $n_Q$) into 1D results ($\mu_B$) by enforcing $\beta$ equilibrium with leptons (e.g., for the electron $\mu_e=-\mu_Q$) and charge neutrality $n_{lep}=n_Q$. 
The Synthesis module then combines the 1D EoSs, allowing the matching to occur through different methods and at different points. 
After matching, Synthesis then checks causality and thermodynamic stability according to Appendix E of \cite{Cruz-Camacho:2024odu}.
The result is a 1D EoS from crust-to-core for a neutron star that can be fed into QLIMR, which then calculates neutron star observables that can be connected to astrophysical data. We use processes to refer to the run of a single module. A chain refers to two or more processes  that  run in sequence, and a group refer to multiple processes or chains that can be run in parallel.

The second set of workflows that we discuss in this paper are shown in  \Cref{fig:workflows_flavor}. 
A given 2D EoS can be fed into the Lepton module, which in this case keeps the 2D results ($\mu_B$ and $\mu_Q$ or $n_B$ and $n_Q$) by enforcing only charge neutrality.
The resulting 2D table is then used as input for the Flavor Equilibration module (alternatively, it could also be outputted directly to be used in, e.g.,~neutron star mergers)\footnote{A 2D EoS could also be produced from a 1D charge neutral, $\beta$-equilibrated EoS through an  expansion in isospin fractions \cite{Yao:2023yda}, although this has not been implemented in MUSES yet.}.
Because the Flavor Equilibration module assumes homogeneous nuclear matter and requires microscopic information including particle fractions and nucleon effective masses, it is not yet possible to use it together with smoothly joined EoSs. 
Rather, we use the Flavor Equilibrium module separately with each EoS that can provide that information, $\chi$EFT and CMF++, at densities
near or above nuclear saturation where matter is expected to be homogeneous.

Although we here focus on workflows that combine all three of our EoSs and calculate neutron star observables (with out-of-equilibrium effects computed in the Flavor Equilibration module for each EoS), simpler workflows are also possible. To give an example of a simpler workflow, one could connect a low $n_B$ EoS table obtained from CompOSE directly to a CMF++ workflow, combine them with Synthesis, and then use that new EoS to calculate neutron star observables. One could also use one or more EoS modules to create CompOSE compatible output tables. 
Interested users also have the possibility to use a single module for their own purposes; for example, using QLIMR coupled to their own EoS to calculate gravitational-wave observables such as tidal deformabilities.

Returning to the primary purpose of this work, we want to systematically study the effects of combining various EoSs across $n_B$ using different matching methods. 
We address questions such as: Is it equivalent to smoothly match EoSs using  pressure in terms of the baryon chemical potential $P(\mu_B)$, versus using the speed of sound squared 
\begin{equation}
\label{eq:cs-def}
c_s^2 \equiv \frac{dP}{d\epsilon}\,,
\end{equation}
and then integrating to obtain the pressure in terms of the energy density, i.e,.~$P(\varepsilon)$?
Is there an ideal location for this smooth matching and what influence does it play on neutron star observables including EoS-independent relations? We also explore for the first time flavor equilibration of the CMF model and $\chi$EFT.

In the next section, we outline the physics involved in our three different neutron-star nuclear (not including the Lepton) EoS modules: Crust-DFT, $\chi$EFT and CMF++.  All three models are well-established and have been used to describe dense matter and neutron stars for many years. 
In our work, we use the standard or ``best fit'' parameters of each of these models, leaving future studies to explore the full parameter space. We also discuss in detail below the functionality of the Lepton and Synthesis modules.

\section{Neutron-Star EoS Modules}\label{sec:ns_eos_modules}

The EoS modules that we used to describe the neutron star EoS are detailed below. We discuss the layers of the neutron star that they cover as well as their regime of validity, the physics contained within each module, and briefly outline the structure of each code.

\begin{figure}
    \centering
    \includegraphics[width=1\linewidth]{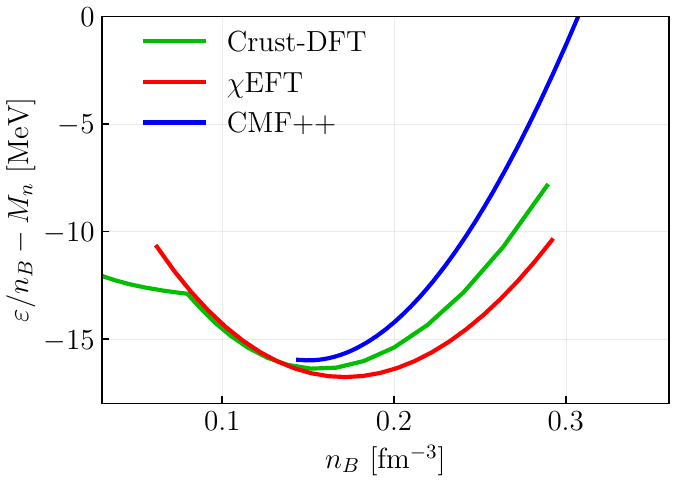}
    \caption{Binding energy per nucleon as a function of the baryon number density for Crust-DFT, $\chi$EFT and CMF++ for isospin symmetric matter $Y_Q=0.5$. We show each module only within their approximate regime of validity that is later considered for matching between EoS.}
    \label{fig:binding}
\end{figure}

Before getting into the details of the individual modules, let us discussion their regime of validity and where we match between the modules to provide some overview of our approach. 
In \Cref{fig:binding}, we plot the binding energy per nucleon vs the baryon number density for isospin-symmetric matter ($Y=0.5$ or $\mu_Q=0$) for all three modules. 
The saturation density, $n_{sat}$, is defined as the baryon number density where the binding energy has a minimum. 
In \Cref{fig:binding} we find that all nuclear EoS modules have $n_{sat}$ in the range of $n_{sat}=\left[0.14,0.17\right]$ fm$^{-3}$, which is compatible with constraints. 
Furthermore, the binding energy per nucleon at $n_{sat}$ for our models should be in the range of $\left[-16.2,-15.7\right]$  MeV, as is also verified in all three EoS modules.  

At low $n_B$, we see in \Cref{fig:binding} that Crust-DFT has a kink, which is an indication of a heavy nuclei-dominated regime (where nucleons are present because $n_B>n_{drip}$, but they are a sub-dominant effect). 
Comparing this to $\chi$EFT, which does not have nuclei in the MUSES CE, a kink is not present at low $n_B$, which indicates that this number density range is outside the regime of validity of $\chi$EFT; we plot this feature just to demonstrate the need for a crust at low $n_B$. 
Close to $n_{sat}$ but slightly below, we see that Crust-DFT and $\chi$EFT are nearly identical because Crust-DFT was previously tuned to reproduce $\chi$EFT. 
At $n_{sat}$, we begin to see subtle differences, where $\chi$EFT has the lowest binding energy and the highest $n_{sat}$.  
Exactly at $n_{sat}$ is where CMF++ becomes valid (although this particular rendering leads to a lower $n_{sat}$ than both Crust-DFT and $\chi$EFT). 
Then slightly above $n_{sat}$, we see stronger differences between our EoS modules, which foreshadows some of the challenges that we find later when matching EoSs in this regime. 
Around $\gtrsim 2n_{sat}$, we expect $\chi$EFT to no longer be valid, so CMF++ must be switched on somewhere between $n_{sat}<n_B<2n_{sat}$ to ensure an overlap in their regime of validity.

Our approach in this paper is to use Crust-DFT at the lowest $n_B$ and smoothly match it to $\chi$EFT slightly below $n_{sat}$, where there is a strong overlap between modules. Then, above $n_{sat}$, we smoothly match the combined Crust-DFT+$\chi$EFT to CMF++ in the range where there is overlapping regime of validity. 
In principle, Crust-DFT could be run at arbitrarily high $n_B$ if one believes there is only nucleonic degrees-of-freedom within neutron stars, but for this paper we are only considering it for the crust part of the EoS. 

\begin{figure}
    \centering
    \includegraphics[width=1\linewidth]{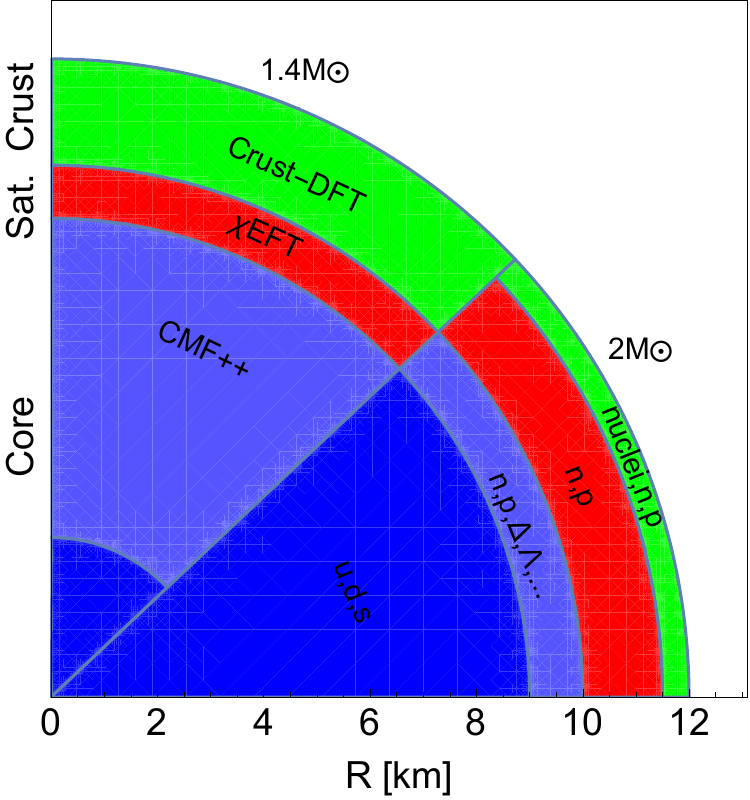}
    \caption{Cartoon showing the layers of two neutron stars with masses of $1.4M_{\odot}$ and $2M_{\odot}$ that may result from a MUSES workflow.  The EoS modules used for each layer are shown in green for crust-DFT, red for $\chi$EFT, and blue for CMF, where the light blue indicates hadrons and the dark blue indicates quarks. On the top-left slice we list EoS module names, while in the bottom-right slice we list the particles each EoS includes.}
    \label{fig:layers}
\end{figure}

In Fig.~\ref{fig:layers} we show a cartoon (the numbers are just estimated and not taken from an exact calculation) of different layers of neutron stars based on the type of EoS modules used in the MUSES CE. 
For a lighter neutron star of $1.4M_{\odot}$, we anticipate a significant contribution from the crust and saturation properties with an overall dominant contribution of neutrons and protons (potentially also including hyperons and/or resonances), but a small quark core (if any). 
For a more massive neutron star of $2M_{\odot}$, we anticipate a very thin crust, with a medium layer around saturation densities, and a significantly larger core (especially with quarks, if they appear). 
The exact boundaries between layers depend strongly on the EoS, the parameters of the EoS, and where/how the matching between layers is performed, so the exact numbers in Fig.~\ref{fig:layers} should not be taken seriously. 
At the end of this paper, we discuss the exact numbers obtained for these layers for a particular, heavy neutron star. 

\subsection{Model for the neutron star crust: Crust-DFT}\label{sec:crustDFT}

Crust-DFT is a phenomenological model that describes nuclei in 
equilibrium with neutrons and protons developed originally in Refs.~\cite{Du:2018vyp,Du:2021rhq}, based on earlier work in Ref.~\cite{Steiner:2012rk}. The Crust-DFT source code can be found in Ref.~\cite{crustdft_code}. For this paper, we focus on the very low $T$ regime of Crust-DFT, although the model allows for high $T$ calculations relevant to supernovae as well. 

At $T\sim 0$, the structure of the outer layers of a neutron star changes rapidly with $n_B$. The atmosphere is described as an ideal Fermi gas of electrons and light nuclei that have $Y_Q\sim 0.5$. As $n_B$ increases, the nuclei become more neutron-rich ($Z\ll A$) and the nuclei play a larger role in the EoS. One must minimize the free energy with respect to the number of protons $Z$ versus the total number of nucleons $A$ within a nucleus to determine which nuclei appear at a given $(T,n_B,Y_Q)$. For lighter nuclei, the masses and their properties have already been well-established experimentally \cite{Moller:2015fba}. However, the nuclei close to the neutron drip line $n_B\sim n_{drip}$ have either not yet been measured, have missing information, or large uncertainties. Thus, we use DFT to calculate the properties of these heavy, neutron-rich nuclei, where experimental data is missing. 
Furthermore, we take into account the spacing of the nuclei on a fixed lattice, as well as electrostatic corrections between the electrons and themselves (repulsive), and the electrons and the nuclei (attractive) using Wigner-Seitz cells. 

Once the neutron drip line is surpassed (when $n_B>n_{drip}$), the system is composed of nuclei, unbound ``free'' nucleons\footnote{By ``free'' nucleons, we mean nucleons no longer confined within a nucleus, not non-interacting nucleons.}, and electrons. 
Past the neutron drip line, we use an excluded volume to describe the size of the nuclei. 
Even when the population of free nucleons is very dilute, we include their 2-body interactions, i.e., nucleon-nucleon (NN) interactions. 
Thus, Crust-DFT can handle the liquid-gas phase transition and is valid even past $n_{sat}$, at least for the regime where NN interactions of nucleons only are still valid. 

In DFT, the nuclei properties are calculated using a virial expansion, which is the best available model when the neutron and proton fugacities ($z_i \equiv \exp[(\mu_i - m_i)/T]$ with $m$ being their masses) are much less than 1, i.e.,~$(z_p,z_n)\ll 1$. Within a virial expansion, one takes a free Fermi gas and then systematically includes $N$-body interactions, where, if $N=2$, these are pairs of nucleons within a nucleus and so forth. 
This method directly computes the EoS from NN scattering phase shifts in a model-independent way~\cite{Horowitz:2005nd}.
The properties of the virial expansion are constrained by nuclear structure experiments around $n_{\mathrm{sat}}$, $\chi$EFT results near $n_{\mathrm{sat}}$ at finite $T$, and neutron star observations at higher $n_B$ \cite{Du:2018vyp}. This model has several parameters that can be modified by the user without spoiling the aforementioned tuning. In this work, we select a few relevant parametrizations. 

Depending on the $(T,n_B,Y_Q)$, the EoS may be in a regime where nuclei and nucleons coexist at the same spatial location. 
In this regime, one requires constraints to determine the population number densities of the nuclei and nucleons. 
In nuclear statistical equilibrium,
reactions, such as the decay of a nucleus $A$ into its constituent nucleons,
$
A(A-Z,Z) \leftrightarrow \left(A-Z\right) n + Z p\,,
$
are in equilibrium when their forward rates match their reverse rates. 
Then, the condition of nuclear statistical equilibrium implies a relation, known as the Saha equation~\cite{saha}, between the associated chemical potentials (see Eq.~23 of Ref.~\cite{Du:2021rhq})
$
\mu_{A} = \left(A-Z\right) \mu_n + Z \mu_p
$.
The Saha equation is the relationship between chemical potentials when the degrees of freedom are only nuclei and nucleons.

In principle, the Crust-DFT EoS can be computed over a large range of temperatures $T \in (0.1,127)~\mathrm{MeV}$, baryon number densities $n_B \in (2\times10^{-12},2)~\mathrm{fm^{-3}}$, and charge fractions $Y_Q \in (0.01,0.7$). However, its regime of validity is only where the degrees of freedom of nuclei and/or nucleons are relevant because quarks, hyperons, nucleonic resonances, and mesons are not included. The exact boundaries of its regime of validity are difficult to precisely determine, but broadly speaking the model is valid  when $T$ is less than a few tens of MeV (when more baryon resonances inevitably appear) and $n_B\lesssim (2$--$4)n_{sat}$ (when nucleons start to overlap at $T=0$).
However, for this paper, 
we focus only on a  vanishing temperature regime, and all calculations are carried out at the fixed temperature of $T=0.1$ MeV.

The five basic Crust-DFT parameters~\cite{Du:2021rhq} are the following: the symmetry energy $E_{sym}$, the slope $L_{sym}$ of $E_{sym}$, two parameters that control the strength of the three-neutron force in the neutron matter EoS, and an index that selects the specific Skyrme model used for nuclear matter from the posteriors in Ref.~\cite{PhysRevC.89.054314} (see Ref.~\cite{Du:2021rhq} for details). Two additional high-density parameters also need to be specified, but they do not impact the equation of state below $n_{sat}$.
The basic output of the Crust-DFT model are thermodynamics properties of matter, including the internal energy, pressure, entropy, and chemical potentials (see \Cref{sec:lepton}). Also included in the output are the fractional abundances of neutrons, protons, five light nuclei (following~\cite{Steiner:2012rk}) and a representative heavy nucleus. In order to be consistent with the MUSES interface, all chemical potentials include their associated rest masses.

The Crust-DFT code is written entirely in C++ and depends on the GSL, HDF5, Boost, and O$_2$scl libraries. Most points in parameter space can be computed within fractions of a second, but this execution time becomes significantly larger at low $T$, near $n_{sat}$, where nuclei and nucleons have nearly equal free energies. 
The Saha equations (see above) are used to reduce the EoS calculation to two equations to be solved using a Newton-Raphson method. The number density of neutrons and protons outside of nuclei are varied to ensure that the desired baryon density and electron fraction are matched. Given a guess for the neutron and proton density, the phenomenological energy density functional is used to compute the thermodynamic properties of neutrons and protons outside of nuclei; then, the Saha equations are used to determine the density of nuclei. Multiple solutions are resolved by choosing the solution that minimizes the free energy.
Crust-DFT also has several precomputed tables available for download in the Zenodo repository \cite{crust_dft_zenodo}.

\begin{figure}[t!]
    \centering
    \includegraphics[width=0.99\linewidth]{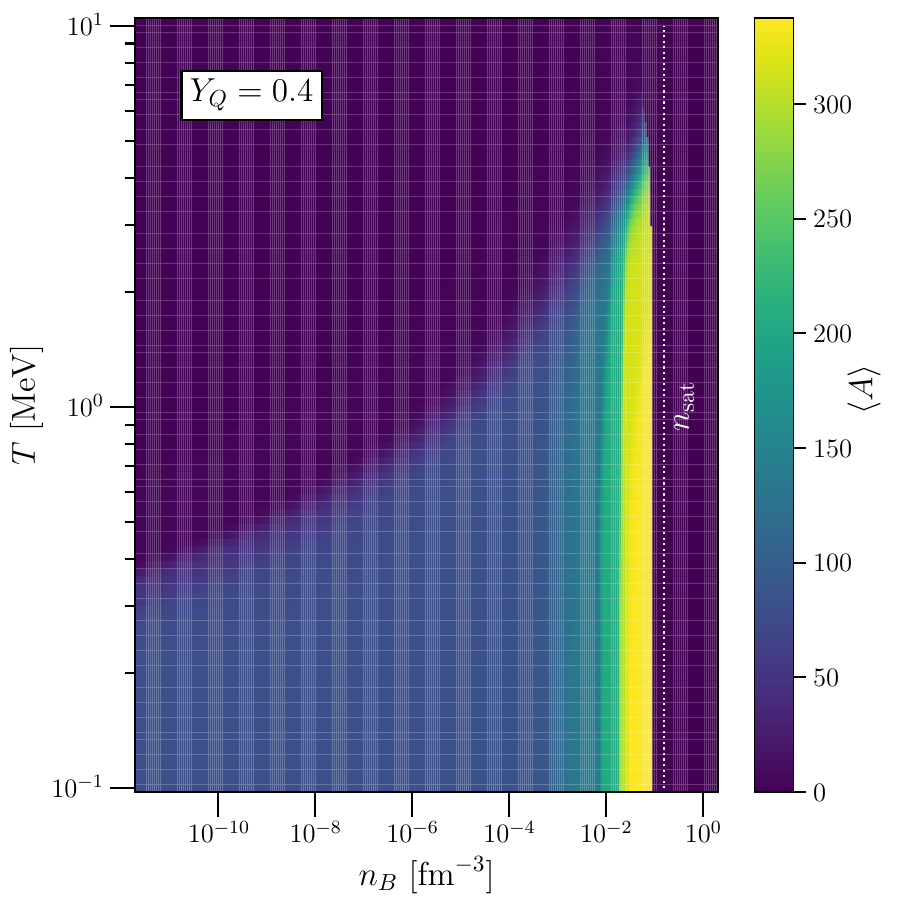}
    \caption{Average nuclear mass number $\langle A\rangle$, see Eq.~(\ref{eqn:avgmass}), as a function of density and temperature at a fixed charge fraction $Y_Q=0.4$ calculated within Crust-DFT. The value of $n_{\mathrm{sat}}$, 0.16~$\mathrm{fm}^{-3}$, is just above the density at which $\langle A\rangle$ drops sharply to zero. 
}
    \label{fig:cdft1}
\end{figure}

At finite $T$, all nuclei appear in the EoS (although, depending on the $n_B$, some nuclei have very small abundances). 
To better understand the relevant nuclei, we can calculate the average nuclear mass number, $\langle A\rangle $, for a given $(T,n_B,Y_Q)$ via~\cite{Du:2021rhq}
\begin{equation}\label{eqn:avgmass}
    \langle A\rangle= \left[\sum_i n_i A_i \right] \left[ \sum_i n_i \right]^{-1}\ ,
\end{equation}
where $A_i$ is the number of nucleons within the nucleus $i$, $n_i$ is the number density of that nucleus, and we sum over all nuclei. 
In \Cref{fig:cdft1} we plot $\langle A\rangle$ as a function of $(T,n_B)$, assuming matter that is close to isospin symmetric, which is similar to what is found in heavy-ion collisions, i.e.,~$Y_Q\sim 0.4$. 
We see that nuclei no longer play a role at high $T$ because one has surpassed the liquid-gas phase transition,  such that nuclei ``melt'' into hadrons. 
However, at temperatures below $T\lesssim 5$ MeV, starting from low $n_B$, there is a direct increase in $\langle A\rangle $ as one increases $n_B$ while keeping $T={\rm{const}}$. 
At high enough $n_B\sim[10^{-2},0.1]$ fm$^{-3}$, there is a sudden sharp rise in $\langle A\rangle\rightarrow 350$, followed by a sudden drop to $\langle A\rangle\rightarrow 0$, which is where $n_{sat}$ occurs. At higher $n_B>n_{sat}$, nuclei are no longer energetically favorable and the EoS consists solely of nucleons. 

\subsection{Model for the neutron star near saturation density: $\chi$EFT}\label{sec:chiEFT}

$\chi$EFT provides a model-independent framework for the study of strongly-interacting matter at the energy scales characteristic of nuclei (for theoretical background, see \cite{Machleidt:2011zz, Epelbaum:2008ga}). The theory describes microscopic nuclear interactions through an order-by-order expansion, which includes both NN and multi-nucleon interactions. The $\chi$EFT EoS is calculated using many-body perturbation theory (MBPT) \cite{Bogner:2005sn} up to second order in homogeneous nuclear matter. The legacy code for $\chi$EFT MBPT was written in \texttt{Fortran 77}, and has been rewritten from scratch into a modern object-oriented \texttt{C++20} version at $T=0$, as well as finite-$T$ nuclear matter. For a detailed description of the theory involved in the $\chi$EFT two- and three-nucleon interactions, $T=0$ MBPT, and finite-$T$ MBPT, see Refs. \cite{Machleidt:2011zz}, \cite{Bogner:2005sn}, \cite{Wellenhofer:2014hya}, respectively. For the source code, see Ref.~\cite{cheft_code}.

Chiral EFT is the low-energy limit of QCD with nucleons and pions as degrees of freedom. The theory offers an order-by-order expansion for both NN and multi-nucleon interactions based on the most general Lagrangian that is consistent with the usual symmetries of low-energy QCD, particularly the spontaneously broken chiral symmetry \cite{Weinberg:1990rz}. The long-range features of the interaction are governed by pion-exchange contributions constrained by chiral symmetry. Short-distance details are encoded in a set of contact interactions with parameters that are fitted to NN scattering phase shifts and bound state properties of light nuclei. Employing $\chi$EFT with nuclear many-body physics then allows for pure predictions of empirical properties of nuclear matter, such as the value of $n_{sat}$ at $T=0$ as well as the liquid-gas phase transition at $T>0$ \cite{Wellenhofer:2014hya}.

Because of the limited energy regime accessible to $\chi$EFT, chiral nuclear interactions are typically regulated at a scale $\Lambda$ lying between the low- and high-energy regimes. For the MUSES $\chi$EFT module, we employ a certain class of NN potentials at next-to-next-to-next-to-leading order (N3LO) in the chiral expansion with resolution scales $\Lambda = (414, 450)\,\text{MeV}$ \cite{Coraggio:2007mc, Coraggio:2012ca}. In addition, we include three-body forces up to N2LO in the chiral expansion through the use of an effective in-medium NN interaction \cite{Holt:2009ty, Holt:2013fwa, Holt:2019bah}. Besides these two potentials fitted to low-energy nuclear scattering data, the module also allows for the low-energy constants and contact parameters of the potential to be freely adjusted. This enables the possibility of fitting these parameters to a different set of empirical quantities related to nuclear matter, such as saturation properties or astrophysical observables. The use of multiple fitted chiral interactions also allows for better uncertainty quantification, as well as the possibility of generating a sampling distribution of EoSs based on $\chi$EFT \cite{Lim:2018bkq, Lim:2018xne, Lim:2019som}.

When working with low-momentum nuclear interactions, MBPT can be used to investigate the nuclear many-body system \cite{Bogner:2005sn}. In the MUSES module, we employ the Kohn-Luttinger-Ward many-body perturbation series \cite{Kohn:1960zz, Luttinger:1960ua} including contributions up to second order to calculate the free-energy per nucleon in infinite homogeneous nuclear matter. Mean-field contributions are also included self-consistently up to first-order in the MBPT expansion. The remaining thermodynamic quantities follow from derivatives of the free-energy by standard thermodynamic relations (see \Cref{app:cheft}).
This allows for a full calculation of the EoS of asymmetric nuclear matter ($0.0 \leq Y_Q \leq 0.5$) at $T=0$ for densities $n_B \sim n_{sat}$ \cite{Wellenhofer:2014hya, Drischler:2021kxf}. For faster evaluation, the EoS at arbitrary isospin asymmetry can also be calculated by first calculating the nuclear symmetry free-energy $\bar{F}_{\text{sym}}$ directly and using a quadratic approximation for the isospin-asymmetry dependence of the EoS \cite{Wellenhofer:2015qba, Wellenhofer:2016lnl}:
\begin{equation}\label{eq:cheft-isospin}
    \bar{F}(n_B, \delta) \approx \bar{F}(n_B, \delta = 0) + \bar{F}_{\text{sym}}(n_B)\,\delta^2 \ ,
\end{equation}
where $\bar{F}(n_B, \delta)$ is the free-energy per nucleon and $\delta = 1 - 2Y_Q$ is the isospin asymmetry parameter.

Due to the limited nature of the EFT, $\chi$EFT can only be used to investigate a limited region of the QCD phase diagram. An accurate treatment of low-density nuclear matter ($n_B<0.5n_{sat}$) must account for nuclear clustering and non-perturbative features of the NN interactions associated with larger scattering lengths. Conversely, any treatment of high density ($n_B>2n_{sat}$) would probe relative energy scales significantly greater than the effective resolution scale $\Lambda$, associated with heavier degrees of freedom not included in the $\chi$EFT expansion. However, within this regime of validity ($0.5<n_B/n_{sat}<2$), $\chi$EFT produces accurate results for the EoS using MBPT \cite{Bogner:2005sn}, as well as the means for quantifying uncertainties \cite{Lim:2018bkq}.

The input parameters in the $\chi$EFT EoS module can be classified into four main categories: $\chi$EFT interaction parameters, physical parameters, computational parameters, and EoS grid parameters. The $\chi$EFT interaction parameters include the low-energy constants (LECs) appearing in the $\chi$EFT Lagrangian (8 $\pi$N LECs, 26 NN LECs, 2 NNN LECs) and the parameters of the low-momentum regulator cutoff (e.g.,~$\Lambda_{\text{NN}}$, $\Lambda_{\text{3N}}$). The physical parameters refer to common physical constants used in the calculation, such as nucleon and pion masses and coupling constants (e.g.,~$g_A$, $f_{\pi}$, $\alpha$). The computational parameters include numerical parameters involved in the calculation, such as the number of integration mesh points for quadrature methods or the number of interpolation points for 1D interpolations, as well as calculation options, such as whether to include NNN interactions or whether to include many-body mean-field interactions. The EoS grid parameters allow the user to specify in what region of the 2D QCD phase diagram to generate the EoS, parametrized by nucleon density $n_B$ and isospin asymmetry parameter $\delta$.

The standard output files generated by the $\chi$EFT EoS module are tables in CSV or HDF5 format, depending on the user's choice of configuration. The full output file contains most of the standard thermodynamic variables in the canonical ensemble: free-energy per nucleon $\bar{F}$, internal energy per nucleon $E/A$, pressure $P$, proton/neutron chemical potential $\mu_p,\, \mu_n$, and $c_s^2$ defined on the 2-dimensional EoS grid of points $(n_B, \delta)$. Also included in this output file are the parameters $({m_p}^{*},\, U_p,\, {m_n}^{*},\, U_n)$ of the non-relativistic single-nucleon energy $\varepsilon(k)$ under the effective mass approximation $\varepsilon(k) = k^2/2m^* + U$, where $m^*$ stands for effective mass and $U$ for the particle potential. There are four other output files that the $\chi$EFT EoS module can generate for users. The first of these output options filters out any unstable and metastable points, leaving only a stable EoS. The stability conditions used at $T=0$ are given by (see App. E in \cite{Cruz-Camacho:2024odu})
\begin{equation}\label{eq:cheft-stability}
    \left.\frac{\partial P}{\partial n_B}\right|_{\delta}  \geq 0\ ,\quad \left.\frac{\partial P}{\partial n_p}\right|_{\delta}  \geq 0\ .
\end{equation}
The second output file option generates a file containing saturation properties and properties related to $E_{sym}$ of the generated EoS \cite{Wellenhofer:2014hya, Wellenhofer:2015qba}. The last two options generate the MUSES standard format (see \Cref{sec:lepton}) and the Flavor Equilibration module format (see \Cref{sec:flavor}), respectively.

The $\chi$EFT EoS module code is mainly written in modern \texttt{C++}, with an additional \texttt{Python} layer for input validation, data post-processing, and assembling output files. The module consists of two main components: the $\chi$EFT interaction potential and the MBPT EoS. The first of these components evaluates the $\chi$EFT NN potential matrix elements for the given set of LECs and regulator parameters. The potential matrix elements are stored in partial-wave format $\langle\,p\,|V_{\text{NN}}^{J\,\ell\,\ell'\,\mathcal{S}\,\mathcal{T}}\,|\,p'\,\rangle$ with a 2-dimensional mesh in both the incoming/outgoing relative particle momentum. The in-medium NN interaction must also be evaluated separately at each point in the phase diagram. The second component of the module uses these stored potential matrix elements to evaluate the appropriate MBPT diagrams. The partial-wave formalism results in integrals over multiple internal momenta in each interaction and relative angles between them. These integrals are carefully evaluated inside the Fermi surface using a standard Gauss-Legendre quadrature method, interpolating over the tabulated matrix elements stored in memory. In the \texttt{Python} post-processing scripts, the resulting free-energy per nucleon $\bar{F}$ calculated by the \texttt{C++} module is used to calculate all remaining thermodynamic variables.

The $\chi$EFT MUSES module is written mainly in modern object-oriented \texttt{C++20}, with a modular structure to allow for additional calculations to be incorporated later. This could include higher order corrections in the MBPT expansion \cite{Holt:2016pjb} or other analytical forms of the $\chi$EFT nuclear potential \cite{Siu:2009nx, Bogner:2009bt}. The many-body integrals are evaluated in parallel using \texttt{OpenMP}, bringing a significant speed-up from the legacy \texttt{Fortran} programs which preceded it. For a single point in the QCD phase diagram, the MUSES module runs 23 times faster than the $\chi$EFT legacy Fortran code, when running on the same machine. As the number of points in the EoS scale to 100, 1000, etc., this rate increases to over 70 times faster. When these runs are multithreaded over 8 cores, the MUSES module can calculate a dense 2-dimensional EoS as much as 250 times faster than before.

The current release of the $\chi$EFT MUSES module calculates a 2-dimensional EoS for $T=0$. In a near-future release, the module will include the finite-$T$ EoS, expanding the potential use cases for astrophysical simulation. The open source version of the $\chi$EFT EoS module, along with comprehensive documentation on building and executing the code, is available in the Zenodo repository~\cite{zenodo_cheft}.

\subsection{Model for the neutron star outer and inner core: CMF}\label{sec:CMF}

The CMF model is a relativistic framework based on a non-linear realization of chiral symmetry, originally developed in~\cite{Weinberg:1968de,Papazoglou:1998vr}, and successfully applied to the description of dense matter in neutron stars~\cite{Dexheimer:2008ax}. This model describes the strong interaction between hadrons and/or quarks mediated by scalar and vector (mean-field) mesons, allowing the study of hadronic and quark matter under extreme conditions, such as those found in neutron stars and heavy-ion collisions. The legacy CMF code, originally written in \texttt{Fortran 77}, has been rewritten from scratch as CMF++, a modern object-oriented implementation in \texttt{C++20} designed (so far) for $T=0$ calculations. For detailed descriptions of the CMF++ theory and implementation, see Ref.~\cite{Cruz-Camacho:2024odu}. The open source code repository is available at~\cite{cmf_code}.

The CMF model describes the physics of dense matter at $n_B \gtrsim n_{\text{sat}}$~\cite{Dexheimer:2008ax}. At moderate densities, the model includes interacting nucleons, while at higher $n_B$ and/or $T$, it incorporates hyperons and other spin-$\frac{3}{2}$ resonances, such as $\Delta$ baryons, eventually transitioning to quark matter at the highest $n_B$ and/or $T$. The model is built from a Lagrangian that respects known symmetries and incorporates the hadronic states of the SU(3) baryon octet, decuplet, and light and strange quarks. The interactions among these fermions are mediated by scalar mesons, which provide attractive forces, and vector mesons, which account for repulsive interactions.  Additionally, isovector interactions are mediated by mesons carrying isospin and strange baryon and quark interactions are mediated by mesons with hidden strangeness. To model the transition between hadrons and quarks, the Lagrangian includes a Polyakov loop-like potential, which induces a first-order phase transition at low $T$~\cite{Dexheimer:2009hi}. In the CMF++ model, the phases of matter are determined by the properties of particles, their couplings, and the interaction ansatz. For instance, at $T=0$, the choice of model parameters can result in various phases, such as one dominated by nucleons, a chiral transition from nucleons to nucleons plus hyperons and/or resonances, a direct transition to quark matter, or a sequence of transitions from nucleons to nucleons plus hyperons and/or resonances, and then to quark matter~\cite{Cruz-Camacho:2024odu}.

The CMF++ code effectively describes the nuclear liquid-gas phase transition and the quark deconfinement phase transition~\cite{Dexheimer:2009hi}. The model explores critical phenomena, such as the QCD critical point, where the deconfinement transition becomes a crossover~\cite{Dexheimer:2009hi,Hempel:2013tfa,Aryal:2020ocm, Kumar:2023qcs}. The model also accounts for chiral symmetry restoration, predicting a reduction in baryon and quark masses under extreme $n_B$ and/or $T$, providing insights into the mechanisms of QCD symmetry-breaking and restoration~\cite{Papazoglou:1997uw, Cruz-Camacho:2024odu}. The CMF model has been calibrated using constraints from lattice QCD, low-energy nuclear experiments, and astrophysical observations~\cite{Dexheimer:2008ax,Dexheimer:2015qha,Kumar:2024owe,Dexheimer:2018dhb}. The model produces EoSs and particle compositions that can be used for a wide range of applications, including heavy-ion collision simulations~\cite{Steinheimer:2009nn,Steinheimer:2007iy,Steinheimer:2022gqb,Steinheimer:2024eha,Steinheimer:2025hsr}, core-collapse supernovae~\cite{Jakobus:2023fru}, stellar cooling~\cite{Negreiros:2010hk,Dexheimer:2012eu}, and neutron star mergers~\cite{Most:2018eaw,Most:2019onn,Most:2022wgo}\footnote{Some of the work described here is based on an alternative version of the CMF model, which includes the chiral partners of the baryons, as detailed in~\cite{Steinheimer:2011ea,Dexheimer:2012eu,Motornenko:2019arp,Steinheimer:2022gqb,Jakobus:2023fru,Steinheimer:2024eha,Steinheimer:2025hsr}. This version is not published as a MUSES module.}.

The primary limitation of the CMF++ module is that it does not yet include finite $T$ or magnetic field effects, although earlier versions of the model do account for these~\cite{Peterson:2023bmr}. Future extensions of CMF++ aim to explore the 4D/5D QCD phase diagram, incorporating both finite $T$ and magnetic fields. The CMF model relies on the mean-field approximation, which excludes quantum fluctuations. This limitation reduces its accuracy in certain regimes, particularly near the QCD critical point, where it cannot capture universal scaling features~\cite{Papazoglou:1998vr}. At high $T$ and $n_B$, achieving greater precision may require incorporating additional degrees of freedom or higher-order effects. For instance, heavier hadronic states are needed to reproduce lattice QCD results at high $T$ and vanishing $n_B$~\cite{Alba:2017hhe,SanMartin:2023zhv}. 

The input parameters in the CMF++ model are divided into two main categories: computational parameters and physical parameters. Computational parameters include options such as the choice of particle sets (e.g., octet and/or decuplet baryons, hyperons, quarks), specifications for output files and formats, domain boundaries, and resolutions for $\mu_B$, $\mu_Q$ and $\mu_S$, as well as initial conditions for the mean-field mesons in the self-consistent solver. Physical parameters encompass quantities such as meson vacuum masses, quark bare masses, quark-to-meson couplings, self-coupling constants for the scalar and vector meson equations, meson and baryon vacuum masses, vector meson-nucleon couplings, and deconfinement potential coupling constants, along with other physically significant variables (all coming from experiments or fitted to experiments/observations). In total, CMF++ has 55 computational parameters and 60 physical parameters, all of which are detailed in~\cite{Cruz-Camacho:2024odu}. In this work, we explore the C4 vector meson parametrization; however, the CMF MUSES module supports full configuration of other parametrizations (C1, C2, and C3).

The standard output files generated by the CMF++ model consist of tables in either CSV or HDF5 format, depending on the user’s input configuration. Since CMF++ accounts for phase transitions, users can obtain data for stable branches alone or include metastable and unstable branches as well. By default, CMF++ outputs three separate EoS files corresponding to the stable, metastable, and unstable branches of the EoS. Additionally, CMF++ provides three optional types of output files. The first generates the required input files for the Lepton module (see \Cref{sec:lepton}), while the second produces formatted files for the Flavor Equilibration module (see \Cref{sec:flavor}). The third option outputs detailed particle properties, including particle densities, chemical potentials, effective chemical potentials, masses, effective masses, and optical potentials for baryons and quarks (covering both stable and metastable branches).

The CMF++ code's main solver is modular and object-oriented, implemented in modern \texttt{C++} with a \texttt{Python} layer for YAML preprocessing and post-processing tasks, such as data cleaning and isolating phase transition branches (stable, metastable, and unstable). The solver employs a self-consistent approach to solve seven non-linear algebraic equations of motion--six for the meson mean fields and one for the deconfinement order parameter--using the \texttt{fsolve} routine~\cite{Burkardt:fsolve}. Compared to the legacy \texttt{Fortran 77} implementation, CMF++ achieves a runtime improvement of over four orders of magnitude for solving coupled equations of motion for the chemical potentials ($\mu_B$, $\mu_S$, $\mu_Q$) at $T=0$ (see Fig.~5 of \cite{Cruz-Camacho:2024odu}). This speed up results from the elimination of deprecated code, a streamlined algorithmic structure, and compiler optimizations. These enhancements not only reduce computational overhead but also extend the code's usability, allowing for precise exploration of the 3D EoS across a broader parameter space. The open source CMF++ module, along with detailed documentation on building and executing the code, is available in the Zenodo repository~\cite{zenodo_cmf}.

\subsection{Lepton}
\label{sec:lepton}

The Lepton module calculates the EoS for a free Fermi gas. The module can be used to either (i) obtain a pure lepton EoS, (ii) calculate the lepton contribution to an existing EoS to ensure charge neutrality, or (iii) calculate the lepton contribution to an existing EoS to ensure charge neutrality, while enforcing $\beta$-equilibrium.
In principle, electrostatic corrections could also be incorporated, being important, for example, in the crust-core transition of neutron stars, but we leave this for future work.

The relevant equations for this module can be found in \Cref{app1}, as well as in~\cite{Peterson:2021teb} with $T$ and magnetic field effects, and in Sec.~IIB of Ref.~\cite{Yao:2023yda} without.
In our formalism, the different flavors of leptons (electron, muon, and tau) have the exact same chemical potential (also applies to neutrinos) because they have the same quantum numbers
\begin{align}
    \mu_e&=\mu_\mu=\mu_\tau \ , \label{eqn:mu_lep} \
\\
    \mu_{\nu_e} & =  \mu_{\nu_\mu}=\mu_{\nu_\tau} \ . 
\end{align}

\subsubsection{Charge neutrality}

To enforce charge neutrality, we impose the following relation by summing over all particles that carry baryon number ($B$) and all that do not ($l)$:
\begin{equation}\label{eq:charge_neutrality}
    \sum_{i \in B} n_{i} Q_i + \sum_{i \in l } n_{i} Q_i =0\ ,
\end{equation}
where $B$ can include nuclei, baryons, and/or quarks, and $l$ includes the leptons $e^-$, $\mu^-$, $\tau^-$, and their respective neutrinos $\nu_l$'s. The quantity $n_{i}$ is the number density of particle $i$ and $Q_i$ is its electric charge.
Using the definition of charge fraction as the ratio of the total electric charge of nuclei/baryons/quarks per total number of baryons (and dividing numerator and denominator by the volume to obtain densities), we obtain
\begin{equation}
    Y_Q=\frac{n_{Q}}{n_B}=\frac{ \sum\limits_{i \in B} n_{i} Q_i}{\sum\limits_{i \in B} n_{i}} \ .
\end{equation}
Using the definition of lepton fraction as the ratio of the total number of leptons per total number of baryons (and dividing again numerator and denominator by the volume), we obtain
\begin{equation}
    Y_l=\frac{ n_ l}{n_B}=\frac{ \sum\limits_{i \in l} n_{i}}{\sum\limits_{i \in B} n_{i}} \ .
    \label{Yl}
\end{equation}

The number density of a lepton is given by \Cref{nlep} and each lepton (that is not a neutrino for which $Q_i=0$) carries electric charge $Q_i=-1$. Adding everything in \Cref{eq:charge_neutrality}, we obtain
\begin{equation}\label{eq:charge_neutrality2}
    n_Q - n_{l^-} =0 \ ,
\end{equation}
or more explicitly
\begin{equation}\label{eq:charge_neutrality3}
   n_Q - \sum_{i \in l^-} \frac{(\mu_i^2-m_i^2)^{3/2}}{3 \pi^2} =0\ ,
\end{equation}
where $m_i$ is the mass of lepton $i$, $l^-$ stands for electrically charged leptons (no neutrinos), and $n_Q$ is an input from the other EoS modules (or from an external table--see the format in \Cref{tab:eos_format}). 

The way that \Cref{eq:charge_neutrality2} is solved depends on the assumptions (or lack of assumptions) of $\beta$-equilibrium and fixed lepton fraction discussed in the following. Once \Cref{eq:charge_neutrality2} is solved, $P$ and $\varepsilon$ contributions from the leptons (\Cref{eqn:ideal}) are added to the ones from other EoS modules (or from an external table) to obtain the total $P$ and $\varepsilon$ according to
\begin{align}
P&= P_B+P_l \ ,\\
\varepsilon&=\varepsilon_{B}+\varepsilon_{l} \ .
\end{align}

\subsubsection{$\beta$-equilibrium}

In $\beta$-equilibrium, weak interaction processes, such as neutron decay and electron capture, are balanced, i.e.,~the interactions
\begin{align}
    n&\rightarrow p+e^{-}+\bar{\nu}_e\ ,
    \\
    p+e^{-}&\rightarrow n+\nu_e\ ,
\end{align}
occur at equal rates.  
Then one can balance the particle chemical potentials, such that 
\begin{equation}\label{1212}
    \mu_n=\mu_p+\mu_e\ ,
\end{equation}
if one assumes that
    \vspace{-0.3cm}
\begin{itemize}
    \item [(a)] neutrinos are in the untrapped, free-streaming regime, so $\mu_{\nu_l}=0$ (as opposed to a fixed or conserved number of leptons);
    \vspace{-0.3cm}
    \item [(b)] the temperature is low enough so that the neutrino energy and momentum can be neglected \cite{Alford:2018lhf}.
\end{itemize}
    \vspace{-0.3cm} 
The above example is all that is required for a simple system of just $npe$ (neutron-proton-electron) matter. 
However, significantly more complicated weak-interactions can be considered both with the possibility of other leptons, as well as other baryons or quarks. 

Given that the CE contains, for example, the CMF++ module, which includes the full SU(3) baryon octet and decuplet, as well as quarks, we must consider all relevant weak interactions to define $\beta$-equilibrium correctly.
Additionally, we consider heavier leptons beyond the electron, such that we define $l^-=\left\{e^-,\mu^-,\tau^-\right\}$.
Then, the $\beta$-equilibrium interactions that must balance are
\begin{align}
n&\rightarrow p+l^{-}+\bar{\nu}_l\label{1313}\ ,\\
\Lambda&\rightarrow p+l^{-}+\bar{\nu}_l\label{1414}\ ,\\
\Sigma^-&\rightarrow n+l^- +\bar{\nu}_l\label{decHAD:sig-_ne-nu}\ ,\\
\Sigma^+ +l^-&\rightarrow  n+\nu_l \label{decHAD:sig+e-_nnu}\ ,\\
\Xi^0 &\rightarrow  \Sigma^+ + l^-+\bar{\nu}_{l}\label{decHAD:xi0_sig+mu-nubar}\,\\
\Xi^- &\rightarrow  \Sigma^0 +l^-+\bar{\nu}_l\label{decHAD:xi-_sige-nubar} \ ,\\
\Xi^- &\rightarrow  \Xi^0 +l^-+\bar{\nu}_l\label{decHAD:xi-xie-nubar} \ ,\\
\Omega^- &\rightarrow  \Xi^0 +l^-+\bar{\nu}_l\label{decHAD:omega-xie-nubar} \ .
\end{align}
Note that $\Sigma^0$ does not present a $\beta$-decay because they rather undergo electromagnetic interactions into a $\Lambda+\gamma$.
The $\Delta$ baryons also do not have a corresponding weak decay because their strong decays into nucleons are too rapid.
Similarly, the hyperons that carry a star, e.g., $\Sigma^*$, are resonances of the $\Sigma$ baryons and would also rapidly decay under the strong force. While the main decay channels of hyperons are to pions (e.g., $ \Lambda \rightarrow p + \pi^- $ and  $ \Lambda \rightarrow n + \pi^0 $)~\cite{Alberico:2004yc}. Here, we restrict our discussion to $\beta$-decays and $\beta$-equilibrium.

To understand the $\beta$-equilibrium constraints for a complex system of particles, we must consider the chemical potential of each baryon or quark, i.e.,
\begin{equation}\label{mu}
    \mu_i = B_i \mu_B + Q_i \mu_Q + S_i \mu_S \,.
\end{equation}
where $B_i$, $Q_i$ and $S_i$ are the baryon number, electric charge, and strangeness of the particle $i$, and $\mu_B$, $\mu_Q$ and $\mu_S$ are their associated chemical potentials (see App.~A of \cite{Aryal:2020ocm} for expressions for the baryon octet and light and strange quarks).
Then, for a system with at least neutrons $n$, protons $p$, Lambdas $\Lambda$, and leptons $l$, we define the  chemical potentials of these particles from \Cref{mu}
\begin{align}
    \mu_n&=\mu_B \label{22}\ ,\\
    \mu_p&=\mu_B+\mu_Q \label{23}\ ,\\
    \mu_\Lambda&=\mu_B-\mu_S\ .
\end{align}
Combining these with the neutrinoless neutron $\beta$-decay \Cref{1212}, we find
\begin{equation}\label{muQE}
    -\mu_Q=\mu_e\ ,
\end{equation}
and combining these with the balanced particle chemical potentials from the neutrinoless Lambda $\beta$-decay Eq.~\eqref{1414} ($\mu_\Lambda=\mu_p+\mu_e$), we find
\begin{equation}
    \mu_S=0
\end{equation}
such that as long as our system includes both $\Lambda$'s and n's in $\beta$-equilibrium, then $\mu_S=0$ (see out-of-beta equilibrium discussion for strangeness in \cite{Alford:2020pld}). 
If the system included only strange-baryons, then this condition would no longer hold. 
Furthermore, we point out that the $\mu_S=0$ condition is specific to a long-lived system, such as neutron stars where weak decays can occur. 
In heavy-ion collisions, the system is significantly too short-lived for weak decays to occur, such that one applies the (net) strangeness neutrality condition, i.e.,~$\langle n_S\rangle=0$, that leads to a finite $\mu_S>0$. 

Replacing Eq.~\eqref{muQE} into \Cref{eq:charge_neutrality3} for all the electrically charged leptons allows one to identify which $\mu_Q$ or $n_Q$ (for a given $\mu_B$ or $n_B$) fulfills charge neutrality \Cref{eq:charge_neutrality2}. In this way, we can eliminate one dimension, transforming the original EoS table (or external table) from 2D to 1D.

\subsubsection{Other cases}

Let us now briefly describe other cases that are not the focus of this paper, but are included in the MUSES module.
If one does not assume $\beta$-equilibrium,
\begin{equation}
    -\mu_Q\neq\mu_e\ ,
\end{equation}
and for each point in the original 2-dimensional EoS table (or external table) 
\Cref{eq:charge_neutrality3} must be solved for the lepton chemical potentials, leaving the EoS table 2-dimensional. This is the output that is read into the Flavor Eq.~module.

In the case of fixed lepton fraction (implemented only in $\beta$-equilibrium),
we can rewrite Eq.~\eqref{1212} as
\begin{equation}
    \mu_n=\mu_p+\mu_e-\mu_{l} ,
\end{equation}
which combined with Eqs.~\eqref{22} and \eqref{23} gives
\begin{equation}
-\mu_Q= \mu_e -\mu_{l} .
\end{equation}
In this case, \Cref{Yl} (fixed to a determined value) has to be solved together with charge neutrality \Cref{eq:charge_neutrality2}. See Eq.~9 of Ref.~\cite{Roark:2018uls} for an alternative (and somehow equivalent) definition of $\mu_Q$ for $\mu_{l}\ne0$.

\begin{table}[t!]
\centering
\begin{tabular}{cll}
\toprule
\textbf{Column} & \textbf{Quantity} & \textbf{Units} \\ 
\midrule
1 & Temperature ($T$) & MeV \\
2 & Baryon chemical potential ($\mu_B$) & MeV \\
3 & Strange chemical potential ($\mu_S$) & MeV \\
4 & Charge chemical potential ($\mu_Q$) & MeV \\
5 & Baryon density ($n_B$) & fm$^{-3}$ \\
7 & Strangeness density ($n_S$) & fm$^{-3}$ \\
7 & Charge density ($n_Q$) & fm$^{-3}$ \\
8 & Energy density ($\varepsilon$) & MeV fm$^{-3}$ \\
9 & Pressure ($P$) & MeV fm$^{-3}$ \\
10 & Entropy density ($s$) & fm$^{-3}$ \\
11 (optional)  & Particle baryon density ($n_B^{\rm particle}$) & fm$^{-3}$ \\
\bottomrule
\end{tabular}
\caption{Default format of the input and output files in MUSES.}
\label{tab:eos_format}
\end{table}

\subsubsection{Usage}

To use the Lepton module with an existing EoS, the minimal setup required is to enable the flags to compute only charge neutrality or charge neutrality with $\beta$-equilibrium, and to specify which leptons enter the computation. 

The input and output data follow the same column convention as the EoS modules described in \Cref{tab:eos_format}. The optional column with $n_B^{\rm particles}$ accommodates models that contain a rearrangement term to the baryon density, arising from $n_B= d P/d \mu_B$, such as the CMF model~\cite{Dexheimer:2009hi}, making it different from the density obtained from summing baryon and quark densities. 
\footnote{For models with rearrangement terms in the density, we calculate lepton and charge fractions (defined as a ratio of quantum numbers, $l/B$ and $Q/B$) with $n_B^{\rm particle}$, so if a tenth column is included in the input EoS table, the lepton fraction is calculated as $Y_{l}= l/B=n_{l}/n_B^{\rm particle}$ and the charge fraction is calculated as $Y_{Q}= Q/B=n_{Q}/n_B^{\rm particle}$~\cite{Dexheimer:2017nse,Roark:2018uls}. See Ref.~\cite{Typel:1999yq} for a discussion on rearrangement terms in a density-dependent (DD) mean-field model or Refs.~\cite{Grams:2018gjs,Pelicer:2021ils} for a discussion of rearrangement terms in a nuclear statistical equilibrium model.} 
All input EoS files must be in \texttt{CSV} format, though there are also options to output the EoS in CompOSE~\cite{Typel:2013rza,CompOSECoreTeam:2022ddl} or \texttt{HDF5} format. The module also produces an output for the Flavor equilibration module, if the proper flags are enabled. The column convention is described in \Cref{tab:flavor_format}.

Additional functionalities are available in the module by enabling the relevant flags. For example, the module can compute derivatives (such as speed of sound and susceptibilities), verify if stability and/or causality are respected, compute the beta-equilibrated matter EoS with trapped neutrinos (in which case $Y_{l}$ must be specified), and produce additional (optional) files with parameters specific to the EoS model, such as particle potentials and effective masses. In this case, an additional file must be provided with the model-specific parameters, which the module interpolates to obtain their values in equilibrated matter. The code can also be used as a free-Fermi gas code by changing the relevant lepton properties in the input particles data file.
Although there is an option to include $\tau$'s within the module, they are not relevant for neutron stars due to their high masses and are not discussed further in this paper.

\begin{figure*}[t!]
    \centering
    \includegraphics[width=0.33\linewidth]{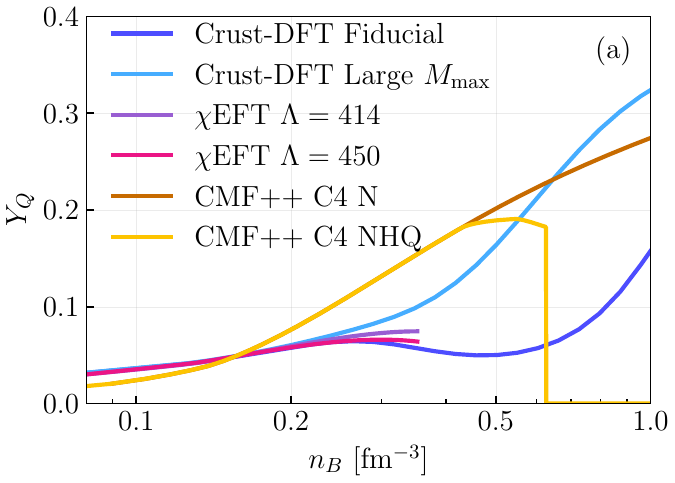}
    \includegraphics[width=0.33\linewidth]{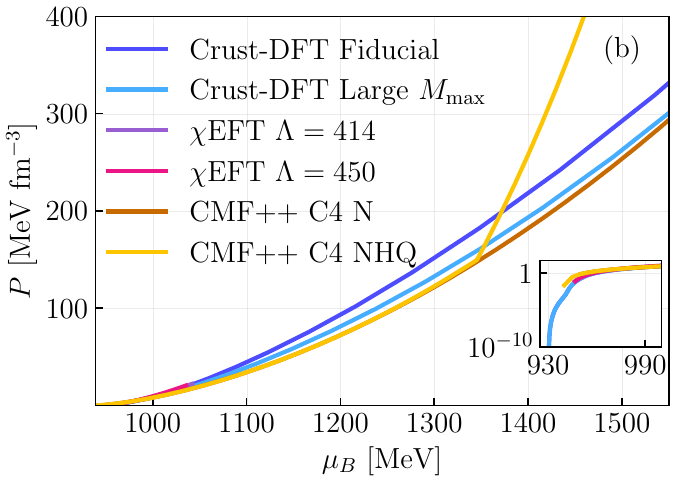}
    \caption{Left: Charge fraction as a function of baryon density.  Right: Pressure as a function of baryon chemical potential. We show two different parametrizations for each of the three nuclear EoSs explored in this paper in the $\beta$-equilibrated charge neutral case. The insert shows the low baryon chemical potential region.
    }
    \label{fig:lepton}
\end{figure*}

To compute the lepton chemical potentials for charge neutral matter (by solving \Cref{eq:charge_neutrality2} and, if required, \Cref{Yl}) and for $\beta$-equilibrium (by additionally imposing \Cref{muQE} and \Cref{eqn:mu_lep}), the Lepton module uses the Ceres library~\cite{Agarwal_Ceres_Solver_2022}. The Ceres library implements different trust regions and line-search methods to solve the non-linear least-square problem, from which the user can choose one. The default is the Levenberg-Marquardt method, which combines ideas from the Gauss-Newton and gradient-descent methods~\cite{KANZOW2004375}.

For the $\beta$-equilibrium computation, the code automatically processes the input data in sub-grids of constant $n_B$ or $\mu_B$, depending on the regularity of the data. Thus, the calculations are more accurate if the data has a regularly spaced grid. However, the module also works with irregular grids, provided the flag for enabling the multidimensional interpolator is set. 
If the grid is regularly spaced, for each $n_B$ or $\mu_B$, the algorithm interpolates in $\mu_Q$ to find $\mu_Q$ at equilibrium. All other quantities are obtained for $\beta$-equilibrated matter by interpolating them as functions of $\mu_Q$. Otherwise, if the grid is irregular, the algorithm generates a regular grid in density, and computes the equilibrium point at each $n_B$ by interpolating in $n_B$ and $\mu_Q$. Using the default configuration, the code takes $\sim 10^{-2}$~s per $n_B$ to calculate the solution to the $\beta$-equilibrium problem. Still, the run time depends on the precision and the methods chosen for the computation.

\subsubsection{Results at $\beta$-equilibrium}
\label{subsubsec:Results-beta}

As discussed in \Cref{sec:MUSES_overview}, the $\beta$-equilibrated, charge neutral EoS is first computed for each nuclear EoS module, before matching different modules. 
For this work, we demonstrate only two default parameter sets within each module that have been well-tuned to various data sets in nuclear and astrophysics. 
Let us discuss the choices we have made for this paper below.
\begin{itemize}
\item For Crust-DFT, we choose two representative sets of parameters that can be changed, but are default in the code. One model is called ``Fiducial'', which is a baseline model with typical values for all of the parameters and is shown in dark blue in the plots. We then compare the Fiducial line to another parameter set within Crust-DFT called ``Large $M_{max}$'' wherein the parameters were specifically chosen to have a large maximum mass for cold neutron stars in beta-equilibrium and is shown in light blue in the plots.
\item Since $\chi$EFT is an effective field theory, the concept of free parameters is not the same as in the other modules. However, one can vary the cutoff scale, $\Lambda$, which is done here. We choose two reasonable values of $\Lambda$ ($\Lambda=414\,\text{MeV}$ in purple and $\Lambda=450\,\text{MeV}$ in pink). Furthermore, $\chi$EFT lines are only shown in their regime of validity, such that they stop at $n_B=0.36$ $\rm{fm}^{-3}$. 
\item For CMF++, we choose the C4 coupling scheme. Within CMF various coupling schemes are possible, which are just ansatze for the interactions.
C4 was chosen here because it gives the most accurate neutron star description out of all CMF coupling schemes \cite{Dexheimer:2015qha}. The dark orange curve (C4 N) includes only nucleons, while the yellow curve (C4 NHQ) includes nucleons, hyperons from both the octet and decuplet, and Delta baryons (though no baryons from the decuplet appear in this particular coupling scheme, parametrization, and conditions), and light and strange quarks.
\end{itemize}

In \Cref{fig:lepton},  panel (a) shows $Y_Q(n_B)$, while panel (b) shows 
$P(\mu_B)$, both for charge neutral $\beta$-equilibrium matter. $\chi$EFT and Crust-DFT have fairly similar $Y_Q(n_B)$ values around $n_{sat}$ ($0.1-0.2$ fm$^{-3}$), which is to be expected because the parameter set used in this work in Crust-DFT was specifically tuned to reproduce our renderings of $\chi$EFT. The charge fraction $Y_Q$ in CMF++ remains slightly higher around $n_{sat}$, generally having a larger $Y_Q(n_B)$ than the other models. 
At high $n_B$, $\chi$EFT reaches the limits of its regime of validity. However, we can compare CMF++, which includes a variety of hadronic and quark states, to Crust-DFT, which is nucleonic only. 
The $Y_Q(n_B)$ for Crust-DFT appears to be strongly dependent on its parameters, such that Fiducial line has a much lower $Y_Q(n_B)$, whereas the parameter set that leads to a large maximum mass has a significantly larger $Y_Q(n_B)$ that even surpasses 0.3 at large enough $n_B$.
We can then compare this to CMF++ where the nucleon only configuration of C4 has a similar $Y_Q(n_B)$ to Crust-DFT in that it also reaches large $Y_Q(n_B)$ at high $n_B$ (although the shape of $Y_Q(n_B)$ is different between the two models).
However, when hyperons and quarks are included in CMF++, we find first a decrease in $Y_Q(n_B)$ when hyperons appear, and then a dramatic drop across the deconfinement phase transition to $Y_Q(n_B)\rightarrow 0$.

In panel (b) of \Cref{fig:lepton} we observe the differences in $P(\mu_B)$ at $\beta$-equilibrium between our different models. 
Before beginning our discussion, we caution readers that comparing models in $P(\mu_B)$ is not the same as comparing models in $P(\varepsilon)$, such that the ordering of EoSs and what is considered ``stiff'' vs ''soft'' is rather non-trivial. This is because of the relation between $n_B$ and $\mu_B$, which can be seen in the Gibbs-Duhem relation (here shown in the $T \to 0$ limit),
\begin{equation}
    P+\varepsilon=n_B\mu_B.
    \label{eq:GD}
\end{equation}
However, $P(\mu_B)$ makes it easier to compare EoSs when first-order phase transitions are present, which happens for CMF++ with quarks. 
At low $\mu_B$, the differences among the different nuclear EoSs are not easily perceivable in $P(\mu_B)$, which is unsurprising. This happens because all models are tuned to reproduce saturation properties, but also because, even though there are differences in $Y_Q(n_B)$, the influence of isospin asymmetry around $n_{sat}$ is only a very small effect (they play a larger role at larger $n_B$).
At high $\mu_B$, we find larger differences between the models, although CMF++ with nucleons only and Crust-DFT with Large $M_{max}$ appear to be similar.
However, we see a sharp divergence of CMF++ with hyperons and quarks, which coincides with the onset of deconfinement. 
The kink seen in $P(\mu_B)$ for CMF++ C4 NHQ is precisely a typical signature of a first-order phase transition, where the rapid rise in $P(\mu_B)$ at higher $\mu_B$ is when the quarks appear. 

\begin{figure*}[t!]
    \centering
    \includegraphics[width=0.328\linewidth]{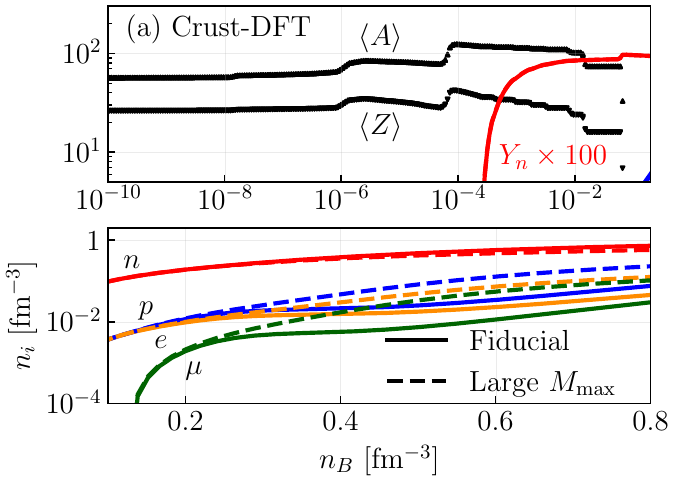}
    \includegraphics[width=0.328\linewidth]{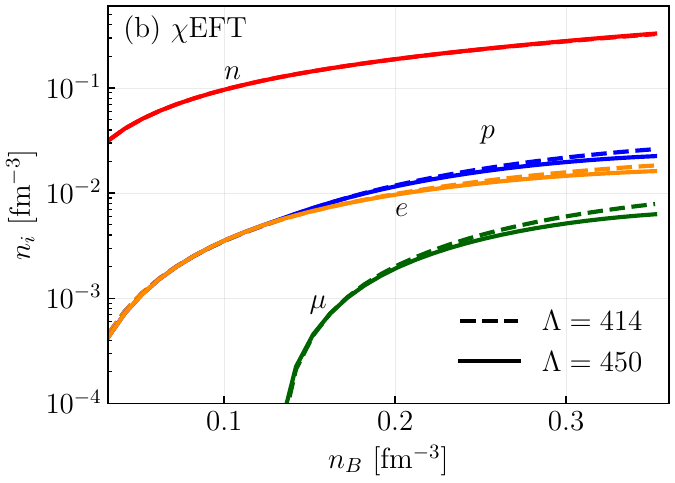}
    \includegraphics[width=0.328\linewidth]{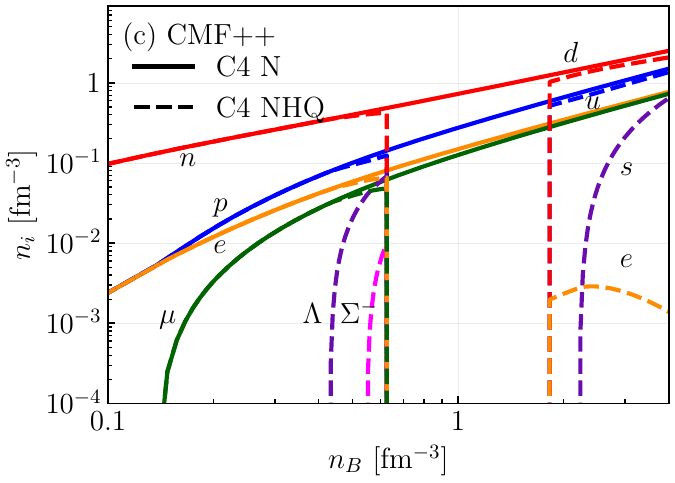}
    \\
    \includegraphics[width=0.328\linewidth]{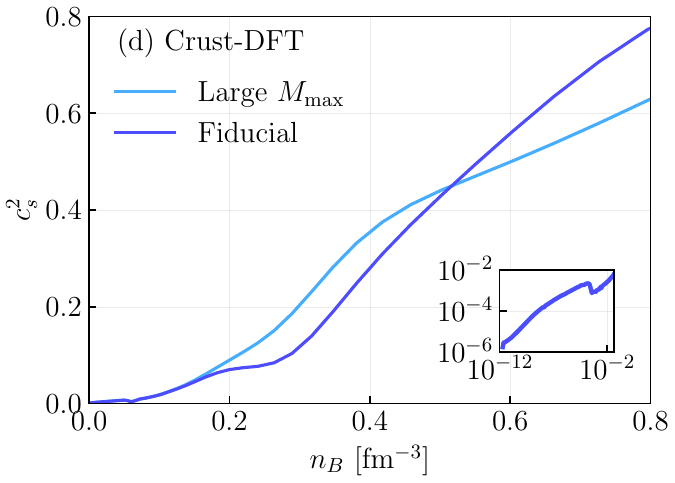}
    \includegraphics[width=0.328\linewidth]{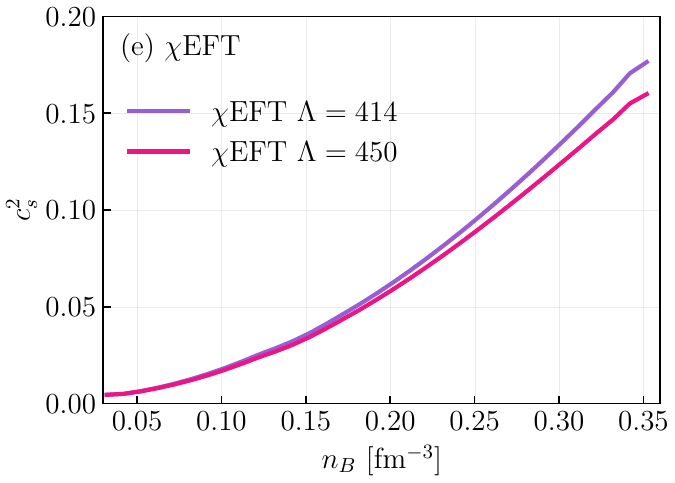}
    \includegraphics[width=0.328\linewidth]{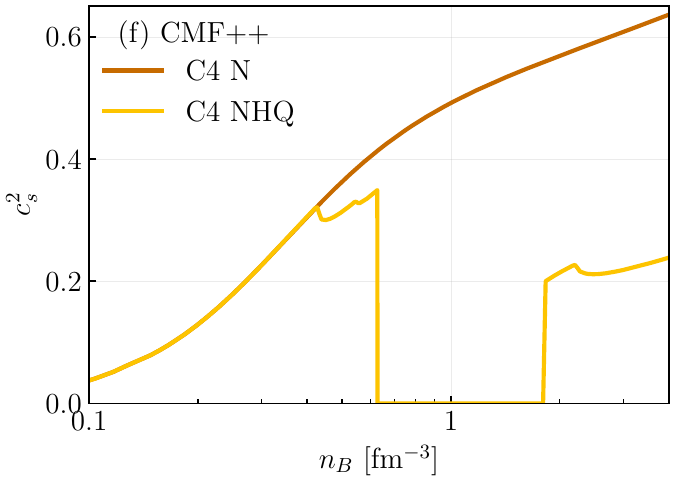}
    \caption{Top: Particle densities as a function of baryon density for Crust-DFT (left), $\chi$EFT (middle), CMF++(right).  
    Bottom: Speed of sound squared as a function of baryon density for Crust-DFT (left), $\chi$EFT (middle), CMF++(right). 
    We show two different parametrizations for each of the three EoSs explored in this paper for the $\beta$-equilibrated charge neutral case. 
    The particle number densities are shown for nucleons and leptons for Crust-DFT as well as the average number of nucleons within nuclei $\langle A\rangle$, the average number of protons within nuclei $\langle Z\rangle$, and the fraction of free neutrons when nuclei are present. 
    For $\chi$EFT the particle number densities of nucleons and leptons are shown. 
    For CMF++ the particle number densities for hadrons, quarks, and leptons are shown. 
}
    \label{fig:pop}
\end{figure*}

In \Cref{fig:pop}, the top panels show the particle populations for each EoS module, while the bottom panels show the slope of the EoS (\Cref{eq:cs-def})~\footnote{Note that at $\beta$-equilibrium there is only one independent chemical potential, such that the usual partial derivative of $c_s^2$ turns into a total derivative.}. Both top and bottom rows are shown as functions of $n_B$ for charge neutral, $\beta$-equilibrated matter. 
It is useful to plot $c_s^2(n_B)$ in comparison to population plots (i.e.,~number densities of individual particles $n_i$ versus $n_B$) because changes in the populations (as in the appearance of new particles) can lead to kinks (third-order phase transitions) or plateaus (first-order phase transitions) in $c_s^2(n_B)$.

On the left side of \Cref{fig:pop} we present results for the Crust-DFT module.
Again, we show two parametrizations: ``Fiducial'' and ``Large $M_{max}$'' for Crust-DFT, which differ on the choice of high-$n_B$ EoS parametrizations from \cite{Steiner:2014pda}. 
Since Crust-DFT begins at very low $n_B$, the properties of nuclei are shown up to $n_{sat}$.  
One can see on the top of panel (a) that the nuclei dominate up until $n_B\gtrsim 10^{-4}$ fm$^{-3}$, at which density free neutrons rapidly appear. 
In the regime where the nuclei dominate, the average number of nucleons and protons are relatively constant starting at the lowest $n_B$ up until $n_B\gtrsim 10^{-6}$ fm$^{-3}$. 
After that point, there is a clear jump that increases both $\langle A\rangle$ and $\langle Z\rangle$. 
However, after that, $\langle A\rangle$ continues to either remain reasonably flat, while $\langle Z\rangle$ decreases. 
There is another similar jump just below $n_B< 10^{-4}$ fm$^{-3}$. 
What is happening in the regime where $\langle A\rangle\sim const$ but $\langle Z\rangle$ decreases is that the system prefers to decrease $Y_Q(n_B)$ rather than add more nucleons to the nuclei. Eventually, a point is reached where it becomes advantageous to have heavier nuclei dominate the system, such that another jump appears in both $\langle A\rangle$ and $\langle Z\rangle$. 

Comparing $\langle A\rangle$ in  \Cref{fig:pop} to \Cref{fig:cdft1}, we find that at $\beta$-equilibrium the maximum possible $\langle A\rangle$ is closer to $\langle A\rangle_{max}\sim 120$ (as compared to $\langle A\rangle_{max}\sim 350$ for $Y_Q=0.4$). 
We see a lower $\langle A\rangle_{max}$ at $\beta$-equilibrium because $Y_Q$ is significantly smaller there; then, there are fewer protons within nuclei, such that the neutron drip line is reached at smaller values of $A$. 
At these very low densities, it is harder to see this structure in $c_s^2$ because it is so small on this scale. 
However, one can see a kink in $c_s^2(n_B)$ when the free neutrons appear (see the inset in panel (d) of \Cref{fig:pop}).

At $n_B>n_{sat}$ the system is always composed of protons, neutrons, and leptons, such that major changes in the $c_s^2(n_B)$ are not anticipated. 
Indeed, one finds that both $c_s^2(n_B)$ from Crust-DFT monotonically increases with $n_B$. 
However, there is a clear difference between the two parametrizations of Crust-DFT. 
The Large $M_{max}$ case has a larger $c_s^{2}(n_B)$ at intermediate $n_B$, which leads to  heavier neutron star maximum masses. The higher $n_B$ behavior of $c_s^2$ is the opposite in that Fiducial has a larger $c_s^2$, but that regime contributes little or not at all to $M_{max}$ (see, e.g., \cite{Tan:2021ahl} for discussion on how the behavior of $c_s^2(n_B)$ affects maximum masses).
In turn, we find that parametrizations that reproduce heavier neutron stars also appear to be more proton-rich (or in other words, have larger $Y_Q(n_B)$--see panel (a) of \Cref{fig:lepton}). 
This observation is consistent with our understanding of the symmetry energy slope \cite{Lopes:2024bvz,Dexheimer:2015qha}.

Next, we discuss $\chi$EFT in terms of its population and $c_s^2(n_B)$, where the only parameter we have to play with is the cut-off scale $\Lambda$. These quantities are shown in the middle panels ((b) and (e), respectively) of \Cref{fig:pop}.
Similar to Crust-DFT, $\chi$EFT only allows for the possibility of protons, neutrons, and leptons (not counting mesons that mediate interactions). 
Generally, the smaller the cut-off scale (N3LO-414), the stiffer the $c_s^2(n_B)$ at large $n_B$, as compared to N3LO-450. 
This can be explained by $E_{sym}$, since the slope $L$ is slightly larger for the N3LO-414 parametrization than the N3LO-450 one.
Once again, we find that an EoS rendering with a large $c_s^2(n_B)$ leads to a more proton-rich system at higher $n_B$ or rather a larger $Y_Q(n_B)$.

Finally, we discuss CMF++, which covers the $n_B\gtrsim n_{sat}$ regime of the EoS, shown in panels (c) and (f) of \Cref{fig:pop}.
For the nucleon only EoS, we find similar results to Crust-DFT and $\chi$EFT in that we have a monotonically increasing $c_s^2$. 
The other CMF++ rendering (C4 NHQ), which contains $\Lambda$, $\Sigma^-$, and quarks, leads to a slightly smaller $M_{max}$ (see Fig.~7 of \cite{Roark:2018boj}). The appearance of the neutral $\Lambda$s at $n_B\sim0.45$ $\rm{fm}^{-3}$ (panel c) of \Cref{fig:pop}) triggers a kink in $Y_Q$ (panel (a) of \Cref{fig:lepton}). This appearance of $\Lambda$'s causes a softening of the EoS (panel (f) of \Cref{fig:pop}), followed by another smaller one for the $\Sigma$s, followed by a large plateau in $c_s^2=0$ for the quarks at $n_B=0.6$ $\rm{fm}^{-3}=4\,n_{sat}$. 
The hyperon kinks are too subtle to be seen in $P(\mu_B)$ in panel (b) of \Cref{fig:lepton}, but can be clearly seen as kinks in its derivative ($c_S^2(n_B)$), shown in panel (f) of \Cref{fig:pop}. There, the zero value shows well the deconfinement first-order phase transition, followed by a small kink due to the appearance of strange quarks. The quark phase presents very few leptons (panel (c) of \Cref{fig:pop}), which explains why $Y_Q\sim0$ during the first-order phase transition (panel (a) of \Cref{fig:lepton}), as expected. 

\subsection{Synthesis}
\label{sec:synthesis}
 
The Synthesis module combines different EoSs into a single EoS across overlapping ranges of validity in $n_B$.  In Synthesis, it is possible to either smoothly match EoSs across a range of a thermodynamic variable (e.g., $n_B$), implement a thermodynamically consistent first-order phase transition between two EoSs, or force a first-order phase transition into another phase of matter, even if it was unstable (we do not discuss this last option further here, but allow it as an option for the user).
All three of our EoSs in this paper overlap both in the range of validity in terms of $n_B$, but also within the same phase of matter. 
For instance, Crust-DFT, in principle, covers very low $n_B$, as well as the liquid-gas phase transition and N-body nucleonic interactions; $\chi$EFT also covers the liquid-gas phase transition and N-body nucleonic interactions; CMF++ covers only the high $n_B$ side of the liquid-gas phase transition (i.e.,~the liquid-nucleonic state) but then describes more realistically the high $n_B$ regime. 
Thus, the type of matching is an option for the user, depending on the type of physics they would like to explore. 

We point out that, while Synthesis was originally designed to smoothly match Crust-DFT, $\chi$EFT, and CMF++, it was also designed to flexibly take in any EoS grid (e.g., a table from CompOSE in MUSES format)  that can replace any (or all) of the EoS modules. In the future, other EoS modules will also be available in MUSES that could be connected via Synthesis. 
In the following, we discuss in detail the procedure for either smoothly matching EoSs across a given range (of a chosen thermodynamic variable) or implement a thermodynamically-consistent first-order phase transition between two EoSs.

\subsubsection{First-order phase transitions}
\label{Synthesis1st}

First-order phase transitions with more than one globally conserved charge are said to be non-congruent. Examples of these are phase transitions in heavy-ion collisions where $B$, $Q$, and $S$ are conserved, or phase transitions in neutron stars where $B$ and $Q$ are conserved, and sometimes also $l$, the lepton number (when neutrinos cannot free stream). Non-congruent phase transitions differ from congruent phase transitions in, e.g., dimensionality of phase diagrams and location of critical points. In non-congruent phase transitions, the concentration of, e.g.,~baryon number and electric charge vary across the phase transformations, creating phase coexistence regions (usually referred to as mixed phase). An exception is isospin-symmetric matter with $Y_Q=0.5$, in which the requirement of a null isospin chemical potential leads necessarily to a congruent phase transition, even when baryon number is also conserved (referred to as azeotropic behavior); see Ref.~\cite{Hempel:2013tfa} with references within for a thorough review of the topic.

As we investigate different phases in the core of neutron stars, where the nuclear force is dominant over the Coulomb force, we treat first-order phase transitions as Coulomb-less. We also ignore surface tension effects and focus on two limits for the description of a first-order phase transition, one in which the surface tension between the phases is infinite, resulting in a forced-congruent phase transition with locally conserved quantities (referred to in astrophysics as a ``Maxwell construction'') and one at which it is zero, resulting in a non-congruent phase transition with globally conserved quantities (referred to in astrophysics as a ``Gibbs construction''). In reality, the surface tension is neither, and a more complicated approach to study the mixture of phases is necessary \cite{Mariani:2023kdu,Grunfeld:2020gnv,Constantinou:2023ged,Avancini:2012ee}. Nevertheless, it has been shown that specific stellar properties (e.g.,~mass and radius) are approximately the same in both approaches, and both limits can provide a band for the possible results with surface tensions~\cite{Glendenning:2001pe,Bhattacharyya:2009fg}. In this work, we discuss the implementation of a Maxwell construction and a Gibbs construction to describe the quark deconfinement first-order phase transition in MUSES within the CMF++ module. See Ref.~\cite{Roark:2018uls} for examples of mixed phases in the CMF model also including fixed lepton fraction.

\begin{enumerate}
\item {\bf Maxwell construction (forced-congruent phase transition):} we impose that $P$ and $\mu_B$ are the same in both phases at the phase transition, i.e.,
\begin{equation}\label{eq:maxwell}
    P^I = P^{II}\,, \quad \mu_B^I = \mu_B^{II}\,,
\end{equation}
whereas we currently assume both EoSs are in thermal equilibrium in the code. This is equivalent to computing the Maxwell construction through the equal-area method (see Appendix A of~\cite{Vovchenko:2015vxa}). Since charge neutrality is imposed locally within each phase (with a sharp boundary between them), this makes the thermodynamic variables of $\mu_Q$, $\varepsilon$, $n_B$, and particle densities $n_i$ discontinuous~\cite{Bhattacharyya:2009fg}. 

\item {\bf Gibbs construction (non-congruent phase transition):} in the Gibbs construction, instead of a sharp boundary, we build a region with a mixed phase, where the constraints of the Maxwell construction (\Cref{eq:maxwell}) are still valid, but charge neutrality is imposed globally instead of locally
\begin{equation}\label{eq:gibbs_charge_neutrality}
    f n_Q^I + (1-f) n_Q^{II} + n_{{\rm leptons}, Q} =0\,.
\end{equation}
Here I and II are labels for the two phases involved in the transition. Phase I occupies a volume fraction $f$ of a volume element and $(1-f)$ is the volume fraction occupied by phase II, where both phases are homogeneously mixed. In $\beta$ equilibrium, it is requires
\begin{equation}
    \mu_Q^I= \mu_Q^{II} = - \mu_e \,,
\end{equation}
where the electric charge potential is determined from solving 
\Cref{eq:gibbs_charge_neutrality}.
\end{enumerate}

The computation of a first-order phase transition creates several EoS branches. The one with the largest $P$ at a fixed $\mu_B$ is (thermodynamically) favored, and considered the stable branch at that $\mu_B$. This is so provided the EoS also obeys the stability conditions of positive compressibility and the chemical hardness~\cite{chandler1987introduction} (equivalent to Eq.~(E27) of~\cite{Cruz-Camacho:2024odu} for baryon number, as in the $\beta$-equilibrated, charge neutral case at $T\sim0$ the problem becomes 1D), namely
\begin{equation}\label{eq:stability}
    \frac{d P}{d n_B} \geq 0\,, \qquad \frac{d \mu_B}{d n_B} \geq 0\, .
\end{equation}
The second inequality is equivalent to demanding that the second susceptibility remains positive.
The other branches are called metastable if they respect \Cref{eq:stability} but have lower $P$, and are unstable otherwise.

\begin{figure}[t!]
\includegraphics[width=0.8\linewidth]{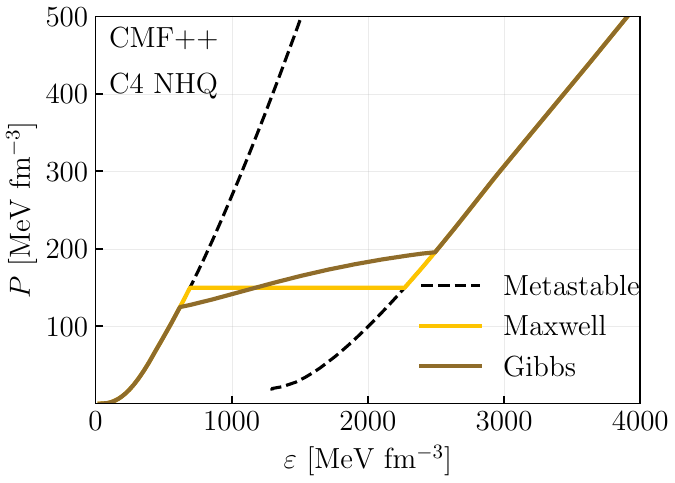}
\caption{Equation of state for the CMF++ module including nucleons, hyperons and quarks. The deconfinement transition is calculated using the Maxwell (yellow) and Gibbs (brown) constructions. The dashed black curve represents metastable phases.  }
\label{fig:cmf-pt}
\end{figure}

The results of applying the Maxwell construction to describe the quark deconfinement first-order phase transitions in the CMF++ module were already shown in \Cref{fig:pop}. Now we extend this discussion and also show the Gibbs construction in \Cref{fig:cmf-pt}. 
In the latter,  we
show the stable branches in yellow connected by a horizontal line, which gives rise to the Maxwell construction. 
We also show two metastable branches as dashed black lines. Finally, the Gibbs construction is shown in brown, where one can easily identify an intermediate mixed phase between the hadronic and quark phases.
The difference between the Gibbs and Maxwell constructions leads to a slightly different $M(R)$ curve, where the Gibbs construction has a slightly higher $M_{max}$ than the Maxwell construction (not shown here but also reproduced using QLIMR), see Fig.~7 of \cite{Roark:2018boj}.  Essentially, the Maxwell construction for this EoS leads to an instability at the onset of deconfinement.

To solve the equations involved, the Synthesis module uses the Ceres Solver library~\cite{Agarwal_Ceres_Solver_2022}, like the Lepton module. It also uses the Levenberg-Marquardt method as default and allows the user to choose different trust region and line search methods to solve the non-linear least-square problem. The code takes between 0.05s and 0.1s to compute the Maxwell construction or one point in the Gibbs construction.

\subsubsection{Smooth matching between EoS}

The method used to ensure smooth transitions between EoSs is the hyperbolic tangent interpolation, such that there is no discontinuity between the EoSs or phases. 
The procedure consists of interpolating some thermodynamic function $Y$ as a function of another variable $x$ such that we combine the EoSs using
\begin{equation}\label{eq:Yx}
    Y(x) =  Y^I(x) f_- (x)+ Y^{II}(x) f_+(x)\,,
\end{equation}
where the interpolating functions $f_\pm(x)$ are defined as
\begin{equation}\label{eq:interpolation_functions}
    f_\pm (x) = \frac{1}{2} \left( 1 \pm \tanh{\left[ \frac{x - \bar x }{\Gamma} \right]} \right)\,.
\end{equation}
The parameter $\bar x$ represents the midpoint of the interpolation and the parameter $\Gamma$ is related to the width of the interpolated region~\cite{Masuda:2012ed}. 

The user can decide to either smoothly match using thermodynamic variables like $P(\mu_B), P(n_B), \varepsilon(n_B)$, or to smoothly match in the derivative of $P(\varepsilon)$, i.e.,~$c_s^2(n_B)$, instead.
There are different approaches to obtain all necessary thermodynamic quantities for both methods.
If one begins with $P$ or $\varepsilon$, 
then a combination of one derivative and the use of the Gibbs-Duhem relation (\Cref{eq:GD}) is required to recover the minimal thermodynamic quantities: $\left\{P,\varepsilon,n_B,\mu_B\right\}$. 
To avoid numerical noise, we compute derivatives analytically when possible, using the chosen $Y(x)$ to recover one thermodynamic quantity. Such a derivative typically carries derivatives of $f_\pm$, so we define for future convenience
\begin{align}
g(x)  & = \frac{df_+}{dx} = - \frac{df_-}{dx} = \frac{2}{\Gamma\left(e^{-(x-\bar{x})/\Gamma}+e^{(x-\bar{x})/\Gamma} \right)^2 }\,.
\end{align}
With that in hand, we then compute the remaining thermodynamic quantity through the Gibbs-Duhem relation.

The following four $Y(x)$ combinations are allowed in the Synthesis module:
\begin{enumerate}
\item $P(\mu_B)$: when interpolating $P$ as a function of $\mu_B$, $n_B$ can be calculated from the relation
\begin{equation}
    n_B(\mu_B)= \frac{d P}{d \mu_B}\,,
\end{equation}
which results in the analytic expression
\begin{equation}
    n_B(\mu_B)= f_- n_B^I + f_+ n_B^{II} + \Delta n_B\,,
\end{equation}
where the correction term for the density is given by
\begin{equation}\label{eq:PmuB_nBrearrangement}
    \Delta n_B= -  g(\mu_B)  \left( P^I - P^{II} \right)\,.
\end{equation}
For the energy density, we use the Gibbs relation, $\varepsilon(\mu_B) = n_B(\mu_B) \mu_B - P(\mu_B)$.

\item $\varepsilon(n_B)$: when interpolating $\varepsilon$ as a function of $n_B$, we can calculate $P$ from 
\begin{equation}\label{eq:pressure}
P(n_B)= n_B^2 \frac{d \left( \varepsilon/n_B\right)}{d n_B}\,,
\end{equation}
such that
\begin{equation}
P (n_B) = f_- P^I + f_+ P^{II} + \Delta P\,,
\end{equation}
with 
\begin{equation}\label{eq:EnB_Prearrangement}
    \Delta P = - g(n_B) \, n_B \left( \varepsilon^I - \varepsilon^{II}\right)\,.
\end{equation}
For the chemical potential, we again use the Gibbs relation, but this time in the form $\mu_B(n_B)  = (\varepsilon(n_B) + P(n_B))/n_B $.

\item $P(n_B)$: when interpolating $P$ as a function of $n_B$, we can compute $\varepsilon$ by integrating \Cref{eq:pressure} by parts, which leads to~\cite{Masuda:2012ed}
\begin{equation}
    \varepsilon(n_B)= \varepsilon^I  f_- + \varepsilon^{II} f_+ + \Delta \varepsilon\,,
\end{equation}
with
\begin{equation}\label{eq:PnB_Erearrangement}
    \Delta \varepsilon= n_B \int_{\bar{n}_B}^{n_B} \frac{d n_B'}{n_B'} g(n_B) \left(\varepsilon^I - \varepsilon^{II} \right)\,.
\end{equation}
For the chemical potential, we use the Gibbs relation in the form $\mu_B(n_B)  = (\varepsilon(n_B) + P(n_B))/n_B $, just as before.
\item $c_s^2(n_B)$: when interpolating $c_S^2$ as a function of $n_B$, we solve the differential equations
\begin{align}
\frac{d\varepsilon}{dn_B} &= \frac{\varepsilon + P}{n_B}\,,
\\
\frac{dP}{dn_B} & = c_s^2(n_B) \frac{\varepsilon + P}{n_B}\,,
\end{align}
which can be done in discrete form using a linear stencil via~\cite{Tan:2021ahl}
\begin{equation}
\begin{split}\label{eq:cs2-tanh}
    &n_{B, i+1} = n_{B, i} + \Delta n_B\,, \\
    &\varepsilon_{i+1} = \varepsilon_i + \Delta n_B \left( \frac{\varepsilon_i + P_i}{n_{B, i}} \right)\,, \\
    &P_{i+1} = P_i + c_s^2(n_{B, i}) \Delta n_B \left( \frac{\varepsilon_i + P_i}{n_{B, i}} \right)\,.
\end{split}
\end{equation}
In principle, one could instead use a Runge-Kutta method for this integration, as was done in \cite{Mroczek:2024sfp}, but we leave that for a future work.
For the chemical potential, we use the Gibbs relation in the form $\mu_B(n_B)  = (\varepsilon(n_B) + P(n_B))/n_B $.
\end{enumerate}
One essential difference between the $P(n_B)$ interpolation and the others, is that the corrections to $\varepsilon$ and $\mu_B$ are not restricted to a region around the interpolation midpoint $\bar{x}$~\cite{Masuda:2012ed} due to the correction being an integral. 

For case 4, $c_s^2(n_B)$, one starts with a derivative, allowing all quantities to be computed through direct integration. The first step is to choose a starting point for the integration, say at $n_B^0$. This must be somewhere below the matching midpoint ($n_B^0<\bar{n}_B$), such that the effect of the $\tanh$ is negligible; such as $n_B^0\lesssim \bar{n}_B-3\Gamma $. 
For instance, if one ``trusts'' the low $n_B$ region more than the high $n_B$ EoS, this low $n_B$ EoS would be our EoS I. In this case, the user would choose a low value for $n_B^0$. Another EoS would be chosen to describe the high $n_B$ region (EoS II), which would be used for the speed of sound interpolation according to \Cref{eq:Yx}. All other quantities are then integrated into the high $n_B$ region following \Cref{eq:cs2-tanh}, where the initial point belongs to EoS I.

\begin{table}[t!]
\begin{tabular}{ccc}
\toprule
    {Crust-DFT + $\chi$EFT} & {$\bar{x}$} & {$\Gamma$} \\ 
    \midrule
    $P(\mu_B)$ & 950.0 & 10.0 \\
    $\varepsilon(n_B)$ & 0.090 & 0.03 \\ 
    $P(n_B)$ & 0.100 & 0.02 \\ 
    $c_s^2(n_B)$ & 0.065 & 0.01 \\ 
    \bottomrule
\end{tabular}
\caption{Hyperbolic-tangent parameters ($\bar{x}, \Gamma$) used to generate the interpolated EoS between Crust-DFT and $\chi$EFT using different thermodynamic variables. }
\label{tab:dft-ceft}
\end{table}

\begin{figure*}[t!]
\centering
\includegraphics[width=0.75\linewidth]{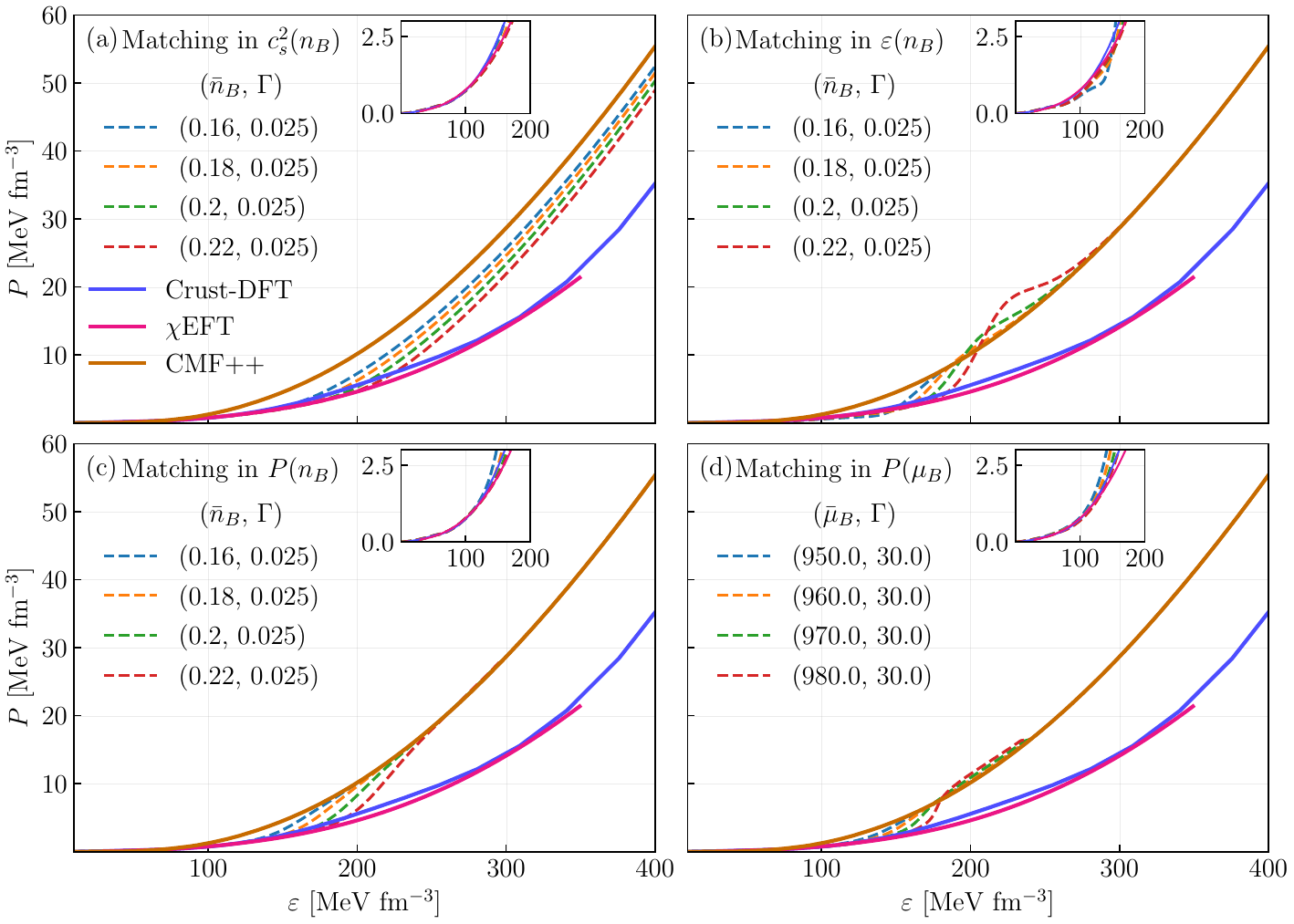}
\caption{Pressure as a function of energy density for EoSs smoothly matched using different thermodynamic variables: (a) speed of sound squared as a function of baryon density; (b) energy density as a function of baryon density; (c) pressure as a function of baryon density; (d) pressure as a function of baryon chemical potential.  
Solid lines represent the original model's EoSs, while dashed lines show smoothly matched EoSs with numbers representing the midpoint and width of the interpolation. The insets show the matching between Crust-DFT and $\chi$EFT.}\label{fig:eos-tanh}
\end{figure*}

Numerically, we use linear interpolation in the EoSs to compute the matching and an adaptive integration~\cite{cubature} to evaluate \Cref{eq:PnB_Erearrangement}. The code takes between 0.05 and 0.1s to output the interpolated EoS using the default configuration for about 150 points in the interpolated region.
Moreover, as in the Lepton module, additional functionalities are available, such as computing the speed of sound and susceptibilities and requesting the output in \texttt{HDF5} or CompOSE format. Extra functionalities are available both for a first-order phase transition and for smooth matching. All input and output files follow the convention described in \Cref{tab:eos_format}.

\subsubsection{Results of smoothly matched EoS with different approaches}
\label{SynthesisSmooth}

\begin{figure*}[t!]
\includegraphics[width=0.75\linewidth]{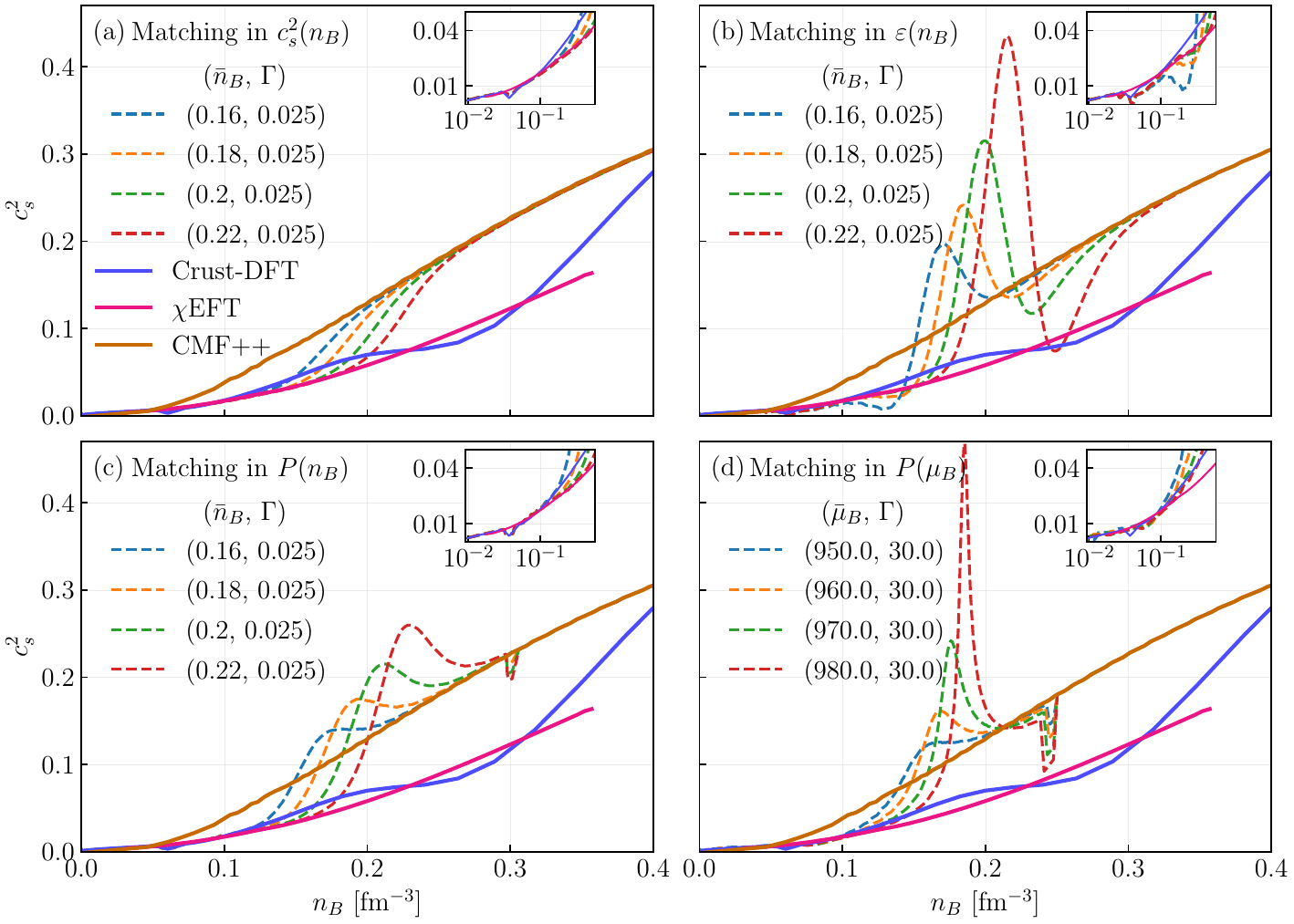}
\caption{Speed of sound squared as a function of baryon density for EoSs smoothly matched using different thermodynamic variables: (a) speed of sound squared as a function of baryon density; (b) energy density as a function of baryon density; (c) pressure as a function of baryon density; (d) pressure as a function of baryon chemical potential.  
Solid lines represent the original model's EoSs, while dashed lines show smoothly matched EoSs with numbers representing the midpoint and width of the interpolation. The insets show the matching between Crust-DFT and $\chi$EFT.}
\label{fig:cs2-tanh}
\end{figure*}

For each choice of (smooth) matching thermodynamic variable $Y(x)$, we first match Crust-DFT and $\chi$EFT, and then we match this combined EoS with the CMF EoS. 
Because the choice of parameters for Crust-DFT and $\chi$EFT do not significantly impact the properties of the neutron star (which we explore in the next section), we fix them to a unique value for each matching case, which are shown in \Cref{tab:dft-ceft}. 
These numbers were chosen 
so that (i) the final EoS is stable and (ii) the transition from Crust-DFT to $\chi$EFT occurs close to the liquid-gas phase transition, (see, e.g.,~the appearance of the homogeneous neutron gas in the Crust-DFT population--panel (a) of \Cref{fig:pop}). 
Our choices produce EoSs that can describe the crust and the outer core of neutron stars.

In the second (smooth) matching, we join the new combined EoS with CMF++. The choices made when matching to CMF++ (unlike the first matching at low $n_B$) have a significant impact on neutron star properties. 
Thus, we explore different parameters for the midpoint of the interpolation ($\bar{x}$), which is more influential than the width ($\Gamma$). 
We choose the midpoint between 0.16 and 0.22 fm$^{-3}$ and a fixed $\Gamma = 0.025$ fm$^{-3}$. This ensures that the CMF++ model dominates above 2~$n_{\rm sat} \approx 0.32$ fm$^{-3}$, meaning that $f_- \rightarrow 0$ in \Cref{eq:Yx}. For the $P(\mu_B)$ interpolation, we choose $\Gamma$ and $\bar{\mu}_B$ values that remain close to the specified $n_B$ region, but each model has a different $\mu_B$ at a fixed $n_B$.

In panel (a) of \Cref{fig:eos-tanh}, we show combined EoSs produced with the $c_s^2(n_B)$ interpolation, in panel (b) we show EoSs produced with $\varepsilon(n_B)$, in panel (c) $P(n_B)$, and in panel (d) $P(\mu_B)$.
While the original CMF++ $P(\varepsilon)$ is recovered in three of the matching thermodynamic variables, the matching in  $c_s^2(n_B)$ never returns to the original $P(\varepsilon)$. 
We can understand this deviation from the original $P(\varepsilon)$ in the following manner.
Since the matching in $c_s^2(n_B)$ begins with a derivative and integrates upwards in $n_B$ to recover the EoS, we can always recover the slope of the EoS in the $P(\varepsilon)$-plane.  
In order to recover the actual pressure, one would need to start the integration at a $P(n_B^0)$ that is identical for both matched EoSs. 
However, that would mean that the matched EoSs would need to be identical at the point of matching, not just in $P(\mu_B)$ but also in $c_s^2(n_B)$, which never occurs for these specific renderings of our modules.
Note that it may be possible to find such a point when varying over the large parameter space of all our modules, but we leave that exercise for a future work.

Each choice of matching variable has its advantages and disadvantages. The $\varepsilon(n_B)$ matching, for example, produces bumps in the pressure due to the rearrangement term~\Cref{eq:EnB_Prearrangement}. Meanwhile, although the $P(n_B)$ matching does not seem to produce any artificial structure in the EoS, there is a slight mismatch between the interpolated $\varepsilon$ and $\mu_B$ with respect to the CMF++ one due to the rearrangement \Cref{eq:PnB_Erearrangement}. Finally, the $P(\mu_B)$ matching modifies $\varepsilon$ and the $n_B$ through \Cref{eq:PmuB_nBrearrangement}. 
Therefore, one must choose what thermodynamic variable they deem the most important to preserve with the acceptance that the other thermodynamic variables deviate from their original values. 
In the case that $c_s^2(n_B)$ is chosen as the thermodynamic variable we prefer to preserve, one then obtains deviations in $P(\varepsilon)$. The advantage of this method is discussed below.

In \Cref{fig:cs2-tanh}, we show $c_s^2$ as a function of $n_B$.
We find that smooth matching in $P(n_B)$ or $P(\mu_B)$ (which seemed to reproduce best the EoSs in the previous figure), actually introduces significant, artificial bumps in $c_s^2$. 
The same effect happens for the $\varepsilon(n_B)$ matching. 
Additionally, when matching in $P$, a dip in $c_s^2$ appears at the end of the overlap between $\chi$EFT and CMF EoSs, which arises due to the discontinuity created by the non-zero rearrangement terms at the boundary. Dips can be mitigated by using a smaller $\Gamma$, which limits the region where the rearrangement is significant. 
However, a smaller $\Gamma$ results in narrower and larger bumps. 

The introduction of large bumps in $c_s^2$ at relatively low $n_B$ likely affects gravitational wave observables (see, e.g., \cite{Tan:2021nat}). 
Thus, the choice is to either preserve $c_s^2(n_B)$ but not reproduce the other thermodynamic relations at high $n_B$ or to preserve $P$ and introduce artificial features into $c_s^2$.
We also find that the mismatch/new features are enhanced if one smoothly matches at higher $n_B$ whereas smoothly matching at lower $n_B$ gives the closest to the original EoS (regardless of matching method).

Matching EoSs containing different microphysics is particularly challenging even in the 1D charge-neutral, $\beta$-equilibrated case.
Depending on the choice of thermodynamic quantity the user chooses to match in, and the midpoint and width of the matching, the EoS might not be stable or causal (one, the other, or both depending on the parameters).  
In this case, the module returns an error message. 

Also, there may not be a sufficiently large overlap region between the two EoSs for the smooth matching to be continuous, as the dips in $c_s^2$ indicate when the matching with CMF++ occurs in $P$.
Therefore, finding an optimal set of parameters is essential for matching the EoSs while respecting causality and stability. It may even be possible that some EoSs cannot be smoothly matched at all if they are sufficiently different from each other in their overlapping regime of validity. 

\section{Observable Modules}\label{sec:observable_modules}

The CE currently contains two observable modules that can be used at $T\sim0$ and are relevant for neutron stars. 
One calculates neutron star properties that are observable either through electromagnetic or gravitational waves, while the other calculates dynamical properties related to flavor equilibration. We discuss each of these below in detail. 

\subsection{QLIMR}\label{sec:QLIMRmethods}

The QLIMR module--an acronym for Quadrupole moment, tidal Love number, Moment of inertia, Mass, and Radius--is a new, optimized \texttt{C++} implementation for calculating macroscopic neutron star observables. Developed using perturbation methods in general relativity, the module assumes slowly-rotating and slightly-deformed stars. These methods draw from foundational techniques proposed by  Hartle and Thorne \cite{Hartle:1967he, Hartle:1968si} for slowly-rotating stars, and by Hinderer and Flanagan \cite{Hinderer:2007mb, Flanagan:2007ix} for static tidal fields that minimally deform stars from a perfect spherical shape.

When a star rotates, its matter-energy distribution rearranges itself to minimize its energy, adopting an oblate shape. Small deformations can be treated as a small perturbation to the spacetime of a non-rotating, spherically-symmetric star. The Hartle-Thorne method employs a slow-rotation expansion of the spacetime metric, using a dimensionless parameter $\epsilon$, defined as the ratio of the star's spin frequency to the spin frequency at the mass-shedding limit ($\epsilon \equiv \Omega/\Omega_{\textrm{shed}}$).
This approach requires the parameter to remain sufficiently small to maintain the validity of the perturbative method. The approximation breaks down when the spin frequency approaches the critical threshold where rapid rotation might cause matter disruption. Notably, comparisons with numerical relativity simulations in full general relativity have demonstrated that the Hartle-Thorne approximation, up to $\mathcal{O}(\epsilon^{2})$, remains remarkably accurate for modeling even the fastest observed pulsars \cite{Berti:2004ny,Yagi:2014bxa}. For example, the quadrupole moment computed with the Hartle-Thorne approximation (at second-order in rotation) has, at most, a 20\% relative fractional error for the fastest observed millisecond pulsars. 

The construction of slowly-rotating neutron-star solutions using the Hartle-Thorne approach relies on several key simplifying assumptions about the source distribution. First, the neutron star is modeled as an ideal perfect fluid characterized by a barotropic EoS, $P=P(\varepsilon)$. Given that the Fermi temperature is substantially lower than the star's overall temperature, thermal agitation effects are considered negligible \cite{Friedman2013}.
Although differential rotation, particularly in newly-born neutron stars, has been explored \cite{Hartle1970diff, Chirenti:2013xm}, it has been shown that normal stars with differential rotation tend to evolve toward an uniform rotation equilibrium state \cite{Duez:2006qe}. Moreover, the contribution of the magnetic field energy to the geometry is considered very small compared to the energy density $\varepsilon$ of the fluid for most neutron stars.

Under these constraints, solutions for unmagnetized, uniformly and slowly-rotating neutron stars are derived through an iterative approach. The perturbed Einstein equations are solved systematically, order by order, in the spin-frequency expansion parameter $\epsilon$. Once the solution is obtained, macroscopic properties can be calculated, including the star's mass, radius, and higher-order multipole moments that emerge from its rotation.
At zeroth order in the star's spin, the gravitational mass $M_{*}$ and radius $R_{*}$ are determined by solving the well-known, Tolman-Oppenheimer-Volkoff (TOV) equations \cite{Tolman:1939jz,Oppenheimer:1939ne}. Incorporating linear corrections in the spin-frequency expansion, the leading-order contributions to the star's spin angular momentum $J$ and moment of inertia $I$ can be calculated. Extending the expansion to quadratic order in the slow-rotation approximation enables the determination of the rotational quadrupole moment $Q^{\textrm{rot}}$, the stellar eccentricity $e_{s}$, the equatorial radius $R_{\text{eq}}$, the first-order mean correction to the radius of the non-rotating star, denoted as $ \langle \delta R \rangle$, and the first-order correction to the TOV gravitational mass $\delta M$.

Notice that when the star is rotating, the radial distance from the center to the star's surface depends in general on the polar angle, $\theta$. To find an invariant parameterization of the stellar surface, we search for the surface in 3D, using spherical coordinates $(\tilde{r}, \theta, \phi)$ with the same intrinsic geometry as the surface of constant density of the star at $R_{*}$ \cite{Hartle:1968si}. This involves embedding the star's geometry into a 3D flat space. Up to order $\mathcal{O}(\epsilon^{2})$, the surface contour is described by
\begin{align}
\tilde{r} = R_{*} + \epsilon^{2}\left[ \xi^{(2)}_{0} + \left( R_{*} k^{(2)}_{2} +  \xi^{(2)}_{2}  \right) P_{2}(\cos\theta)  \right]_{R_{*}}  
\label{eq:surface}
\end{align}
where $\xi^{(2)}_{0}$, $k^{(2)}_{2}$, and $\xi^{(2)}_{2}$ are metric perturbation functions of $\mathcal{O}(\epsilon^{2})$, and $P_{2}$ is the Legendre polynomial of degree 2. By taking an integral average over the 2-sphere,  the total mean radius of the star can be obtained from Eq.~\eqref{eq:surface} as
\begin{equation}
\langle \tilde{r} \rangle = R_{*} + \epsilon^{2} \langle \delta R \rangle 
\end{equation}
with $ \langle \delta R \rangle \equiv  \xi^{(2)}_{0}(R_{*})$. We refer to the quantity $\langle \delta R \rangle$ as the first mean contribution to the TOV radius $R_{*}$. The equatorial and polar radii of the rotating star are defined using Eq.~\eqref{eq:surface} as $R_{\textrm{eq}} = \tilde{r}(\theta=\pi/2)$ and $R_{\textrm{pol}} = \tilde{r}(\theta=0)$, respectively. Thus,
\begin{align}
\label{eq:Req}
    R_{\textrm{eq}} &= R_{*} +  \epsilon^{2}\delta R_{\textrm{eq}} \, , \\
    R_{\textrm{pol}} &=  R_{*} +  \epsilon^{2} \delta R_{\textrm{pol}} \, .
\end{align}
where
\begin{align}
\delta R_{\textrm{eq}} &=  \left[ \xi^{(2)}_{0} -\dfrac{1}{2} \left( R_{*} k^{(2)}_{2} +  \xi^{(2)}_{2}  \right)  \right]_{R_{*}} \, , \\ 
\delta R_{\textrm{pol}} &=  \left[ \xi^{(2)}_{0} +  R_{*} k^{(2)}_{2} +  \xi^{(2)}_{2}     \right]_{R_{*}} \, .
\end{align}
From these quantities, a useful measure associated with the rotation of the star is the stellar eccentricity $e_{s}$, which serves as a measure of the deviation from a perfect sphere. The stellar eccentricity is defined as
\begin{align}
e_{s} &\equiv \left[ \left( R_{\textrm{eq}}/R_{\textrm{pol}} \right)^{2} -1 \right]^{1/2}  \\
&= \left[ -\dfrac{3 \epsilon^{2}}{R_{*}}\left( R_{*} k^{(2)}_{2} + \xi^{(2)}_{2}  \right) \right]^{1/2}_{R_{*}} \, .
\label{eq:ecc}
\end{align}
The first correction to the TOV mass, denoted as $\delta M$, is the first additional contribution to the total mass monopole of the star, and it is related to the total mass by
\begin{equation}
M = M_{*} + \epsilon^{2}\delta M \ .
\label{eq:mass-correction}
\end{equation}

Beyond their rotational properties, neutron stars are also characterized by their response to external gravitational fields. One such measure is the static tidal Love number, $\lambda^{\textrm{tidal}}$, which quantifies the quadrupole deformation of a non-rotating star in a static gravitational tidal field. This quantity has been studied in general relativity in recent years, beginning with the pioneering work of Hinderer and Flanagan \cite{Hinderer:2007mb, Flanagan:2007ix}, and later formalized in greater detail by Damour and Nagar \cite{Damour:2009vw}, and by Binnington and Poisson \cite{Binnington:2009bb}. The central idea involves analyzing linear perturbations that capture deviations from sphericity of a static and spherically-symmetric spacetime describing a non-rotating star. By solving the perturbed Einstein equations throughout the entire space, it is possible to extract $\lambda^{\textrm{tidal}}$ in the \textit{buffer zone}. 
An equivalent approach to obtain the perturbed Einstein equations and $\lambda^{\textrm{tidal}}$
implies setting the dragging function $\omega$, which accounts for rotational spacetime effects, to zero in the Hartle-Thorne approximation to 
$\mathcal{O}(\epsilon^{2})$. This latter approach 
was the one used 
by Yagi and Yunes \cite{Yagi:2013awa} when studying universal relations of neutron stars~\cite{Yagi:2013bca,Yagi:2013awa}, including both, slowly rotation and static tidal deformations in the same perturbation scheme. 

A notable consequence of computing the macroscopic observables previously discussed is the emergence of dimensionless quantities $\bar{I}$, $\bar{\lambda}^{\textrm{tidal}}$, and $\bar{Q}^{\mathrm{rot}}$, defined as
\begin{equation}
\bar{I} \equiv \dfrac{I}{M_{*}^{3}} \hspace{0.25cm} ; \hspace{0.25cm} \bar{\lambda}^{\textrm{tidal}} \equiv \dfrac{\lambda^{\textrm{tidal}}}{M_{*}^{5}} \hspace{0.25cm} ; \hspace{0.25cm} \bar{Q}^{\mathrm{rot}} \equiv - \dfrac{Q^{\mathrm{rot}}M_{*}} {J^{2}}\,,
\end{equation}
which follow \textit{quasi-universal} relations independent of the EoSs~\cite{Yagi:2013bca,Yagi:2013awa}. These relations enable tests of general relativity in the strong-field regime and can help break parameter degeneracies in astrophysical observations \cite{Yagi:2013bca}.

\begin{figure*}[t!]
    \includegraphics[width=0.75\linewidth]{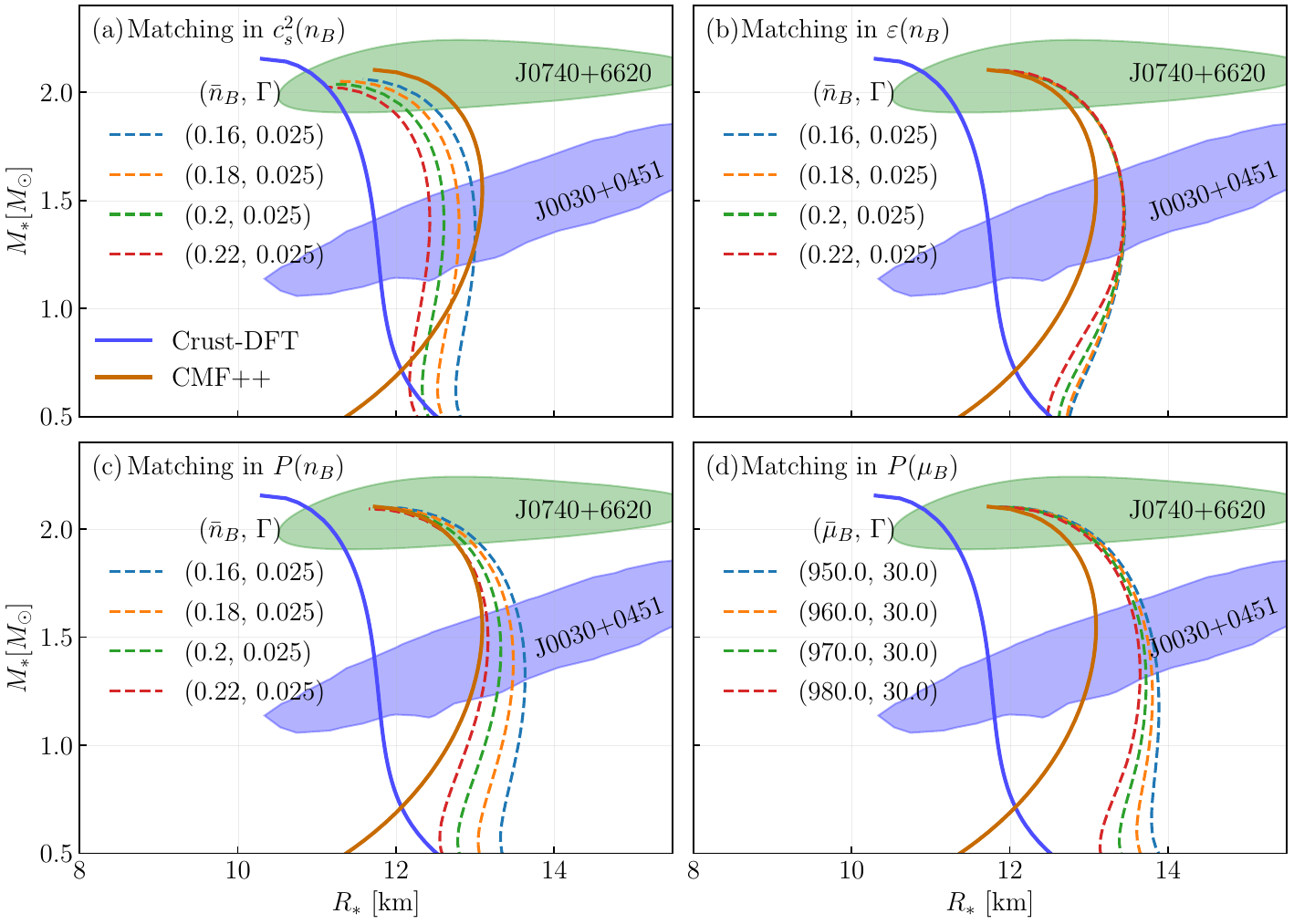}
    \caption{Mass-radius diagrams obtained from QLIMR.
    Solid lines represent the original EoSs, while dashed lines show results from our different equation of state smooth matching techniques shown in \Cref{fig:eos-tanh,fig:cs2-tanh}. The numbers in the labels indicate the midpoint and width of the matching. The shaded areas identify the 2$\sigma$ confidence mass-radius measurements from NICER: (i) the green area represents J0740+6620~\cite{salmi_2024_10519473,Salmi:2024aum}; (ii) the purple area represents pulsar J0030+0451~\cite{Miller:2019cac,miller_2019_3473466}. }
    \label{fig:qlimr_sep}
\end{figure*}

Building upon the theoretical framework discussed above, the QLIMR module is designed to compute macroscopic observables of slowly-rotating or slightly-deformed neutron stars. Given a barotropic two-column EoS table $P(\varepsilon)$, a central $\varepsilon$ range $[\varepsilon^{\textrm{initial}}_{c},\varepsilon^{\textrm{final}}_{c}]$, and a specified mass resolution $\Delta M$ over the mass-radius plane, the QLIMR module can compute neutron star sequences of macroscopic quantities along various stable branches. This is achieved by numerically integrating the perturbed Einstein equations from a specified small initial radius to the star's surface. Users can customize input parameters to select the desired macroscopic quantities and choose whether to compute a single neutron star solution or an entire sequence. Upon execution, the QLIMR module produces a data file containing the requested observables for the sequence, formatted according to the following column convention,
\begin{equation}
\left\{ \varepsilon_{c}, \,  R_{*}, \,  M_{*}, \, \bar{I}, \bar{\lambda}^{\textrm{tidal}}, \, \bar{Q}^{\textrm{rot}} , \, \dfrac{e_{s}}{\Omega}, \, \dfrac{ \delta R_{\textrm{eq}}}{\Omega^{2}}, \dfrac{\langle \delta R \rangle}{\Omega^{2}}, \, \dfrac{\delta M}{\Omega^{2}} \right\}\,.
\label{eq:ILQ}
\end{equation}
Additionally, users can specify the desired output file format, choosing between a \texttt{.csv} file for compatibility with general-purpose data tools, an HDF5 file for efficient storage and handling of large datasets, or a CompOSE compatible output \cite{Dexheimer:2022qhn}.

By default, QLIMR computes stable neutron star mass-radius curves unless the user wants to include the unstable region or calculate additional macroscopic quantities. The central energy density $\varepsilon_{c}$ is given in units of $[\text{MeV}/\text{fm}^3]$, radius $R_\ast$ in kilometers [km] and gravitational mass $M_{*}$ in solar masses $[M_{\odot}]$. The stellar eccentricity, $e_{s}$, is normalized with respect to the angular spin frequency of the star, $\Omega$, and the resulting units are seconds $[s]$. The equatorial radial correction, $\delta R_{\textrm{eq}}$, the mean correction to the radius, $\langle \delta R \rangle$, and the correction to the mass, $\delta M$, are normalized with respect to the star’s angular spin frequency squared, $\Omega^{2}$. The units for these quantities are kilometer-seconds squared [km s$^{2}$] for $\delta R_{\textrm{eq}}$ and $\langle \delta R \rangle$, and solar mass-seconds squared [$M_{\odot}$ s$^{2}$] for $\delta M$. These corrections depend on the star's angular spin frequency, $\Omega$. To compute them for a user-selected angular spin frequency, denoted as $\Omega_{\textsf{user}}$, the stellar eccentricity $e_{s}$ must be rescaled by multiplying the entire column by $\Omega_{\textsf{user}}$, while $\delta R_{\textrm{eq}}$, $\langle \delta R \rangle$, and $\delta M$ should each be multiplied by $\Omega_{\textsf{user}}^{2}$. Similarly, the quantity $Q^{\text{rot}}$ depends also on the angular spin-frequency of the star 
and its value for some angular spin-frequency specified by the user $\Omega_{\textsf{user}}$, can be obtained from the dimensionless quantity $\bar{Q}^{\text{rot}}$ from QLIMR's output in Eq.~\eqref{eq:ILQ} as
\begin{align}
Q^{\text{rot}}_{\textsf{user}}&=-\dfrac{\bar{Q}^{\text{rot}}M_{*}}{I^{2}\Omega^{2}_{\textsf{user}}} \ .
\end{align}
Additional macroscopic quantities, such as the compactness $C=M_\ast/R_\ast$ and the rotational tidal Love number $\lambda^{\text{rot}} \equiv \bar{I}^{2} \bar{Q}^{\text{rot}}$, can also be computed with QLIMR. Beyond these, QLIMR provides the option to output local quantities as a function of the radial (Hartle-Thorne) coordinate $r$ inside the star. These include the profiles $P(r)$, $\varepsilon(r)$, and the enclosed mass $m(r)$, as well as the solutions for all metric functions to $\mathcal{O}(\epsilon^{2})$. 

\begin{figure*}[t!]
    \includegraphics[width=0.35\linewidth]{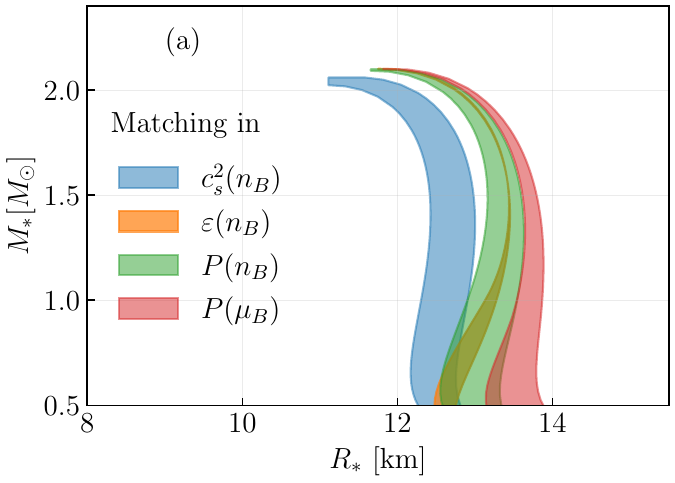}
    \includegraphics[width=0.35\linewidth]{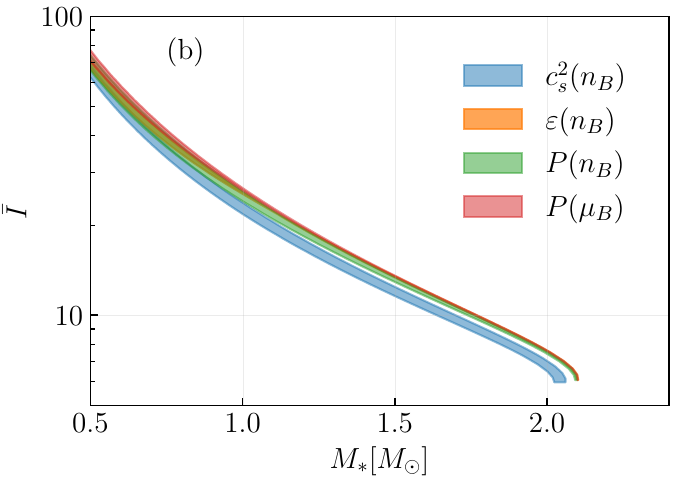}
    \includegraphics[width=0.35\linewidth]{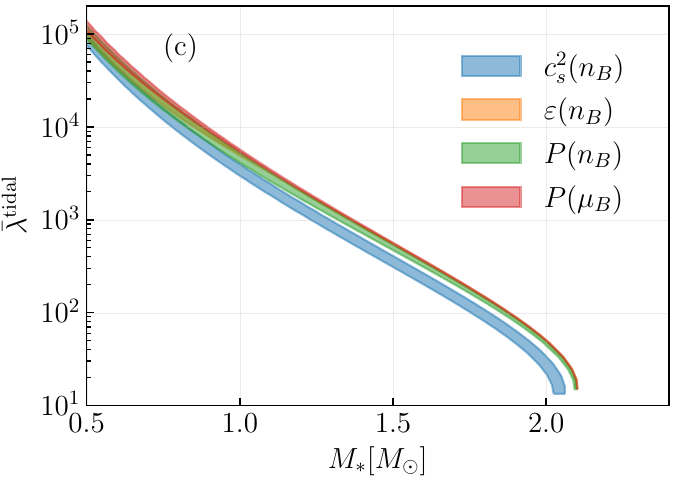}
    \includegraphics[width=0.35\linewidth]{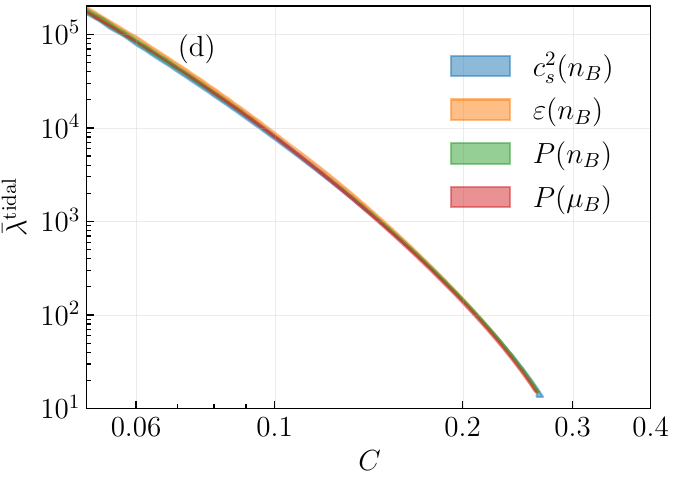}
    \caption{Panel (a) shows regions made up by the mass-radius lines obtained from QLIMR and shown in \Cref{fig:qlimr_sep}; panel (b) shows the corresponding dimensionless moment of inertia as a function of neutron star masses; panel (c) shows the corresponding dimensionless tidal Love number as a function of neutron star masses; panel (d) shows the corresponding dimensionless tidal Love number as a function of compactness $C=M_\ast/R_\ast$.
    The shaded regions show the range in results obtained using the four different smooth matching techniques we used in this work. The region widths indicate the differences that arise from (physically meaningful) parameter sweeping.}
    \label{fig:I-Love-Qrelations}
\end{figure*}

The QLIMR code features a modular architecture and is containerized using Docker. It includes a Python layer responsible for preprocessing input data, which involves validation through YAML files, as well as managing output file-formatting conventions. The core source code is written in \texttt{C++} with an object-oriented approach, relying on a single class dependency to maintain simplicity and scalability. For numerical integration, the module employs the GSL library, offering sufficient efficiency to handle a wide range of EoSs. 

For a neutron star sequence comprising 100 points, the module requires 20 milliseconds to integrate the TOV equations and generate the mass-radius curve, and 0.6 seconds to compute all macroscopic observables up to $\mathcal{O}(\epsilon^{2})$ for a mass resolution of $\Delta M_\ast = 0.025 M_{\odot}$. The algorithm is optimized to solve only the necessary set of perturbed Einstein equations based on the desired input observables combination, reducing computational overhead. To accurately locate the star's surface, the module employs Lindblom's approach \cite{lindblom1992}, or the pseudo-enthalpy method, which minimizes errors compared to the traditional form of the TOV equations\footnote{This precision is important, as small variations in the radius significantly affect the tidal Love number, which scales as $\lambda^{\text{tidal}} \sim C^{-5}$.}. The open source QLIMR module, accompanied by detailed instructions for building and executing the code, is accessible via the Zenodo repository and the MUSES documentation website~\cite{conde_ocazionez_2024_14525356}.

\subsubsection{Results from QLIMR using different matching methods}

\Cref{fig:qlimr_sep} shows neutron star sequences produced by QLIMR using different, complete (crust to core) EoSs. Every point along any curve represents a single, stable, neutron star solution.
Each panel shows (smooth) matching using a different thermodynamic variable $Y(x)$ in \Cref{eq:Yx}. 
Only the original module EoSs Crust-DFT and CMF++ are shown, as $\chi$EFT does not extend high enough in $n_B$ to produce an entire curve. 
In a given panel, the different dashed lines come from  matching EoSs using different choices of $(\bar{x},\Gamma)$ to match Crust-DFT+$\chi$EFT with CMF++. 
One can immediately see that the different parameter sets $(\bar{x},\Gamma)$ lead to different stellar radii, but not necessarily different masses. %
Although the sweep in parameters $(\bar{x},\Gamma)$ in the $c_s^2 (n_B)$ matching produces a small change in $M_{max}$ of $\sim2\%$, matching in the other thermodynamic variables does not have a significant change in $M_{max}$ (orders of magnitude below $1\%$). On the other hand, matching (\emph{within a given procedure}) using the four different thermodynamic variables produces a change in the stellar radius of up to $\sim5\%$ for a neutron star with mass $1.4 M_{\odot}$, with the smallest change of $\sim0.02\%$ for the $\varepsilon(n_B)$ matching case. See Table \ref{tab:summary}  in \Cref{appQLIMR} for a numerical summary.
If we then compare across all smooth matching procedures we see a change up to $\sim10\%$ in the stellar radius for a neutron star with mass $1.4\, M_{\odot}$.

\cref{fig:I-Love-Qrelations} shows the mass-radius diagram plus additional macroscopic observables computed using QLIMR. 
Each panel displays four shaded regions, corresponding to the phase space occupied by the lines shown in \Cref{fig:qlimr_sep} for the different EoS matching techniques employed in this work. The shaded regions are bounded by configurations defined by specific sets of matching parameters, $(\bar{x}, \Gamma)$, where $\bar{x}$ is varied while $\Gamma$ is held fixed. When $x=n_{B}$ serves as the independent variable for matching in $c^{2}_{s}$, $\varepsilon$ and $P$, the boundaries are defined by the parameter sets
\begin{align}
    \boldsymbol{\theta}_{a}(n_{B}) &\equiv ( \bar{n}_{B} = 0.16 \, \text{fm}^{-1}, \,  \Gamma = 0.025 \, \text{fm}^{-1} ) \, , \\
    \boldsymbol{\theta}_{b}(n_{B}) &\equiv ( \bar{n}_{B} = 0.22 \, \text{fm}^{-1}, \, \Gamma = 0.025 \, \text{fm}^{-1} ) \, .
\end{align}
Alternatively, when $x=\mu_{B}$
is chosen as the independent variable and matching is performed in
$P$, the corresponding boundary parameter sets are 
\begin{align}
    \boldsymbol{\theta}_{a}(\mu_{B}) &\equiv ( \bar{\mu}_{B} = 950 \, \text{MeV}, \, \Gamma = 30 \, \text{MeV} ) \, , \\
    \boldsymbol{\theta}_{b}(\mu_{B}) &\equiv (  \bar{\mu}_{B} = 980 \, \text{MeV}, \, \Gamma = 30 \, \text{MeV} ) \, .
\end{align}

In panel (a), the shaded regions depict the mass-radius curves corresponding to the different matching procedures bounded by $\boldsymbol{\theta}_{a}$ and $\boldsymbol{\theta}_{b}$. 
For the matching cases performed in $c_{s}^{2}(n_{B})$, $P(n_{B})$ and $P(\mu_{B})$, the upper-right boundary curves of each shaded region correspond to $\boldsymbol{\theta}_{a}$, while the lower-left boundary curves correspond to $\boldsymbol{\theta}_{b}$. 
By fixing $\Gamma$ in each case, we observe that increasing the mean values $\bar{n}_{B}$ and $\bar{\mu}_{B}$ results in stars with smaller radii for the same mass, indicating higher compactness $C=M_\ast/R_\ast$, as also shown in \Cref{fig:qlimr_sep}. 
Similarly, in panels (b) and (c), the upper boundary curves are defined by $\boldsymbol{\theta}_{a}$, while the lower ones correspond to 
$\boldsymbol{\theta}_{b}$. 
For a constant mass, $\bar{I}$ is smaller at the limiting curve defined by $\boldsymbol{\theta}_{b}$ and larger at $\boldsymbol{\theta}_{a}$. 
This behavior is consistent since $\bar{I} \sim C^{-2}$ and the compactness is higher at the curve defined by $\boldsymbol{\theta}_{b}$ for a fixed mass in panel (b). 
A similar argument can be given for $\bar{\lambda}^{\textrm{tidal}}$ in panel (c) since such quantity scales as $\bar{\lambda}^{\textrm{tidal}} \sim C^{-5}$. 
The matching case $\varepsilon(n_{B})$ does not show a significant difference in radii for masses higher than $\sim 1.4M_{\odot}$ in panel (a), but it follows the same behavior as the other matching cases for masses lower than $1.4 M_{\odot}$. 

For any observable $A$, we use the curves obtained from $\boldsymbol{\theta}_{b}$ as a reference and define the average fractional error as 
\begin{equation}
\label{AAA}
\left \langle\Delta A(\%) \right\rangle \equiv \left\langle \frac{ | A(\boldsymbol{\theta}_{a}) - A(\boldsymbol{\theta}_{b}) | }{A(\boldsymbol{\theta}_{b})} \times 100 \% \right\rangle \, .
\end{equation}
The average fractional errors are computed for the observables $R_\ast$, $\bar{I}$ and $\bar{\lambda}_{M}^{\textrm{tidal}} \equiv \bar{\lambda}^{\textrm{tidal}}(M_\ast) $ based on the curves in panels (a), (b), and (c), respectively. For panel (d), we calculate the average fractional error of $\bar{\lambda}_{C}^{\textrm{tidal}} \equiv \bar{\lambda}^{\textrm{tidal}}(C)$. The results are summarized in
\Cref{tab:fractional_errors}. 

\begin{table}[t!]
    \centering
    \begin{tabular}{l S S S S S}\toprule
        {$Y(x)$} & {$\left \langle\Delta R_\ast(\%) \right\rangle$} & {$\left \langle\Delta \bar{I}(\%) \right\rangle$} & {$\left \langle\Delta \bar{\lambda}_{M}^{\textrm{tidal}}(\%) \right\rangle$} & {$\left \langle\Delta \bar{\lambda}_{C}^{\textrm{tidal}}(\%) \right\rangle$} \\ 
    \midrule
        {$c_{s}^{2}(n_{B})$} & 4.87 & 8.59  & 36.55 & 3.80  \\
        {$\varepsilon(n_{B})$} & 0.87 & 2.16  & 7.47  & 3.48  \\
        {$P(n_B)$}            & 4.09 & 6.22  & 23.79 & 4.86  \\
        {$P(\mu_{B})$}        & 1.90 & 1.97 & 6.88  & 3.95  \\
    \bottomrule
    \end{tabular}
    \caption{Average fractional errors for $R_\ast$, $\bar{I}$, $\bar{\lambda}^{\textrm{tidal}}_{M}$, and $\bar{\lambda}^{\textrm{tidal}}_{C}$ obtained from panels (a), (b), (c) and (d) in \cref{fig:I-Love-Qrelations}.  }
    \label{tab:fractional_errors}
\end{table}

\begin{figure*}
    \centering
    \includegraphics[width=0.4\linewidth]{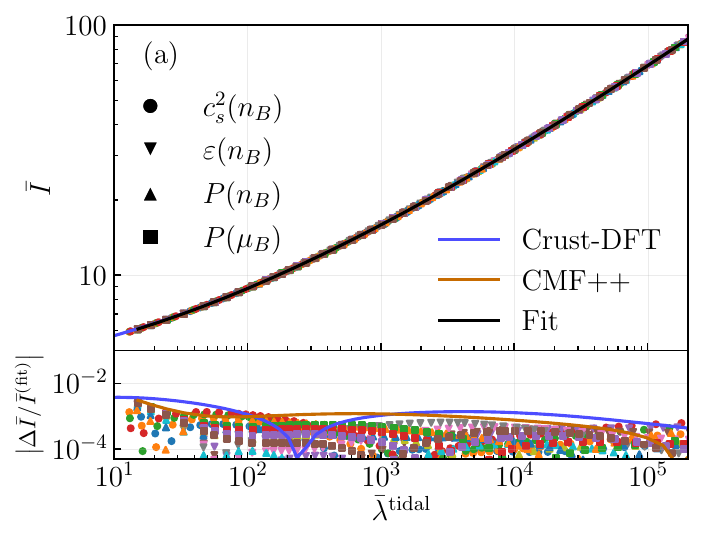}
    \includegraphics[width=0.4\linewidth]{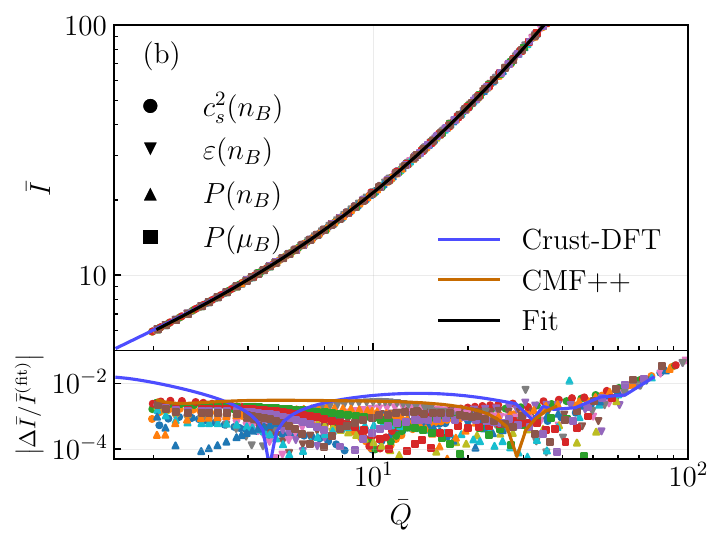}
    \includegraphics[width=0.4\linewidth]{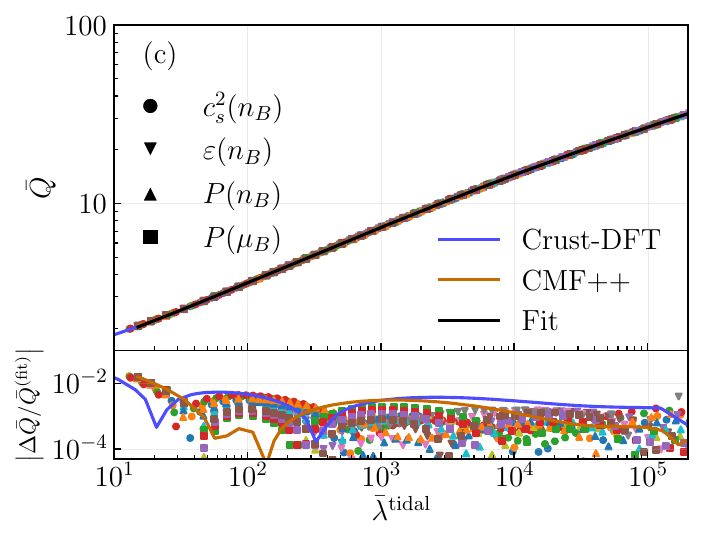}
    \includegraphics[width=0.4\linewidth]{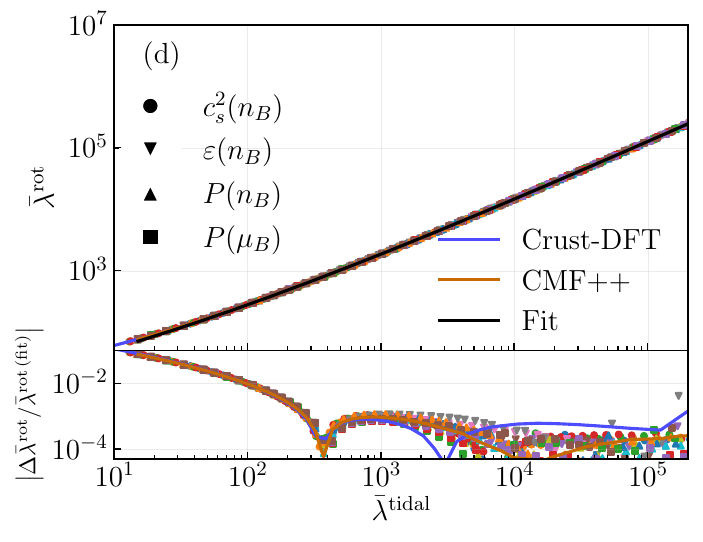}    
    \caption{Quasi-universal relations from QLIMR module for the different smoothed matched EoS from the different panels in \Cref{fig:qlimr_sep}. Panel (a) shows the I-Love relation, panel (b) the I-Q relation, panel (c) the Q-Love relation, and panel (d) the Love-Love relation. The lower sub-panels show the normalized difference between the fits (explained in the text) and the computed quantities. The colors of the markers follow the convention of \Cref{fig:qlimr_sep}. }
    \label{fig:iloveq}
\end{figure*}

Additionally, we obtained error estimates for $R_\ast$, $\bar{I}$, and $\bar{\lambda}^{\textrm{tidal}}$ for a neutron star with mass equal to $1.4 M_{\odot}$, as explained in \Cref{appQLIMR} and Table \ref{tab:summary} therein.
Regarding the \textit{quasi-universal} relations, \Cref{fig:iloveq} consists of four main panels (a) through (d). In each panel, there are two sub-panels: The upper sub-panels illustrate the  I-Love-Q relations, while the lower sub-panels show the normalized differences between the I-Love-Q quantities derived from various EoSs and a reference fitting curve. This is represented by $\Delta A / A^{fit}$ where $\Delta A = A - A^{fit}$ and $A$ could be either $\bar{I}$, $\bar{\lambda}^{\textrm{tidal}}$ or $ \bar{Q} \equiv \bar{Q}^{\textrm{rot}} $ according to each panel. For every matching case in the thermodynamic variable $Y(x)$, four different curves are shown, corresponding to the same set of parameters for the smoothed matched EoSs shown in \Cref{fig:qlimr_sep}. The reference curve is obtained by fitting the EoSs from all matching cases, including the original Crust-DFT and CMF models, using a polynomial fitting function given by
\begin{equation}
\begin{split}
    \ln y_i =  a_i &+ b_i \left( \ln x_i \right) + c_i \left( \ln x_i \right)^2 \\
    & + d_i \left( \ln x_i \right)^3 + e_i \left( \ln x_i \right)^4 \, .
\end{split}
\label{eq:fitting_formula}
\end{equation}
The corresponding numerical coefficients for this fitting procedure are presented in Table \ref{tab:fit_iloveq}.

\begin{table}[t!]
    \centering
    \begin{tabular}{l l S@{\hspace{-5pt}}S@{\hspace{3pt}}S@{\hspace{-22pt}}S@{\hspace{-25pt}}S}\toprule
             {$y_i$} & {$x_i$} & {$a_i$} & {$b_i$} & {$c_i$} & {$d_i$} & {$e_i$}\\
    \midrule
    {$\bar{I}$} & {$\bar{\lambda}^{\rm tidal}$} & 1.49 & 0.058 & 0.023 & -7.3e-4 & 9.0e-4 \\
    {$\bar{I}$} & {$\bar{Q}$}                    & 1.35 & 0.651 & -0.063 & 5.7e-2 & -5.0e-3 \\
    {$\bar{Q}$} & {$\bar{\lambda}^{\rm tidal}$} & 0.51 & 0.184 & 0.028   & -2.4e-3 & 6.5e-5 \\
    {$\bar{\lambda}^{\rm rot}$} & {$\bar{\lambda}^{\rm tidal}$} 
                                                & 2.76 & 0.428 & 0.052 & -2.41e-3 & 4.5e-5 \\
        \bottomrule
    \end{tabular}
    \caption{Numerical coefficients for the fitting formula in Eq.~\eqref{eq:fitting_formula} which are used to describe the reference curve for each panel in \Cref{fig:iloveq}. }
    \label{tab:fit_iloveq}
\end{table}

We found that the EoSs used in all panels of \Cref{fig:iloveq} follow universal relations with errors accurate to $\mathcal{O}(0.3)\%$, $\mathcal{O}(1)\%$, $\mathcal{O}(1.6)\%$, and $\mathcal{O}(7.6)\%$ for panels (a), (b), (c), and (d), respectively. As discussed in \cite{Yagi:2013awa}, the small normalized difference errors can be attributed to the fact that the I-Love-Q trio is more sensitive to the outer layers of the neutron star. Since we are fixing the matching parameters at the intersection of $\chi$EFT and Crust-DFT, which corresponds to these outer layers, the I-Love-Q relations remain highly accurate.

\subsubsection{Mass-radius spin correction of a MUSES neutron star}

\begin{figure*}[t!]
    \includegraphics[width=0.75\linewidth]{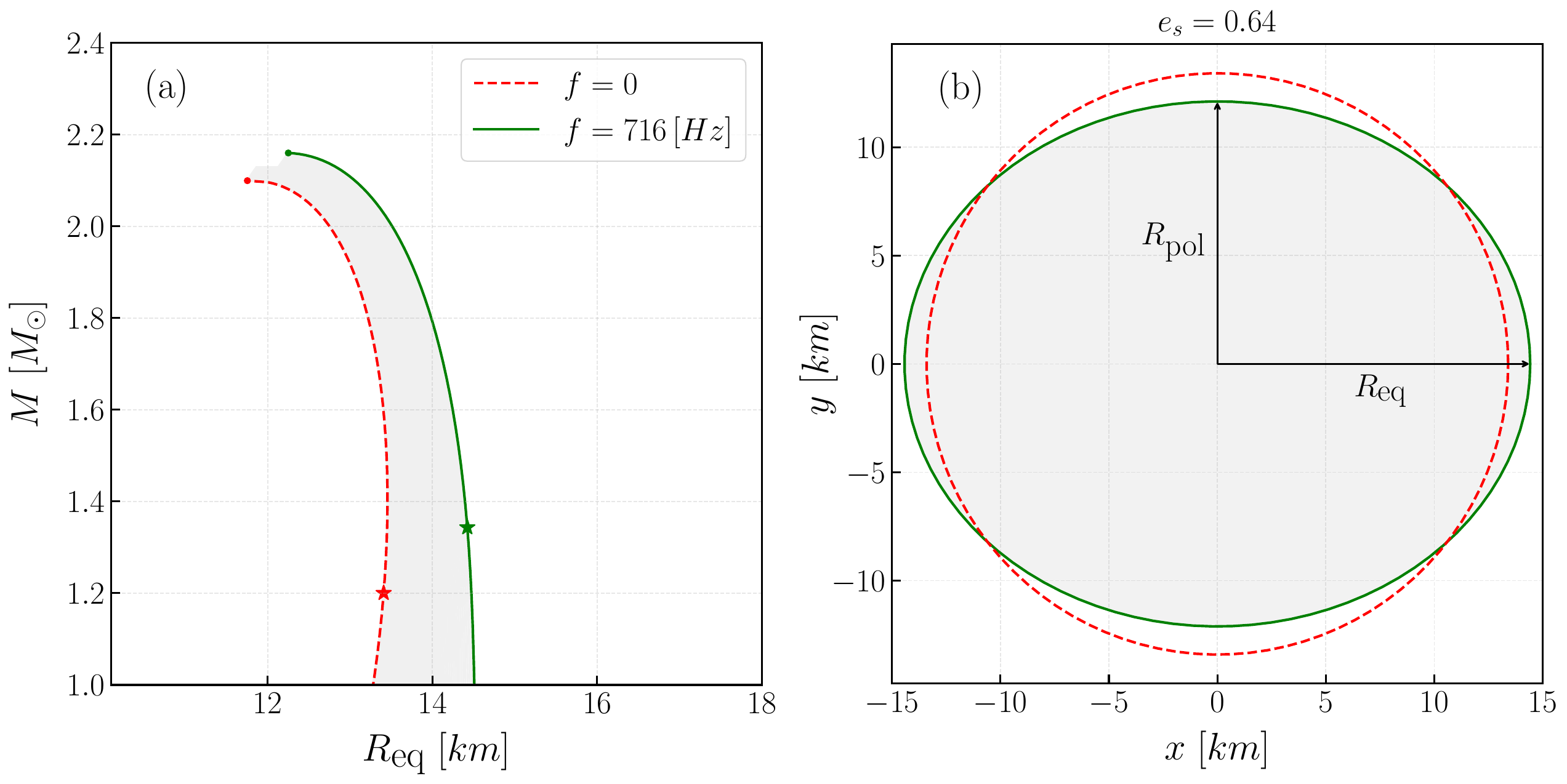}
    \caption{Panel (a) shows the total mass $M$ as a function of the equatorial radius $R_{\textrm{eq}}$ for neutron star sequences at different spin frequencies $f$. In panel (b), a cross-section of a non-rotating star (red) and a rotating star (green) is shown for a star with $f = 716$ Hz and central energy density $\varepsilon_{c} = 300.44$ $\text{MeV}/\text{fm}^3$.
     }
    \label{fig:eccentricity}
\end{figure*}

When a neutron star rotates, its rotational energy contributes to its total mass and the radial distance from the center to the surface increases at the equator, as first shown in~\cite{Hartle:1967he,Hartle:1968si,Hartle1970diff}. The spin-correction to the non-rotating (TOV) mass at $\mathcal{O}(\epsilon^{2})$ is given by Eq.~\eqref{eq:mass-correction} and the equatorial radius is described by Eq.~\eqref{eq:Req}. In panel (a) of \Cref{fig:eccentricity}, we calculate the total mass $M$ of the star versus the equatorial radius $R_{\textrm{eq}}$ using QLIMR for an EoS obtained based on the matching procedure in $\varepsilon(n_{B})$ with $\bar{n}_{B}=0.16 \, \mathrm{fm}^{-3}$ and $\Gamma = 0.025$. We vary the spin frequency $f$ from 0 to 716 Hz, which corresponds to the maximum frequency observed for a pulsar \cite{Hessels:2006ze}, in order to generate neutron star, mass-radius sequences. The red curve in panel (a) represents the non-rotating case, while the green curve represents the case at 716 Hz. The gray shaded region between the two curves corresponds to sequences with spin frequencies lower than 716 Hz.

Observe that, as the frequency increases, both the total mass 
$M$ and the equatorial radius $R_{\textrm{eq}}$ also increase. For each point in the sequence corresponding to the non-rotating configuration at a given $\varepsilon_c$, the non-rotating system shifts along a diagonal line toward higher $M$ and higher $R_{\textrm{eq}}$ as $f$ increases. For reference, we highlight in panel (a) two points (with star symbols) that have the same central energy density, $\varepsilon_{c} = 300.44$ $[\text{MeV}/\text{fm}^3]$.

In panel (b) of \Cref{fig:eccentricity}, we show the cross-sectional shape of the neutron star for the same central energy density $\varepsilon_{c}$. The dashed red line represents the circular contour of the non-rotating (TOV) case, while the solid green curve depicts the surface contour of a rotating star at $f = 716$ Hz. The rotating star adopts an ellipsoidal shape, where the equatorial radius, $R_{\textrm{eq}}$, is the distance from the center to the surface at an angle $\theta = \pi/2$, and the polar radius, $R_{\textrm{pol}}$, is the distance from the center to the surface at $\theta = 0$. The eccentricity $e_{s}$ of the rotating star (see Eq.~\eqref{eq:ecc}), is found to be $e_{s} = 0.64$ in this case.

\subsubsection{Slice of a MUSES neutron star}

\begin{figure}[t!]
    \centering
        \includegraphics[width=1\linewidth]{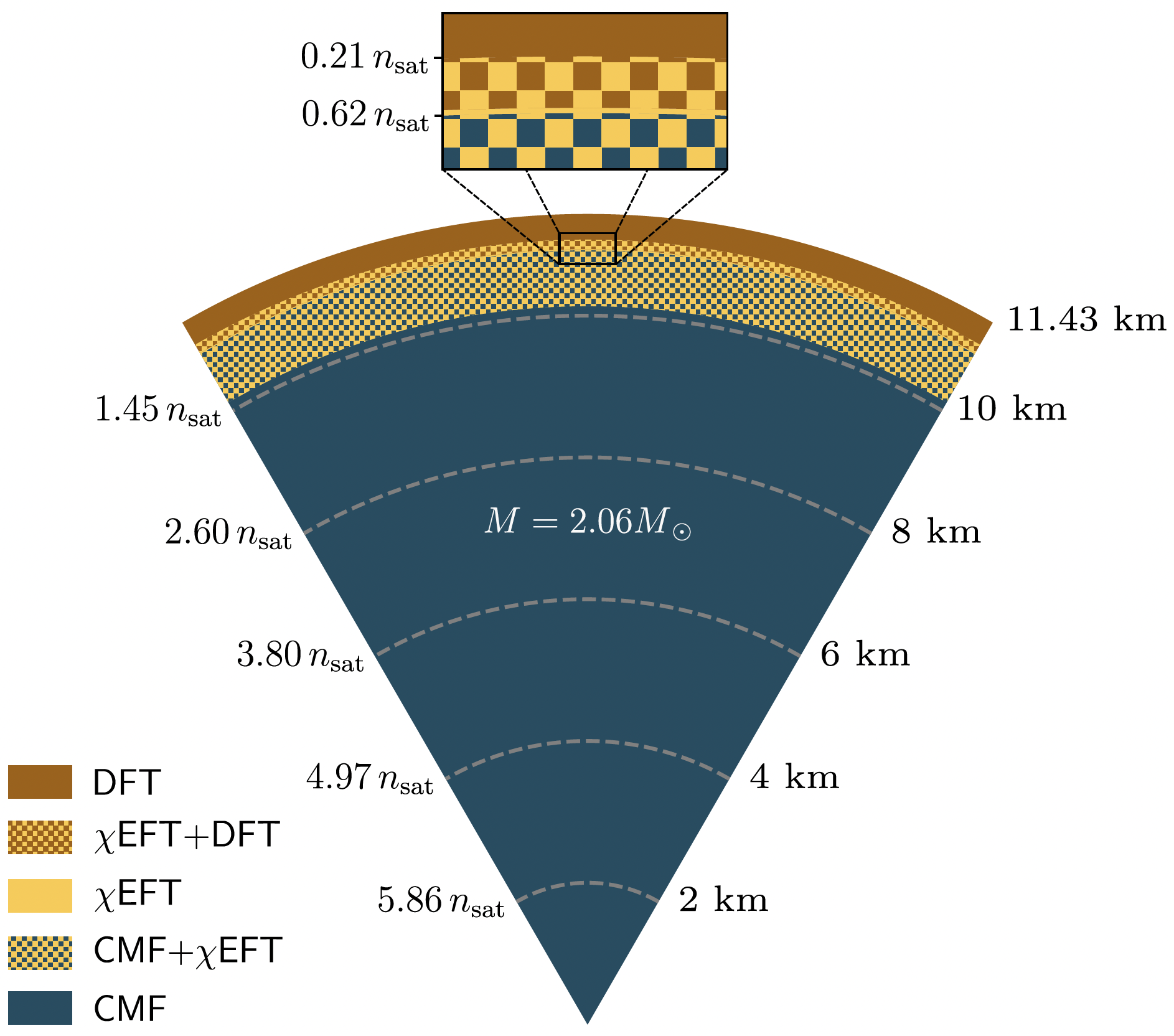}
    \caption{Neutron-star slice produced using the different EoS modules in this paper. The left side of the figure shows the baryon density in units of saturation density, while the right size shows the distance from the center. Smooth matching using speed of sound squared with parameters from \Cref{tab:dft-ceft} between Crust-DFT+$\chi$EFT and $\chi$EFT and parameters ${\bar{n}_B}=0.16$ fm$^{-3}$ and $\Gamma=0.025$ $\mathrm{fm^{-3}}$ between $\chi$EFT and CMF were used. }
       \label{fig:NS_slice}
\end{figure}

\Cref{fig:NS_slice} shows an example of a non-rotating neutron star slice built when smoothly matching in $c_s^2$ (with a particular set of parameters described in the figure label). 
The size of each layer of a neutron star strongly depends on the maximum central density, which in turn produces a given mass and radius of the star. 
In other words, heavier neutron stars have larger contributions from the core, whereas for lighter neutron stars the crust/outer core plays a larger role. 
Here, we show a slice of the neutron star that has the largest possible mass in its given mass-radius sequence ($M=2.06$ M$_\odot$), which leads to a picture where most of the star has $n_B>n_{sat}$, such that the core dominates.  
We find there is a  thin outer shell described by the Crust-DFT model (nuclei+nucleons), followed by the $\chi$EFT, and then CMF++ at the core (approximately $10$ km in radius). 
Because of the matching between Crust-DFT+$\chi$EFT+CMF++, the physical radius that $\chi$EFT occupies alone is quite small (highlighted in the inset of \Cref{fig:NS_slice}) but the region of matching to ensure a smooth change between the EoSs is quite a bit larger, occupying roughly $1$ km in the radius. 
With this picture in mind, it becomes easier to understand why the different approaches in the EoS smooth matching can play such a large role when it comes to stellar properties such as the mass, radius, and tidal deformability that we discussed in the previous section. 

\subsection{Flavor Equilibration}
\label{sec:flavor}

Let us now describe a different workflow in the CE, presented in \Cref{fig:workflows_flavor}, involving the Flavor Equilibration module. This module is essential for analyzing the out-of-equilibrium behavior of matter (with respect to the weak force) in neutron stars when density oscillations occur.
Density oscillations on the millisecond timescale occur at low amplitude in isolated neutron stars \cite{Andersson:2000mf,Alford:2013pma,Kantor:2020dex} and at high amplitude in neutron-star mergers \cite{Alford:2017rxf,Most:2022yhe}. 
These oscillations temporarily drive matter out of $\beta$ (flavor) equilibrium, after which matter relaxes back via weak interactions.  The Flavor Equilibration module calculates a set of quantities that characterize this relaxation, as described below. The module calculates these relaxation quantities at the user's requested values of $(T,n_B)$ for an EoS that is supplied in the form of a 3D table in $(T, n_B, Y_Q)$, either from another MUSES module or from an external source.

\begin{table}[t!]
\centering
\begin{tabular}{cll}
\toprule
\textbf{Column} & \textbf{Quantity} & \textbf{Units} \\ 
\midrule
1 & Temperature ($T$) & MeV \\
2 & Baryon chemical potential ($\mu_B$) & MeV \\
3 & Strange chemical potential ($\mu_S$) & MeV \\
4 & Electron chemical potential ($\mu_e$) & MeV \\
5 & Baryon density ($n_B$) & fm$^{-3}$ \\
7 & Strangeness density ($n_S$) & fm$^{-3}$ \\
7 & Charge density ($n_Q$) & fm$^{-3}$ \\
8 & Energy density ($\varepsilon$) & MeV fm$^{-3}$ \\
9 & Pressure ($P$) & MeV fm$^{-3}$ \\
10 & Entropy density ($s$) & fm$^{-3}$ \\
11 & Proton effective mass ($m^\ast_p$) & MeV \\
12 & Neutron effective mass ($m^\ast_n)$ & MeV \\
13 & Proton chemical potential ($\mu_p$) & MeV \\
14 & Neutron chemical potential ($\mu_n$) & MeV \\
15 & Proton density ($n_p$) & fm$^{-3}$ \\
16 & Neutron density ($n_n$) & fm$^{-3}$ \\
17 & Proton energy shift ($U_p$) & MeV \\
18 & Neutron energy shift ($U_n$) & MeV \\
\bottomrule
\end{tabular}
\caption{Default input format for the Flavor Equilibration module in MUSES.}
\label{tab:flavor_format}
\end{table}

Relaxation rates go to zero at $T=0$, but the nuclear EoS
is only weakly temperature-dependent at $T\lesssim 5$\,MeV because the system is well below its Fermi temperature. 
Thus, if an EoS is only available in the form of a $T=0$ table, it is still possible to obtain reasonable results for flavor equilibration at a temperature  $T\lesssim 5$\,MeV by specifying that $T$ in the input file but supplying the $T=0$ EoS as the data in the file.
That is how the results discussed here were generated.

The Flavor equilibration module is currently formulated for neutrinoless $npe$ matter, so
processes that drive flavor equilibration are neutron decay $n\to p\,e^-\,\bar\nu_e$ and electron capture $p\,e^-\to n\,\nu_e$. The net rate of conversion of neutrons to protons is
\begin{equation}
 \Gamma_I(T, n_B,Y_p) \equiv
 \Gamma_{n\to p\,e^-\,\bar\nu_e} - \Gamma_{p\,e^-\to n\,\nu_e} \, .
 \label{eq:Gamma-I}
\end{equation}
These rates are determined by the EoS and the dispersion relations of the nucleonic excitations. 
For any EoS model of the strong interaction that can provide this information (not, e.g., Crust-DFT),
the Flavor Equilibration module 
can compute the following characteristics of flavor equilibration, which are all functions of $T$ and $n_B$:
\begin{enumerate}
\item The equilibrium value of the proton fraction $Y^\text{eq}_p$, defined by $\Gamma_I(T, n_B,Y^\text{eq}_p)=0$ \cite{Alford:2018lhf}. 

\item The (isothermal) flavor relaxation rate, which is evaluated at $Y_p=Y^\text{eq}_p(T, n_B)$,
\begin{equation}
    \gamma \equiv - \dfrac{1}{n_B} \dfrac{\partial\Gamma_I}{\partial Y_p}\Bigr|_{T,n_B}\ .
    \label{eq:gamma-I}
\end{equation}

\item The static bulk viscosity $\zeta_0$ evaluated at $Y_p=Y^\text{eq}_p(n_B,T)$, from which one can obtain the full frequency dependence of the bulk viscosity
\begin{equation}
  \zeta(\omega) = \zeta_0 \dfrac{\gamma^2}{\gamma^2 + \omega^2}\ .
\end{equation}

\item The static isothermal incompressibility $K$, from which the damping time $\tau_d(\omega)$ for low-amplitude density oscillations of frequency $\omega$ can be obtained.
\end{enumerate}
For a derivation and in-depth discussion of these quantities, see Refs.~\cite{Alford:2021ogv, Alford:2023gxq,Yang:2023ogo}. 

\begin{figure*}[t!]
    \centering
    \includegraphics[width=0.45\linewidth]{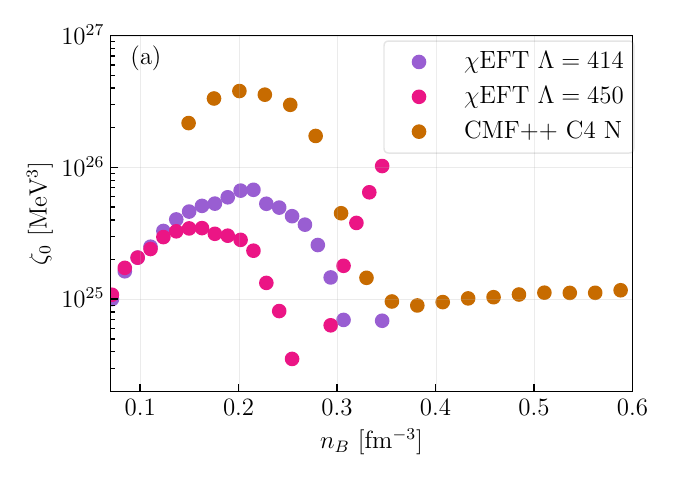}
    \includegraphics[width=0.45\linewidth]{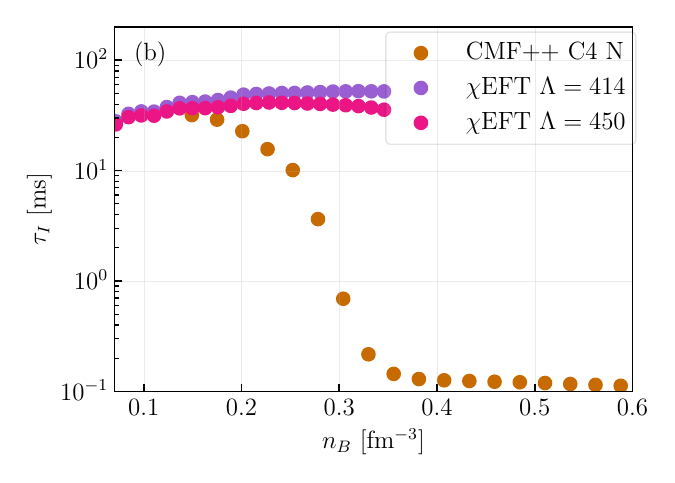}
    \caption{The left panel shows the static bulk viscosity, while the right panel show the isothermal flavor relaxation rate using two of the nuclear EoSs explored in this paper ($\chi$EFT and CMF) at a fixed temperature of $T=2$ MeV.
    }
    \label{fig:flavor}
\end{figure*}
 
The required input consists of a set of configuration parameters and a 3D EoS table.
The configuration parameters include:
\hspace{.1cm}(1) a ``mission'' flag that instructs the code what to do (either just determine  $Y_Q^{eq}$, or do that and also calculate $\zeta_0$, etc.);
\hspace{.1cm}(2) parameters that specify the range of $(T,n_B)$ to be explored;
\hspace{.1cm}(3) numerical parameters, such as the accuracy with which the equilibrium proton fraction is to be determined.

The input EoS table must contain standard thermodynamic information ($P$, $\varepsilon$, $\mu$'s, etc) for matter with a range of $Y_Q$ at a given $(T,n_B)$, and also the information needed to reconstruct the dispersion relations of the nucleonic excitations, namely effective masses and energy shifts  (additive contributions to the dispersion relations arising from interaction with the meson mean fields \cite{Roberts:2016mwj})
for the proton and neutron (see the Flavor input format in \Cref{tab:flavor_format}).
The input EoS table must be on a regular cubic grid in ($T$,$n_B$,$Y_p$) space. The rates are strongly $T$-dependent and $\gamma\rightarrow 0$ at $T=0$, so a chosen $T>0$ must be used as an input for Flavor Equilibration. While $\gamma$ depends strongly on $T$ even at low $T$, the EoS quantities required for the Flavor Equilibration module are nearly $T$-independent in this regime. 

The module creates an output file whose content depends on the mission flag. If full flavor-equilibration information is requested, for each of the $(T,n_B)$ requested in the configuration parameters the module calculates $Y_p^\text{eq}$ and the corresponding isospin chemical potential $\mu_I= \mu_p - \mu_n + \mu_e$, the parameters $\gamma$, $\zeta_0$, and $K$ described above,
and various other quantities, such as relevant susceptibilities like $\partial P/\partial Y_p |_{T,n_B}$.  All output quantities are in units of MeV to the appropriate power.

The Flavor Equilibration module is written in Python.
When the EoS is first read in, the code goes through the entire grid calculating relevant susceptibilities at every point, taking derivatives via nearest-neighbor finite difference.
The code then loops through the requested values of $(T,n_B)$. For each of these, the code finds $Y^\text{eq}$ by a root-finding search, looking for net Urca rate $\Gamma_I(T,n_B,Y_p)=0$, and then calculates the other equilibration quantities at that $Y_p$. The Urca rate has contributions from the direct Urca and modified Urca processes. The direct Urca rate is calculated by evaluating the full phase space integral \cite{Alford:2021ogv, Alford:2023gxq} using the {\tt Vegas} Monte-Carlo integrator \cite{Lepage:2020tgj}. The modified Urca rates are calculated using standard approximate expressions \cite{Haensel:2001mw}. The code is about twice as fast as the Mathematica scripts from which it was developed.

\Cref{fig:flavor} shows an example of the output from the Flavor Equilibration Module. This figure shows the static bulk viscosity and the isothermal flavor relaxation time $\tau_I=1/\gamma$ \cref{eq:gamma-I} as a function of $n_B$ for the $\chi\text{EFT}$ EoSs with $\Lambda$=414 and 450 EoS and for the CMF++ C4 N (see discussion in Sec.~\ref{subsubsec:Results-beta}) EoS. 
Here we compare EoS from $\chi\text{EFT}$ and CMF++ that only include nucleons (the inclusion of other particles such as hyperons and quarks would require further adaptations to Flavor Eq.~, see, e.g.,~\cite{Alford:2020pld,Alford:2024tyj}).
For the $\chi\text{EFT}$ EoSs, we see that the static bulk viscosity drops to zero (for $\Lambda=414$) or a small value (for $\Lambda=450$) at the density where the $Y_Q$ approaches a stationary point with respect to density (see Fig.~\ref{fig:lepton}). This is because the static bulk viscosity is proportional to $dY_Q/dn_B$. Intuitively, if compression does not change the equilibrium proton fraction, then there is no re-equilibration and hence no bulk viscosity. The $Y_Q$  for CMF++\,C4\,N EoS, on the other hand, does not approach a stationary point with respect to density, and, therefore, its static bulk viscosity does not go to zero.

For the relaxation time, the dominant feature is the direct Urca threshold density. Below that threshold, Urca rates are slow, so the relaxation time is tens of milliseconds. Above the threshold, direct Urca becomes unsuppressed, and relaxation times are shorter (fraction of a millisecond). For muon-less matter, the direct Urca threshold is at proton fraction $Y_Q^\text{dU}=0.11$. We see from Fig.~\ref{fig:lepton} that, for the $\chi\text{EFT}$ EoSs, the proton fraction never reaches that threshold, so their relaxation time remains at around $40\,\text{ms}$. In contrast, the proton fraction of the CMF++\,C4\,N EoS crosses the direct Urca threshold at a number density around $0.3\,\text{fm}^{-3}$, so as we cross that density, the relaxation time drops to a much smaller value because direct Urca processes drive fast conversion between neutrons and protons to establish $\beta$ equilibrium quickly.

\section{Summary and Outlook}\label{sec:conclusions}

In this paper, we have studied neutron stars in and out of $\beta$ equilibrium and at approximately zero temperature with different nuclear physics descriptions of the 2D EoS at low (crustDFT in the crust), intermediate ($\chi\text{EFT}$ in the crust-outer core), and high baryon densities (CMF++ in the outer and inner cores). We have studied how thermodynamic and observable properties of neutron stars are affected by the choice of matching used to connect the different nuclear physics descriptions within their overlapping regimes of validity. 
For example, simple choices of smooth matching variable can easily introduce artificially-large structure in $c_s^2$, first-order phase transitions, or artificial regimes within the star that are unstable (with imaginary speeds of sound) or superluminal (with speeds of sound larger than $c$). 
We devised smooth matching procedures that ensure these artificial features are not present, at the cost of slightly modifying the combined EoS with respect to the parent EoSs outside the matched regime. Because of these slight modifications, observable properties of neutron stars can depend on the details of the matching. In particular, we find that the radius of neutron stars is the most affected observable; the maximum percentage variations are $\Delta M_{\ast \, \rm max} = 3.73$\% and $\Delta R_{\ast \, 1.4 M_\odot}=9.21$\%. Our results are generic for any combined EoS that is built by smoothly matching EoSs (built from different models) in an overlapping regime of validity.  

We also studied other physics features of our matched EoSs that are worth summarizing. While our focus has been on smooth matching between different EoS models, one can also build in first-order phase transitions (on purpose) through a Maxwell or Gibbs construction. Moreover, while we have described in detail only the calculation of the mass and radius of stars, we have also studied the effect of rotation with our matched EoSs. In particular, we have computed the moment of inertia, the quadrupole moment and the tidal Love number, verifying that the I-Love-Q relations remain approximately EoS independent (irrespective of the details of the original EoS models used or the smooth matching procedure). Finally, we have calculated the bulk viscosity in different regimes of the EoS, such as in the crust-outer core interface (with $\chi$EFT) and in the outer and inner cores (with CMF++). We found that the bulk viscosity is around $\zeta\sim\left[10^{24},10^{27}\right] {\rm{MeV}}^3$, but dependent on the parameters chosen within each model (e.g.,~the EFT cut-off scale $\Lambda$ in the $\chi$EFT case). 

In addition to these physics results, this paper also releases a suite of new, open-source software (the MUSES CE), consisting of a set of novel scientific calculation modules, an application for executing workflows composed of these modules, and a framework allowing the community to contribute in the future with new modules that extend its capabilities. 
Together, these software products provide a variety of models for the EoS applicable to heavy-ion collisions, a crust-to-core model of the neutron star EoS, and the calculation of neutron star observables.
We have presented the first results from combining different workflows using the MUSES CE, with a description of the physics equations and their software implementation underlying the various calculation modules. The physics results we obtained using this new software were described above. Computational notebooks showing how to use the CE to obtain the results of this article can be found in \cite{reinke_pelicer_2025_14841267}.

This paper paves the way for future statistical studies that will allow not only nuclear theorists but the entire neutron star community--including gravitational-wave physicists, astrophysicists, statisticians, and experimentalists--to directly sample over nuclear physics parameters instead of relying on phenomenological toy models of the EoS to infer information from gravitational-wave data and/or X-Ray observations of neutron stars. Future work could also concentrate on connecting different regimes of the QCD phase diagram using our open-source tools. 
While our primary focus here has been the $T=0$ EoS, we have already begun working on connecting heavy-ion collision modules (based on lattice QCD, holography, and the hadron resonance gas with Thermal-FIST \cite{Vovchenko:2019pjl}) to provide EoSs across a 4D phase space in $T,\mu_B,\mu_S,\mu_Q$ that will be needed in relativistic viscous hydrodynamics codes with BSQ conserved charges \cite{Plumberg:2024leb,Monnai:2024pvy,Karpenko:2013wva}. 
Additionally, we plan to also incorporate finite $T$ effects in all the EoS modules discussed in detail here (Crust-DFT, $\chi$EFT, CMF++, Lepton, Synthesis, and Flavor). 
Once completed, this will allow us to provide 3D EoS tables that can be used in numerical relativity simulations to study the post-merger dynamics, e.g.,~\cite{Most:2018eaw,Most:2022yhe}. 
Eventually, it will be possible to connect the EoSs from heavy-ion collisions and neutron stars, opening up the possibility for systematically constraining the dense matter EoS simultaneously using both heavy-ion flow data and neutron star observations (see a recent example \cite{Yao:2023yda}). 

\begin{acknowledgments}
We would like to thank the wider MUSES collaboration for many discussions during our collaboration meetings and all colleagues who helped with testing the MUSES CE.  
This work was supported in part by the National Science Foundation (NSF) within the framework of the MUSES collaboration, under grant number OAC-2103680.
This work used Jetstream2 at Indiana University and Open Storage Network at NCSA through allocation PHY230156 from the Advanced Cyberinfrastructure Coordination Ecosystem: Services \& Support (ACCESS) program \citep{NSFACCESS}, which is supported by National Science Foundation grants \#2138259, \#2138286, \#2138307, \#2137603, and \#2138296.
\end{acknowledgments}
\bibliography{inspire,not_inspire}

\appendix

\section{$\chi$EFT: Thermodynamic Relations}
\label{app:cheft}

Using many-body perturbation theory in the framework of the canonical ensemble results in a calculation of the free-energy per nucleon $\bar{F}(n_B,Y_Q)$. The remaining thermodynamic quantities necessary for an EoS can be determined from the standard thermodynamic relations,
\begin{align}\label{eq:cheft-app-thermo}
    P &= n_B^2\,\frac{\partial\bar{F}}{\partial n_B}\ ,\\
    \mu_B &= \frac{\partial\left(n_B\bar{F}\right)}{\partial n_B} - Y_{Q}\,\frac{\partial\bar{F}}{\partial Y_Q}\ ,\\
    \mu_Q &= \frac{\partial\bar{F}}{\partial Y_Q}\ ,
\end{align}
where $P(n_B,Y_Q)$, $\mu_B(n_B,Y_Q)$, $\mu_Q(n_B,Y_Q)$ are the pressure, baryon chemical potential, and charge chemical potential, respectively.

\section{Free Fermi Gas of Leptons}
\label{app1}

\begin{table*}[t!]
    \centering
    \begin{tabular}{l c c c c c c c c c c}\toprule
        {$Y(x)$} & {($\bar{x}, \Gamma$)} & {$M_{\ast \, \rm max}$} & {$\varepsilon_{\rm C \; Mmax}$} & {$R_{\ast \, 1.4 M_\odot}$} & {$\bar{I}_{1.4 M_\odot}$} & {$\bar{\lambda}^{\rm tid}_{1.4 M_\odot}$} & {$\Delta M_{\ast \, \rm max} (\%)$} & {$\Delta R_{\ast \, 1.4 M_\odot} (\%) $} &  {$\Delta \bar{I}_{1.4 M_\odot} (\%)$} & {$\Delta \bar{\lambda}^{\rm tid}_{1.4 M_\odot} (\%)$}  \\\midrule
	 $c_s^2$($n_B$)& (0.16, 0.025) & 2.061 & 1140.21 & 13.01 & 13.97 & 628.17 & 1.8 & 4.62 & 7.65 & 31.89 \\
	 & (0.22, 0.025) & 2.024 & 1273.60 & 12.43 & 12.98 & 476.28 & 1.8 & 4.62 & 7.65 & 31.89 \\\midrule
	 $\varepsilon$($n_B$)& (0.16, 0.025) & 2.100 & 1172.05 & 13.45 & 15.06 & 826.30 & 0.14 & 0.12 & 0.90 & 3.04 \\
	 & (0.22, 0.025) & 2.103 & 1172.05 & 13.44 & 15.20 & 852.22 & 0.14 & 0.12 & 0.90 & 3.04 \\\midrule
	 $P$($n_B$)& (0.16, 0.025) & 2.100 & 1172.05 & 13.62 & 15.12 & 838.38 & 0.35 & 3.57 & 4.52 & 17.75 \\
	 & (0.22, 0.025) & 2.093 & 1172.05 & 13.15 & 14.46 & 712.01 & 0.35 & 3.57 & 4.52 & 17.75 \\\midrule
	 $P$($\mu_B$)& (950.0, 30.0) & 2.102 & 1172.05 & 13.69 & 15.31 & 877.66 & 0.025 & 0.75 & 0.68 & 2.66 \\
	 & (980.0, 30.0) & 2.101 & 1172.05 & 13.59 & 15.20 & 854.94 & 0.025 & 0.75 & 0.68 & 2.66 \\
    \bottomrule
    \end{tabular}
    \caption{Matching variables, matching parameters, reproduced maximum masses and corresponding central energy densities, radii, dimensionless moment of inertia, and dimensionless tidal deformabilities for a 1.4 M$_\odot$ neutron star, calculated at the boundaries of the parameter sweep $\bar{x}=\Theta_a$ and $\bar{x}=\Theta_b$ shown in \Cref{fig:qlimr_sep} for each smooth matching variable. The table also shows the percentage differences in some of these quantities between the maximum and minimum values obtained within the parameter sweep.}
    \label{tab:summary}  
\end{table*}

For a relativistic free Fermi gas at $T=0$, we can write the number density of lepton $i$ in terms of the Fermi momentum $k_{F,i}$,
\begin{equation}
    n_i  = \frac{d_i k_{F,i}^3}{6 \pi^2} = \frac{k_{F,l}^3}{3 \pi^2}\ ,
\end{equation}
where the spin degeneracy for the fermions is $d_{i}=2J_{i}+1$, with the spin of all leptons $J_{i}=1/2$, such that $d_{i}=2$. At $T=0$, the Fermi energy $E_{F,i}$ of the lepton is identical to its chemical potential $\mu_i$ and we can write
\begin{equation}
    E_{F,i}  = \sqrt{m_i^2+k_{F,i}^2}=\mu_i\ ,
\end{equation}
with $m_i$ being the lepton mass.
Thus, we can then rewrite everything in terms of the lepton chemical potential, such that
\begin{equation}
    k_{F,i}=\sqrt{\mu_i^2-m_i^2} \ ,
\end{equation}
and
\begin{equation}
    n_i=\frac{(\mu_i^2-m_i^2)^{3/2}}{3 \pi^2} \ .
    \label{nlep}
\end{equation}
Finally, we can then write down the corresponding energy density $\varepsilon_l$ and pressure $p_l$
\begin{align}
\varepsilon_i&=\frac{1}{\pi^2}\Bigg[\left(\frac{1}{8}m_i^2\sqrt{\mu_i^2-m_i^2}+\frac{1}{4}(\mu_i^2-m_i^2)^{3/2}\right)\mu_i
\nonumber\\
&-\frac{1}{8}m_i^4\ln{\frac{\sqrt{\mu_i^2-m_i^2}+\mu_i}{m_i}}\Bigg]\ ,
\nonumber\\
p_i&=\frac{1}{3\pi^2}\Bigg[\left(\frac{1}{4}(\mu_i^2-m_i^2)^{3/2}-\frac{3}{8}m_i^2\sqrt{\mu_i^2-m_i^2}\right)\mu_i
\nonumber\\
&+\frac{3}{8}m_i^4\ln{\frac{\sqrt{\mu_i^2-m_i^2}+\mu_i}{m_i}}\Bigg]\ ,
\label{eqn:ideal}
\end{align}
where $l=e^-,\mu^-,\tau^-$, $\nu_l$. 

\section{Effect of smooth matching on neutron star observables}
\label{appQLIMR}

\Cref{tab:summary} shows values for non-rotating neutron star maximum masses and their corresponding central energy densities, together with radii, dimensionless moment of inertia and dimensionless tidal deformability of an $1.4$ M$_\odot$ star for the boundaries of the parameter sweep ($\bar{x}=\theta_{a}$, $\bar{x}=\theta_{b}$) shown in the different panels of \Cref{fig:qlimr_sep}. 
The table also shows the difference in these quantities between the maximum and minimum values obtained in the parameter sweep.
These differences are calculated for quantity $A$ as in \Cref{AAA}.

\end{document}